\documentclass[longauth]{aa}  
\usepackage{graphicx}
\usepackage{txfonts}
\usepackage{hyperref}
\usepackage{makecell}
\usepackage{natbib}
\usepackage{lscape}
\usepackage{multirow}
\usepackage{enumitem}
\usepackage{subcaption}
\usepackage{xcolor}
\usepackage{placeins}
\usepackage[switch]{lineno}

\hypersetup{
    colorlinks=true,
    linkcolor=blue,
    filecolor=blue,      
    urlcolor=blue,
    citecolor=blue
    }

\newcommand{\linkorcid}[1]{\href{https://orcid.org/#1}{\includegraphics[width=8pt]{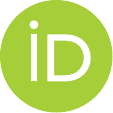}}}

\begin{document}

   \title{Candidate strongly lensed type Ia supernovae in the Zwicky Transient Facility archive}

   \titlerunning{Candidate strongly lensed type Ia supernovae in the ZTF archive}
   \authorrunning{A. Townsend et al. }

   \author{A.~Townsend\inst{\ref{hu}}\fnmsep\thanks{Email: alice.townsend@physik.hu-berlin.de}\linkorcid{0000-0001-6343-3362} \and
   J.~Nordin\inst{\ref{hu}}\linkorcid{0000-0001-8342-6274} \and
   A.~Sagu\'es~Carracedo\inst{\ref{okc}}\linkorcid{0000-0002-3498-2167} \and
   M.~Kowalski\inst{\ref{hu},\ref{desy}}\linkorcid{0000-0001-8594-8666} \and
   N.~Arendse\inst{\ref{okc}}\linkorcid{0000-0001-5409-6480} \and
   S.~Dhawan\inst{\ref{camb}}\linkorcid{0000-0002-2376-6979} \and
   A.~Goobar\inst{\ref{okc}}\linkorcid{0000-0002-4163-4996} \and
   J.~Johansson\inst{\ref{okc}}\linkorcid{0000-0001-5975-290X} \and
   E.~M\"ortsell\inst{\ref{okc}}\linkorcid{0000-0002-8380-6143} \and
   S.~Schulze\inst{\ref{northwestern}}\linkorcid{0000-0001-6797-1889} \and
   I.~Andreoni\inst{\ref{spacesci_maryland},\ref{uni_maryland},\ref{nasa_goddard}}\linkorcid{0000-0002-8977-1498} \and
   E.~Fern\'andez\inst{\ref{csic_spain}} \and
   A.~G.~Kim\inst{\ref{law_berkeley}}\linkorcid{0000-0001-6315-8743} \and
   P.~E.~Nugent\inst{\ref{law_berkeley},\ref{cal_berkeley}}\linkorcid{0000-0002-3389-0586} \and
   F.~Prada\inst{\ref{csic_spain}}\linkorcid{0000-0001-7145-8674} \and
   M.~Rigault\inst{\ref{lyon}}\linkorcid{0000-0002-8121-2560} \and
   N.~Sarin\inst{\ref{okc},\ref{nordita}}\linkorcid{0000-0003-2700-1030} \and
   D.~Sharma\inst{\ref{cal_berkeley}} \and
   E.~C.~Bellm\inst{\ref{dirac}}\linkorcid{0000-0001-8018-5348} \and
   M.~W.~Coughlin\inst{\ref{minnesota}}\linkorcid{0000-0002-8262-2924}\and
   R.~Dekany\inst{\ref{caltech_observ}}\linkorcid{0000-0002-5884-7867}\and
   S.~L.~Groom\inst{\ref{ipac}}\linkorcid{0000-0001-5668-3507} \and
   L.~Lacroix\inst{\ref{okc},\ref{lpnhe}}\linkorcid{0000-0003-0629-5746} \and
   R.~R.~Laher\inst{\ref{ipac}}\linkorcid{0000-0003-2451-5482} \and
   R.~Riddle\inst{\ref{ipac}}\linkorcid{0000-0002-0387-370X} \and
   J.~Aguilar\inst{\ref{law_berkeley}} \and
   S.~Ahlen\inst{\ref{boston}}\linkorcid{0000-0001-6098-7247} \and
   S.~Bailey\inst{\ref{law_berkeley}}\linkorcid{0000-0003-4162-6619} \and
   D.~Brooks\inst{\ref{ucl}} \and
   T.~Claybaugh\inst{\ref{cal_berkeley}} \and
   A.~de la Macorra\inst{\ref{mexico}}\linkorcid{0000-0002-1769-1640} \and
   A.~Dey\inst{\ref{noirlab}}\linkorcid{0000-0002-4928-4003} \and
   B.~Dey\inst{\ref{pitts}}\linkorcid{0000-0002-5665-7912} \and
   P.~Doel\inst{\ref{ucl}} \and
   K.~Fanning\inst{\ref{kavli},\ref{slac}}\linkorcid{0000-0003-2371-3356} \and
   J.~E.~Forero-Romero\inst{\ref{colombia1}.\ref{colombia2}}\linkorcid{0000-0002-2890-3725}
   E.~Gazta{\~n}aga\inst{\ref{ieec_spain},\ref{portsmouth},\ref{icecsic_spain}} \and
   S.~Gontcho~A~Gontcho\inst{\ref{law_berkeley}}\linkorcid{0000-0003-3142-233X} \and
   K.~Honscheid\inst{\ref{cosmo_ohio},\ref{phys_ohio},\ref{ohio}} \and
   C.~Howlett\inst{\ref{queensland}}\linkorcid{0000-0002-1081-9410} \and
   T.~Kisner\inst{\ref{law_berkeley}}\linkorcid{0000-0003-3510-7134} \and
   A.~Kremin\inst{\ref{law_berkeley}}\linkorcid{0000-0001-6356-7424} \and
   A.~Lambert\inst{\ref{law_berkeley}} \and
   M.~Landriau\inst{\ref{law_berkeley}}\linkorcid{0000-0003-1838-8528} \and
   L.~Le~Guillou\inst{\ref{lpnhe}}\linkorcid{0000-0001-7178-8868} \and
   M.~E.~Levi\inst{\ref{law_berkeley}}\linkorcid{0000-0003-1887-1018} \and
   M.~Manera\inst{\ref{physic_barcelona},\ref{ifae_spain}}\linkorcid{0000-0003-4962-8934} \and
   A.~Meisner\inst{\ref{noirlab}}\linkorcid{0000-0002-1125-7384} \and
   R.~Miquel\inst{\ref{ifae_spain},\ref{catalana_barcelona}} \and
   J.~Moustakas\inst{\ref{loudonville}}\linkorcid{0000-0002-2733-4559} \and
   E.~Mueller\inst{\ref{brighton}} \and
   A.~D.~Myers\inst{\ref{wyoming}} \and
   J.~Nie\inst{\ref{china}}\linkorcid{0000-0001-6590-8122} \and
   N.~Palanque-Delabrouille\inst{\ref{law_berkeley},\ref{irfu}}\linkorcid{0000-0003-3188-784X} \and
   C.~Poppett\inst{\ref{law_berkeley},\ref{cal_berkeley},\ref{space_berkeley}} \and
   M.~Rezaie\inst{\ref{kansas}}\linkorcid{0000-0001-5589-7116} \and
   G.~Rossi\inst{\ref{korea}} \and
   E.~Sanchez\inst{\ref{ciemat_spain}}\linkorcid{0000-0002-9646-8198} \and
   D.~Schlegel\inst{\ref{law_berkeley}} \and
   M.~Schubnell\inst{\ref{phys_michigan},\ref{michigan}} \and
   H.~Seo\inst{\ref{uniOFohio}}\linkorcid{0000-0002-6588-3508} \and
   D.~Sprayberry\inst{\ref{noirlab}} \and
   G.~Tarl\'{e}\inst{\ref{michigan}}\linkorcid{0000-0003-1704-0781} \and
   H.~Zou\inst{\ref{china}}\linkorcid{0000-0002-6684-3997}
          }

   \institute{Institut f\"ur Physik, Humboldt-Universit\"at zu Berlin, Newtonstr. 15, D-12489 Berlin, Germany \label{hu}
         \and
         Oskar Klein Centre, Department of Physics, Stockholm University, Albanova University Center, SE 106 91 Stockholm, Sweden \label{okc}
         \and
         Deutsches Elektronen-Synchrotron, D-15735 Zeuthen, Germany \label{desy}
         \and
         Institute of Astronomy and Kavli Institute for Cosmology, University of Cambridge, Madingley Road, Cambridge CB3 0HA, UK \label{camb}
         \and
         Center for Interdisciplinary Exploration and Research in Astrophysics (CIERA), Northwestern University, 1800 Sherman Ave., Evanston, IL 60201, USA \label{northwestern}
         \and
         Joint Space-Science Institute, University of Maryland, College Park, MD 20742, USA \label{spacesci_maryland}
         \and
         Department of Astronomy, University of Maryland, College Park, MD 20742, USA \label{uni_maryland}
         \and
         Astrophysics Science Division, NASA Goddard Space Flight Center, Mail Code 661, Greenbelt, MD 20771, USA \label{nasa_goddard}
         \and
         Instituto de Astrof\'{i}sica de Andaluc\'{i}a (CSIC), Glorieta de la Astronom\'{i}a, s/n, E-18008 Granada, Spain \label{csic_spain}
         \and
         Lawrence Berkeley National Laboratory, 1 Cyclotron Road, Berkeley, CA 94720, USA \label{law_berkeley}
         \and
         Department of Astronomy, University of California, Berkeley, 501 Campbell Hall, Berkeley, CA 94720, USA \label{cal_berkeley}
         \and
         Univ Lyon, Univ Claude Bernard Lyon 1, CNRS, IP2I Lyon/IN2P3, UMR 5822, F-69622, Villeurbanne, France \label{lyon}
         \and
         Nordita, Stockholm University and KTH Royal Institute of Technology, Hannes Alfv\'ens v\"ag 12, SE-106 91 Stockholm, Sweden \label{nordita}
         \and
         DIRAC Institute, Department of Astronomy, University of Washington, 3910 15th Avenue NE, Seattle, WA 98195, USA \label{dirac}
         \and
         School of Physics and Astronomy, University of Minnesota, Minneapolis, Minnesota 55455, USA \label{minnesota}
         \and
         Caltech Optical Observatories, California Institute of Technology, Pasadena, CA 91125, USA \label{caltech_observ}
         \and
         IPAC, California Institute of Technology, 1200 E. California Boulevard, Pasadena, CA 91125, USA \label{ipac}
         \and 
         LPNHE, CNRS/IN2P3, Sorbonne Université, Laboratoire de Physique Nucléaire et de Hautes Énergies, F-75005 Paris, France\label{lpnhe}
         \and
         Physics Dept., Boston University, 590 Commonwealth Avenue, Boston, MA 02215, USA \label{boston}
         \and
         Department of Physics \& Astronomy, University College London, Gower Street, London, WC1E 6BT, UK \label{ucl}
         \and
         Instituto de F\'{\i}sica, Universidad Nacional Aut\'{o}noma de M\'{e}xico,  Cd. de M\'{e}xico  C.P. 04510,  M\'{e}xico \label{mexico}
         \and
         NSF's NOIRLab, 950 N. Cherry Ave., Tucson, AZ 85719, USA \label{noirlab}
         \and
         Department of Physics \& Astronomy and Pittsburgh Particle Physics, Astrophysics, and Cosmology Center (PITT PACC), University of Pittsburgh, 3941 O'Hara Street, Pittsburgh, PA 15260, USA \label{pitts}
         \and
         Kavli Institute for Particle Astrophysics and Cosmology, Stanford University, Menlo Park, CA 94305, USA \label{kavli}
         \and
         SLAC National Accelerator Laboratory, Menlo Park, CA 94305, USA \label{slac}
         \and
         Departamento de F\'isica, Universidad de los Andes, Cra. 1 No. 18A-10, Edificio Ip, CP 111711, Bogot\'a, Colombia \label{colombia1}
         \and
         Observatorio Astron\'omico, Universidad de los Andes, Cra. 1 No. 18A-10, Edificio H, CP 111711 Bogot\'a, Colombia \label{colombia2}
         \and
         Institut d'Estudis Espacials de Catalunya (IEEC), 08034 Barcelona, Spain \label{ieec_spain}
         \and
         Institute of Cosmology and Gravitation, University of Portsmouth, Dennis Sciama Building, Portsmouth, PO1 3FX, UK \label{portsmouth}
         \and
         Institute of Space Sciences, ICE-CSIC, Campus UAB, Carrer de Can Magrans s/n, 08913 Bellaterra, Barcelona, Spain \label{icecsic_spain}
         \and
         Center for Cosmology and AstroParticle Physics, Ohio State University, 191 West Woodruff Avenue, Columbus, OH 43210, USA \label{cosmo_ohio}
         \and
         Department of Physics, Ohio State University, 191 West Woodruff Avenue, Columbus, OH 43210, USA \label{phys_ohio}
         \and
         The Ohio State University, Columbus, 43210 OH, USA \label{ohio}
         \and
         School of Mathematics and Physics, University of Queensland, 4072, Australia \label{queensland}
         \and
         Departament de F\'{i}sica, Serra H\'{u}nter, Universitat Aut\`{o}noma de Barcelona, 08193 Bellaterra, Barcelona, Spain \label{physic_barcelona}
         \and
         Institut de F\'{i}sica d’Altes Energies (IFAE), Barcelona Institute of Science and Technology, Campus UAB, 08193 Bellaterra, Barcelona, Spain \label{ifae_spain}
         \and
         Instituci\'{o} Catalana de Recerca i Estudis Avan\c{c}ats, Passeig de Llu\'{\i}s Companys, 23, 08010 Barcelona, Spain \label{catalana_barcelona}
         \and
         Department of Physics and Astronomy, Siena College, 515 Loudon Road, Loudonville, NY 12211, USA \label{loudonville}
         \and
         Department of Physics and Astronomy, University of Sussex, Brighton BN1 9QH, UK \label{brighton}
         \and
         Department of Physics \& Astronomy, University  of Wyoming, 1000 E. University, Dept.~3905, Laramie, WY 82071, USA \label{wyoming}
         \and
         National Astronomical Observatories, Chinese Academy of Sciences, A20 Datun Rd., Chaoyang District, Beijing, 100012, P.R. China \label{china}
         \and
         IRFU, CEA, Universit\'{e} Paris-Saclay, F-91191 Gif-sur-Yvette, France \label{irfu}
         \and
         Space Sciences Laboratory, University of California, Berkeley, 7 Gauss Way, Berkeley, CA 94720, USA \label{space_berkeley}
         \and
         Department of Physics, Kansas State University, 116 Cardwell Hall, Manhattan, KS 66506, USA \label{kansas}
         \and
         Department of Physics and Astronomy, Sejong University, Seoul, 143-747, Korea \label{korea}
         \and
         CIEMAT, Avenida Complutense 40, E-28040 Madrid, Spain \label{ciemat_spain}
         \and
         Department of Physics, University of Michigan, Ann Arbor, MI 48109, USA \label{phys_michigan}
         \and
         University of Michigan, Ann Arbor, MI 48109, USA \label{michigan}
         \and
         Department of Physics \& Astronomy, Ohio University, Athens, OH 45701, USA \label{uniOFohio}
             }

   \date{Received 12 June, 2024 / Accepted 1 December 2024}
 
  \abstract
   {Gravitationally lensed type Ia supernovae (glSNe Ia) are unique astronomical tools that can be used to study cosmological parameters, distributions of dark matter, the astrophysics of the supernovae, and the intervening lensing galaxies themselves. A small number of highly magnified glSNe Ia have been discovered by ground-based telescopes such as the Zwicky Transient Facility (ZTF), but simulations predict that a fainter, undetected population may also exist. }
   {We present a systematic search for glSNe Ia in the ZTF archive of alerts distributed from June 1 2019 to September 1 2022.}
   {Using the AMPEL platform, we developed a pipeline that distinguishes candidate glSNe Ia from other variable sources. Initial cuts were applied to the ZTF alert photometry (with constraints on the peak absolute magnitude and the distance to a catalogue-matched galaxy, as examples) before forced photometry was obtained for the remaining candidates. Additional cuts were applied to refine the candidates based on their light curve colours, lens galaxy colours, and the resulting parameters from fits to the SALT2 SN Ia template. The candidates were also cross-matched with the DESI spectroscopic catalogue.}
   {Seven transients were identified that passed all the cuts and had an associated galaxy DESI redshift, which we present as glSN Ia candidates. Although superluminous supernovae (SLSNe) cannot be fully rejected as contaminants, two events, ZTF19abpjicm and ZTF22aahmovu, are significantly different from typical SLSNe and their light curves can be modelled as two-image glSN Ia systems. From this two-image modelling, we estimate time delays of $22\pm3$ and $34\pm1$ days for the two events, respectively, which suggests that we have uncovered a population of glSNe Ia with longer time delays.}
  %
   {The pipeline is efficient and sensitive enough to parse full alert streams. It is currently being applied to the live ZTF alert stream to identify and follow-up future candidates while active. This pipeline could be the foundation for glSNe Ia searches in future surveys, such as the Rubin Observatory Legacy Survey of Space and Time.}

   \keywords{supernovae: general --
                gravitational lensing: strong --
                methods: observational -- techniques: photometric
               }

   \maketitle
%

\section{Introduction}
\label{section:intro}
    Strong gravitational lensing is a consequence of general relativity, where light from a distant source is deflected and magnified due to the mass of an intervening object. The intervening massive object is the lens, and it acts like a gravitational telescope to observe the high-redshift universe. Over the last few decades, there have been many observations of strongly lensed galaxies and quasars, but a rapidly developing area of research is the observation of lensed transients. Type Ia supernovae (SNe Ia), the explosions of white dwarf stars in binary star systems, are bright, standardisable candles. Therefore, strongly lensed type Ia supernovae (glSNe Ia) are extremely valuable astrophysical tools.

    In terms of supernova physics, glSNe Ia allow us to probe distant SNe Ia and study their redshift evolution \citep{Petrushevska2017,Johansson2021}. This is important for verifying that high-redshift SNe Ia are standardisable and can be included in the Hubble diagram. In terms of lensing physics, glSNe Ia can be used to determine the matter distribution in lensing systems. Furthermore, from the recent discovery of SN Zwicky \citep{Goobar2023}, the authors suggested that glSNe Ia could be vital for the study of previously unknown compact lens systems. In terms of cosmology, glSNe Ia can be utilised in the method of time-delay cosmography to measure the Hubble constant ($H_0$), which was first proposed by \citet{Refsdal1964}. Strong lensing produces multiple images of the supernova explosion, which travel different paths with different gravitational potentials to reach the observer, leading to time delays between the images. From measurements of the time delays and a model of the lens galaxy's gravitational potential, an absolute distance (known as the ‘time-delay distance’) for the system can be determined. This distance provides constraints on $H_0$ and, to a lesser degree, the mass density parameter, $\Omega_M$, and the dark energy equation of state, $w(z)$ \citep{Goobar2002}.
    
    The method of time-delay cosmography has previously been employed with quasar sources \citep[e.g.][]{Wong2020,Shajib2020,Millon2020}. One of the benefits of using time-delay cosmography with glSNe Ia is that SNe Ia are standardisable candles, with a known peak absolute magnitude that provides absolute magnification constraints. This allows us to constrain the mass-sheet degeneracy \citep{Falco1985}; a systematic effect whereby adding a constant surface mass density to the lens system results in a degeneracy in which all of the lensing observables remain unchanged, except for the time delays. This degeneracy can be broken by knowledge of the unlensed apparent brightness \citep{Birrer2022}. Another benefit is that SNe Ia have predictable light curve shapes with a duration of $\sim 30$ days, whereas the timescale required to monitor time delays from lensed quasars is several years due to their stochastic variability. Additionally, the host and lens galaxy can be studied without any contamination from the supernova once it has faded, which is not always possible for lensed quasars.

    At the time of publication, only three glSNe Ia have been detected with ground-based telescopes: PS1-10afx with the Panoramic Survey Telescope and Rapid Response System \citep[Pan-STARRS;][]{Chornock2013,Quimby2013}, iPTF16geu with the Intermediate Palomar Transient Factory \citep[iPTF;][]{Goobar2017}, and SN Zwicky with the Zwicky Transient Facility \citep[ZTF;][]{Goobar2023, Pierel2023}. All of these events had a high magnification (with a total magnification factor, $\mu$, of greater than $20$), allowing them to be followed up and classified by surveys that obtain spectra of nearby transients. This implies that a larger sample of less magnified glSNe Ia could exist, which may have already been observed by surveys such as ZTF, but remain undiscovered because they would not be targeted by spectroscopic follow-up programmes. Additionally, with upcoming large-scale optical surveys such as the Vera C. Rubin Observatory's Legacy Survey of Space and Time \citep[LSST;][]{Ivezic2019}, it is necessary for us to develop a systematic approach for detecting all glSNe Ia, instead of discovering only the extremely luminous objects. Most transients found by LSST will be too faint for spectroscopy, even with $8\,\mathrm{m}$-class telescopes. Thus, with the lack of available spectroscopic follow-up resources, it would be valuable to devise a method to identify candidate glSNe Ia before a spectrum of the transient is obtained.

    In this work, we present a systematic search of the ZTF archive for glSNe Ia. We developed a pipeline that is capable of filtering alerts and distinguishing candidate glSNe Ia from other variable sources. Initial cuts were applied to the ZTF alert photometry (e.g. constraints on peak absolute magnitude) before forced photometry was obtained for the remaining candidates. Additional cuts were applied to the remaining candidates from their light curve colours, lens galaxy colours, and parameters from light curve fits. We also cross-matched the candidates to the Dark Energy Spectroscopic Instrument (DESI) survey catalogue.

    The structure of this article is as follows. Section \ref{section:detecting} discusses the methods previously employed to search for lensed supernovae, how these methods were used as guidelines when implementing the search pipeline, and predictions for what we should expect to find in the ZTF archive. The data from ZTF used in this analysis is summarised in Sect. \ref{subsection:ztf}. Section \ref{subsection:ampel} outlines the real-time alert processing platform AMPEL, which allowed us to analyse the ZTF data and structure our pipeline. In Sect. \ref{section:method}, we describe our method to identify probable glSNe Ia, starting with the initial alert pipeline described in Sect. \ref{subsection:pipeline}. The candidates remaining after the initial selection method were analysed in parallel in Sect. \ref{subsection:DESI}, in which we cross-match to the DESI spectroscopic catalogue, and Sect. \ref{subsection:cuts}, in which we continue to filter the candidates based on more restrictive cuts. Section \ref{subsection:candidates} summarises the sample of candidates that were selected by the two methods. In Sect. \ref{section:comparison}, we report on the properties of this sample by discussing the possible contaminants (Sect. \ref{subsection:contamination}) and by comparing to the Bright Transient Survey (Sect. \ref{subsection:bts}) and to previously observed glSNe Ia (Sect. \ref{subsection:sample}). Our seven most likely candidates (labelled the `gold sample') are analysed in more detail in Section \ref{section:gold}. The two objects we present as likely glSNe Ia, ZTF19abpjicm and ZTF22aahmovu, are examined in Sects. \ref{subsubsection:ZTF19abpjicm} and \ref{subsubsection:ZTF22aahmovu}, respectively. Finally, in Sect. \ref{section:conc}, we elucidate the findings of our study and what further work is required to confirm that our candidates are truly glSNe Ia.
    
\section{Strategy for detecting lensed supernovae}
\label{section:detecting}
   The strategy for finding lensed supernovae depends on whether the instrument used is capable of resolving the individual images. Lensed supernovae discovered from spatially resolved data include SN Refsdal \citep[type II,][]{Kelly2015} and SN Requiem \citep[likely a type Ia, but unconfirmed,][]{Rodney2021}, which were recovered from Hubble Space Telescope (HST) data that distinguished the separate images of each SN. More recently, the James Webb Space Telescope (JWST) discovered the triply imaged SN Ia H0pe \citep{Frye2023} and the lensed SN Ia Encore \citep[found in the same host galaxy as SN Requiem;][]{Pierel2024}.
   
   To detect unresolved lensed supernovae, a variety of strategies can be employed. \citet{Goldstein2017} put forward the idea of searching for magnified sources (with a peak $B$-band absolute magnitude greater than the typical $-19.4 \: \mathrm{mag}$ for SNe Ia) close to elliptical galaxies. Their reasoning is that normal SNe Ia are the brightest objects in quiescent galaxies, so there is a high likelihood that the magnified object is a lensed transient from a background galaxy. \citet{Wojtak2019} calculated glSNe discovery rates as a function of survey depth and found that shallower, pre-LSST surveys such as ZTF would not be able to detect multiple images from glSNe Ia and the only viable method would be to find magnified sources.
   
   Other proposed methods include detection from light curve information alone \citep{Bag2021,Denissenya2022}, monitoring known galaxy-galaxy strong-lens systems \citep{Shu2018,Craig2024}, or known cluster systems \citep{Saini2000,Sullivan2000}. However, these methods are not utilised in our following analysis. Firstly, we would struggle to characterise glSNe Ia from light curves alone due to occasional gaps in the light curve and limited filters (usually the $g$- and $r$-bands are available, whereas the $i$-band is limited in ZTF), so instead we must combine the parameters we extract from the light curve with additional information (e.g. from a catalogue-matched lens or host galaxy). Secondly, the three glSNe Ia discovered by ground-based telescopes had lens galaxies that were previously unknown, as they were compact (with an Einstein radius, $\theta_E$, of less than 0.5 arcseconds) and could not be detected from ground-based galaxy surveys. Therefore, monitoring known galaxy-galaxy strong-lens systems might not be the optimal search method as it would bias the sample to massive lenses. Thirdly, monitoring cluster systems with shallow ground-based surveys has so far only been successful with the discovery of weakly lensed supernovae \citep[e.g.][]{Goobar2009,Patel2014, Nordin2014,Rodney2015,Petrushevska2016}, which are not the focus of this work.
   
   \citet{Quimby2014} were the first to suggest using light curve colours as a metric to distinguish lensed supernovae from unlensed. Specifically, they proposed that the observed $r-i$ versus $i$-band distribution of unlensed SNe during the pre-peak phase differed significantly from the colours of PS1-10afx (the glSNe Ia detected by Pan-STARRS). We expect glSNe to be observed at higher redshifts (due to the probability of lensing and the number of SNe increasing with redshift) and, as a result, they will be redder than nearby unlensed SNe at similar apparent magnitudes \citep{Quimby2014}. This concept is also demonstrated in simulations by \citet{SaguesCarracedo2024}, which we shall hereafter refer to as SC24. By generating realistic light curves from SNe Ia templates, the authors of SC24 show how the distributions of $g-r$, $g-i$, and $r-i$ at different light curve epochs differ for lensed and unlensed SNe Ia.
   
   In SC24, the authors examine the sample of glSNe Ia that would have been detected by ZTF specifically, by incorporating the observing logs for ZTF, so the results can be directly applied to our systematic search. The findings of SC24 formed a part of our pipeline; in particular, the light curve colours discussed above, but also the simulated distributions of host and lens redshifts. From the distributions of the host and lens galaxy redshifts in SC24, we do not expect to find any glSNe Ia in host galaxies in the range $z <0.1$. Additionally, less than 3\% of identifiable glSNe have lens galaxies in the range $z<0.1$.
   
   SC24 also performed SALT2 light curve fits to simulated glSNe Ia light curves. The SALT2 model is a light curve template for SNe Ia based on two observable parameters: the stretch, $x_1$, and the colour, $c$. This model is based on empirical observations, due to the current lack of understanding about SN Ia progenitor explosions \citep{Guy2007}. For typical SNe Ia (e.g., the ones that are used in cosmological analyses), we expect $x_1 < \left| 3 \right|$ and $c < \left| 0.3 \right|$. As is shown in SC24, simulated glSNe Ia display a wider distribution of stretch values (approximately $-3 \lessapprox x_1 \lessapprox 10$) and are redder in colour ($c\gtrapprox 0$). They demonstrate that a cut of $c>0$ removes approximately half of the unlensed SN Ia sample and most of the lensed core-collapse supernova contaminants, while also retaining 90\% of the glSN Ia sample. SC24 also determined the peak $B$-band absolute magnitude distribution from the SALT2 fits, assuming the redshift of the lens galaxy is the supernova redshift (the motivation behind this assumption will be explained in Sect. \ref{subsection:pipeline}). They show that we can exclude almost all unlensed SNe Ia with a cut of $M_B> -20\: \mathrm{mag}$, although this removes approximately 20\% of the glSN Ia sample.
   
   In addition to the simulations of SC24, our pipeline utilised a relationship between $x_1$ and peak brightness for glSNe Ia. There are two effects that contributed to the width and peak brightness of the observed light curves in our study. The first effect is due to the superposition of the unresolved images of the lensed transient. We expect that objects with shorter time delays will have narrower unresolved light curves with greater peak brightness because the peak values of the individual images will be observed at approximately the same observation epoch. Conversely, we expect that objects with longer time delays will have wider unresolved light curves with lower peak brightness.\footnote{We note that there could be a degeneracy in the case of, for example, glSNe Ia with intrinsically large $x_1$ (i.e., the $x_1$ of each individual image) and small time delays, resulting in a wider unresolved light curve. For simplicity, we do not account for this degeneracy.} The second effect is due to lensing in the system causing an increase in flux, which is known as magnification. In a strong lensing system, the time delays and magnifications are negatively correlated. The combination of these two effects results in a trend in which unresolved glSN Ia light curves with a smaller $x_1$ value will have a larger peak absolute brightness. We implemented a cut for this relationship in our methodology to be sensitive to objects with short and long time delays.

   The most challenging part of an archival study is estimating the true brightness of the candidates in the sample. Our analysis pipeline allowed us to process and filter the entire ZTF alert stream, which was necessary to consider faint, unclassified transients that have not been studied previously. Because we did not have spectra for these unclassified transients, we utilised the redshifts from catalogue-matched nearby galaxies (which, in the case of an actual lensed supernova, would likely be the lensing galaxy). Current spectroscopic surveys are limited to brighter, lower redshift galaxies, meaning that we rely on photometric redshift estimates for distant galaxies, which have larger uncertainties but are more complete.
   
   In a previous work by \citet{Magee2023}, the authors performed a similar archival search for lensed supernovae from the sample of transients that were observed by ZTF and reported to the Transient Name Server (TNS). However, this public sample was limited to brighter objects that would have been reported by the ZTF brokers, and the analysis of the transients did not utilise ZTF forced photometry to study objects at deeper magnitudes. The methods of \citet{Magee2023} included cross-matching with known lens systems, applying colour cuts to identify red objects as in \citet{Quimby2014}, selecting light curves that are inconsistent with SN Ia near elliptical galaxies as is suggested in \citet{Goldstein2017}, and identifying intrinsically luminous objects according to the host redshift. In this work, we focus on two of these methods: applying colour cuts and selecting for intrinsically luminous objects. \citet{Magee2023} identified 132 unique transients across their selection methods, but all of their candidates are consistent with non-lensed transients (and more than half are already spectroscopically classified). We find no overlap between their 132 candidates and our shortlisted candidates, which is likely due to their limited sample selected from public brokers.

   Several studies have forecasted the rates of gravitationally lensed supernovae that ZTF would find per year. For example, in a study by \citet{Goldstein2019}, the authors predicted a discovery rate of 1.23 glSNe Ia per year and at least 7.37 other types of supernovae per year. Therefore, across the 3.25 years in this archival search, we expect there to be approximately 28 glSNe in our sample, with four of those being SNe Ia. However, \citet{Goldstein2019} assumed that observations taken in the same filter in a single night would be stacked, while our analysis uses forced photometry data instead. Additionally, \citet{Wojtak2019} predicted that ZTF would observe approximately two glSNe Ia per year (that is, there should be six or seven total objects in our sample). More conservatively, \citet{SainzdeMurieta2023} forecasted that ZTF should have observed approximately 0.2 glSNe Ia per year, which is consistent with SN Zwicky being the only discovery. However, when incorporating the ZTF observing logs, SC24 predicted that ZTF would detect glSNe Ia with at least five 5-sigma detections around the peak at a rate of approximately four per year \citep{Goobar2024}. After requiring that the object is measured as over-luminous (assuming the object is at the lens redshift), the simulations of SC24 predict that approximately 1.4 glSNe Ia per year will remain. As a result of this study, we can expect to find up to four archival candidates in our analysis.

\section{Analysing ZTF data using AMPEL}
\label{section:data}

\subsection{The ZTF archive}
\label{subsection:ztf}
    The Zwicky Transient Facility is a wide-field, optical survey for time-domain astronomy \citep{Bellm2019a,Graham2019}. The survey utilises a camera with a field of view of 47 degrees squared on the Palomar 48-inch (P48) Telescope that observes in three filters ($g$-, $r$- and $i$-bands) \citep{Dekany2020}. The observation plan consists of a public survey in the $g$- and $r$-bands that scans the northern sky every two days and a partnership survey with a higher cadence and additional $i$-band photometry over a smaller region of the sky \citep{Bellm2019b}. In this study, we had access to both public and partnership data. On average, ZTF produces between 600,000 and 1.2 million alerts on a complete night of observation \citep{Masci2019,Patterson2019}.
    
    Owing to the rapid scanning and extensive sky coverage of the survey, ZTF is an excellent precursor to LSST and is capable of detecting rare transients such as lensed supernovae. The Bright Transient Survey \citep[BTS,][]{Fremling2020,Perley2020} is a spectroscopic survey within ZTF that aims to acquire spectra of all transients brighter than 18.5 magnitudes, which constitute the vast majority of classified objects in ZTF. Previously observed glSNe Ia may be unclassified in the ZTF archive if they were significantly less magnified than discoveries such as SN Zwicky. Therefore, a systematic search of the ZTF archive is a promising avenue to identify less magnified glSNe Ia.
    
    In this analysis, we have demonstrated a method to systematically identify probable unresolved glSNe Ia, by reducing a total of $98\,567\,163$ unique alerts in ZTF between June 1 2019 and September 1 2022 to a sample of seven glSN Ia candidates. This approach will also be utilised in the final months of ZTF for a live search for lensed transients.

\subsection{The AMPEL platform: Alert processing and catalogue-matching}
\label{subsection:ampel}
    To systematically process all archival ZTF data, we require a platform that can ingest and filter large datasets. We also require the ability to cross-match our large sample of alerts to various photometric and spectroscopic redshift catalogues, so that we can estimate the redshift of each transient's lens or host galaxy. To achieve these objectives, we used AMPEL\footnote{\href{https://github.com/AmpelProject}{github.com/AmpelProject}}, which is a publicly available real-time alert processing platform developed by \citet{Nordin2019}. As well as being a broker for ZTF, it will also be one of the brokers for LSST, and is designed to be flexible enough to process a variety of astronomical data streams.
    
    Within AMPEL, it is possible to filter alerts before they are ingested into a database, perform matches to numerous catalogues, and apply analysis methods (such as fitting to the SALT2 template) in a single pipeline. It is also possible to apply this pipeline to both alert and forced photometry from ZTF. We developed our glSNe Ia identification pipeline using the AMPEL infrastructure, specifically the astronomy analysis units from the \texttt{AMPEL-HU-astro} respository\footnote{\href{https://github.com/AmpelAstro/Ampel-HU-astro/}{github.com/AmpelAstro/Ampel-HU-astro/}}. An example Python notebook that matches the structure of the pipeline is publicly available\footnote{\href{https://github.com/AmpelAstro/Ampel-HU-astro/blob/main/notebooks/run\_lensing_query.ipynb}{github.com/AmpelAstro/Ampel-HU-astro/.../run\_lensing\_query.ipynb}}.
   
   Obtaining accurate lens (and host) redshifts for our candidates was the most crucial part of the study because the redshifts are necessary to estimate the absolute luminosity (and therefore the magnification). Several catalogues are available within AMPEL and the complete list is publicly available via the API\footnote{\href{https://ampel.zeuthen.desy.de/api/catalogmatch/docs}{ampel.zeuthen.desy.de/api/catalogmatch/docs}}. The specific catalogues that we queried are as follows: Sloan Digital Sky Survey (SDSS) DR10 \citep{Brescia2015}; NASA/IPAC Extragalactic Database (NED\footnote{\href{https://ned.ipac.caltech.edu}{ned.ipac.caltech.edu}}), accessed through the \texttt{catsHTM} tool \citep{Soumagnac2018}; Galaxy List for the Advanced Detector Era (GLADE) v2.3 \citep{Dalya2018}; WISExSCOS Photometric Redshift Catalogue (WISExSCOSPZ) \citep{Bilicki2016}; 2MASS Photometric Redshift catalogue (2MPZ) \citep{Bilicki2014}; Legacy Survey (LS) DR8 \citep{Duncan2022}; Pan-STARRS1 Source Types and Redshifts with Machine Learning (PS1-STRM) \citep{Beck2021}.

    From the LS catalogue, we obtained the galaxy colours in the $g$-, $z$-, and $W_1$-bands as well as the photometric redshift for each object. In addition to the catalogues offered by AMPEL, we cross-matched our more promising transients with the DESI spectroscopic survey. This is currently the largest spectroscopic survey operating, with the ability to observe high-redshift galaxies. Our catalogue matching to this survey is discussed in more detail in Sect. \ref{subsection:DESI}.

\section{Candidate glSNe Ia from the ZTF archive}
\label{section:method}

\subsection{Alert pipeline}
\label{subsection:pipeline}
\begin{figure}[h!]
   \centering
   \includegraphics[width=0.95\hsize]{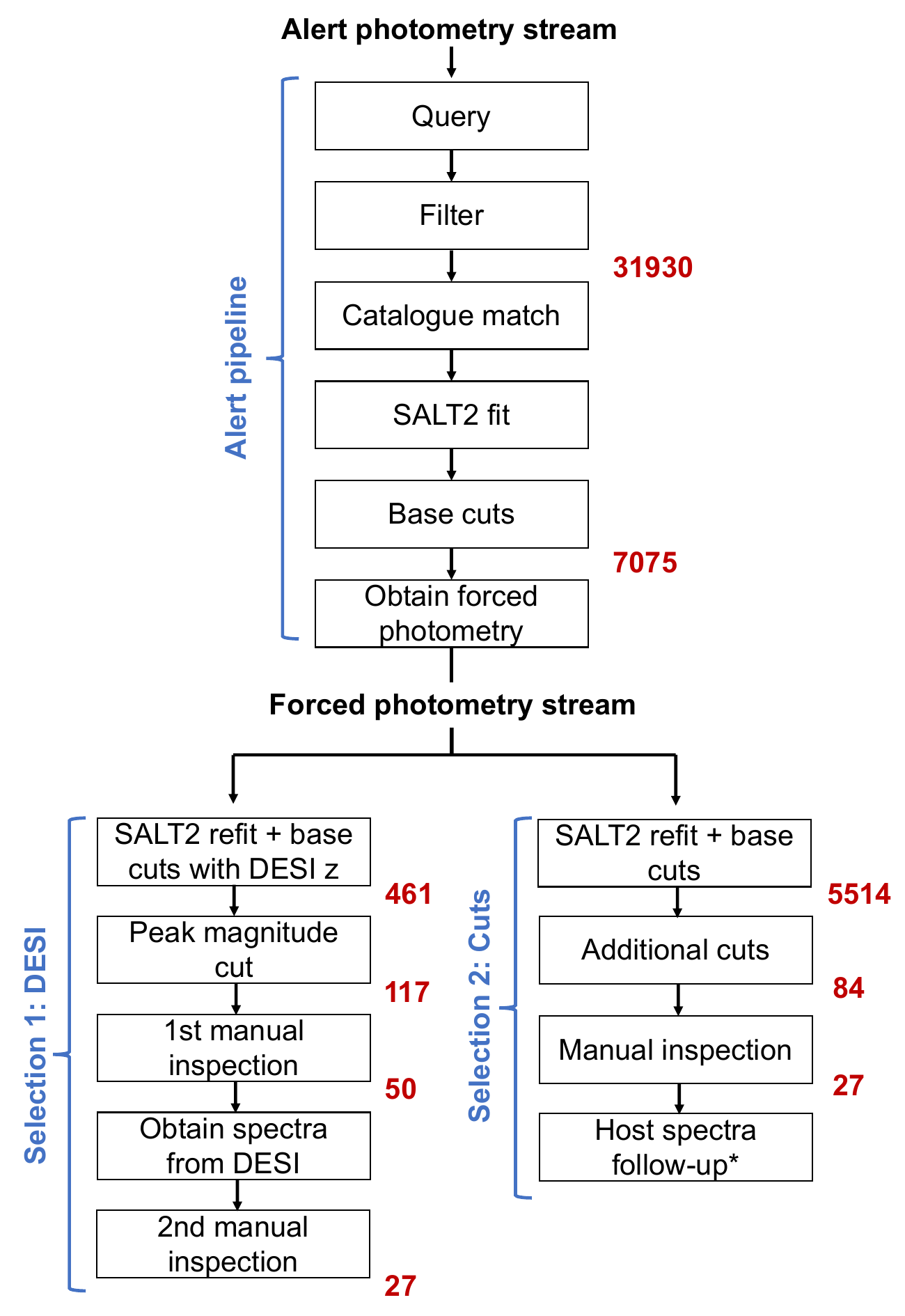}
      \caption{Flowchart to illustrate the stages of the method, as is described in Sect. \ref{section:method}. The red numbers to the bottom right of the boxes indicate the number of transients that remained after that stage of the analysis. The asterisk represents a stage that has not been completed during our study and is left for future work.
              }
         \label{figure:flowchart}
   \end{figure}

    Here, we describe the functional steps of all modules of the workflow (as was mentioned above, this pipeline can be reproduced using the publicly available AMPEL interface). A flowchart to summarise the entire pipeline is shown in Fig. \ref{figure:flowchart}. We note that the term `alert' refers to a transient event, whereas `object' refers to a persistent source that is associated with the alert.

    Alert photometry from June 1 2019 to September 1 2022 was ingested based on a query with basic initial criteria. These criteria were: a point spread function (PSF) apparent magnitude of greater than $18$; greater than six previous detections but fewer than 30; and a positive flux value.
    
    We applied a filter that allowed for SN-like alerts. The filtering criteria included a real bogus (RB) value of greater than 0.3, loose cuts on image quality (such as the full width at half maximum, elongation, and number of bad pixels), distance to known solar system objects, and an alert time history of less than 60 days. Furthermore, probable stars were rejected using a match to the Pan-STARRS PS1 and Gaia DR2 catalogues. The details of this filter are given in the \texttt{DecentFilter} unit of AMPEL\footnote{\href{https://github.com/AmpelAstro/Ampel-ZTF/blob/master/ampel/ztf/t0/DecentFilter.py}{github.com/AmpelAstro/Ampel-ZTF/.../DecentFilter.py}}. The number of alerts remaining after this stage was 31930.
    
    All remaining alerts were cross-matched to the catalogues listed in Sect. \ref{subsection:ampel} to establish whether they are associated with a galaxy. From the catalogues, we determined properties of the associated galaxy, such as the redshift and the angular separation to the alert location. If there was more than one catalogue match for a particular alert, the catalogue with the better ranking (based on accuracy\footnote{Details of the ranking of each catalogue are given in the \href{https://github.com/AmpelAstro/Ampel-HU-astro/blob/main/ampel/contrib/hu/t2/T2DigestRedshifts.py}{\texttt{T2DigestRedshifts}} unit of AMPEL.}) was chosen. If the catalogues are ranked similarly, an average of the redshifts and distances to the galaxy were used. We note that the catalogue-matched redshift might belong to either the host or lensing galaxy. However, it will most likely belong to the lensing galaxy since this is closer and more luminous (it is unlikely the host galaxy will be sufficiently magnified). This was assumed in the following analysis.

    Using the Python library \texttt{SNCosmo} \citep{sncosmo}\footnote{\href{https://sncosmo.readthedocs.io}{sncosmo.readthedocs.io}}, we performed a preliminary SALT2 fit to the alert light curves \citep[using version 2.4, trained by][]{Betoule2014}. From this fit, we obtained a lower limit for the peak $B$-band absolute magnitude. This value is a lower limit because we used the catalogue-matched redshift, which we assume belongs to the potential lens.
    
\begin{figure}[t!]
        \centering
        \includegraphics[width=\hsize]{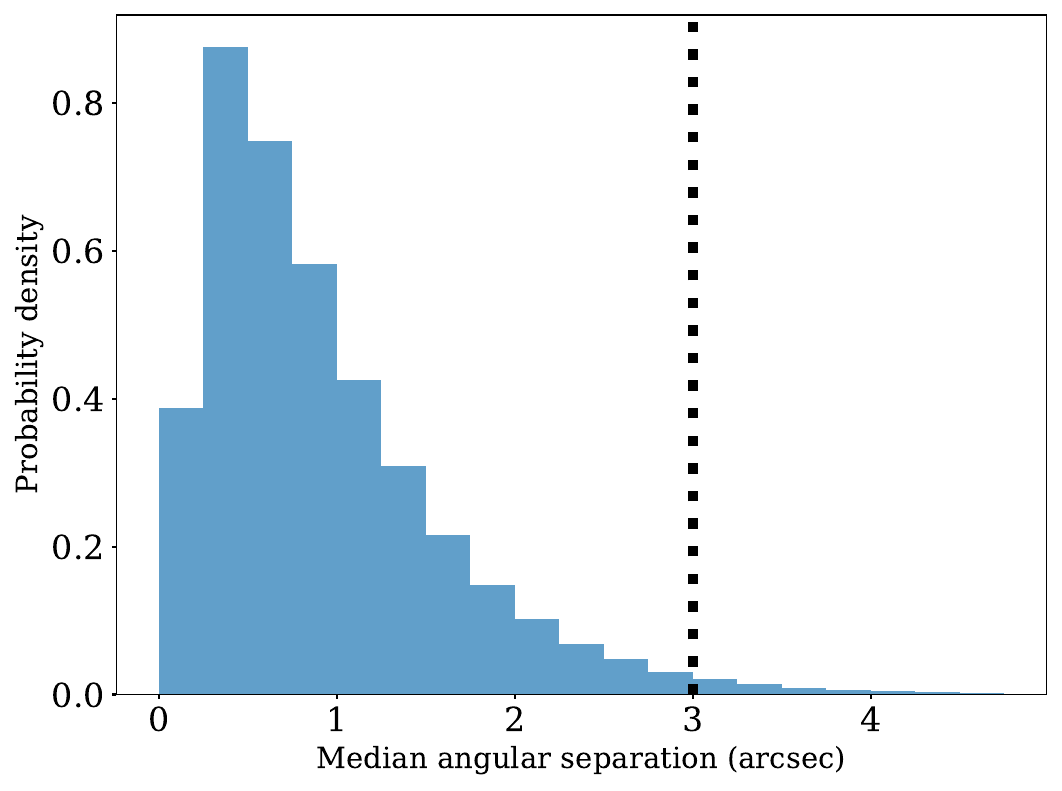}
          \caption{Histogram illustrating the distribution of median angular image separation of simulated glSNe Ia, using data from the catalogues of \citet{Goldstein2019}. The dashed black line indicates a $3\arcsec$ cut that selects 98.4\% of the distribution.
              }
         \label{figure:sep_hist}
         \end{figure}

    We applied `base cuts' to find transients that are most likely to be lensed. Initially, all alerts already classified as something other than a glSN by the ZTF Bright Transient Survey (BTS) were removed. We applied the following cuts on the remaining candidates:
        \begin{enumerate}[label=(\roman*)]
            \item Peak $B$-band absolute magnitude brighter than $-19.5 \: \mathrm{mag}$. This cut selects transients brighter than normal SNe Ia. The simulations of SC24 show that most glSNe Ia are in this magnitude region (regardless of whether the catalogue-matched redshift belongs to the host or the lens galaxy).
            \item A redshift greater than $0.1$ (also based on simulations of lens and galaxy distributions from SC24). Additionally, the probability of lensing increases at higher redshifts, due to the larger number of possible lensing galaxies, and thus a higher chance of strong lensing alignment.
            \item Less than $3\arcsec$ radius to a catalogue-matched galaxy, which ensures that the candidate is close to the lens galaxy core, where strong lensing is likely to occur. This is also supported by the simulated catalogues of \citet{Goldstein2019}, which we utilise in Fig. \ref{figure:sep_hist} to show the distribution of the median angular image separation. From this simulation study, we calculate that 98\% of the glSNe Ia have a median angular image separation of less than 3 arcseconds. For a lensing system in perfect alignment, the image separation is double the separation of the image from the lens. Therefore, we can assume that a $3\arcsec$ radius is a reasonable upper limit for the galaxy separation.
            \item At least five detections around the peak (within a range of 10 days before and 20 days after the peak). This is required to distinguish the nature of the transient from the light curve.
        \end{enumerate}

        The number of glSN Ia candidates remaining at this stage was 7075. Next, we obtained forced photometry (FP) to recover data points at the lower detection limit of ZTF. This was done using PSF fitting \citep{Masci2023}, with the forced photometry pipeline \texttt{fpbot} developed by \citet{Reusch2023}\footnote{\href{https://github.com/simeonreusch/fpbot}{github.com/simeonreusch/fpbot}}. Forced photometry allows us to differentiate between supernovae and long-lasting transients such as active galactic nuclei (AGN) and follow the tail of said supernovae. We removed the baseline of the forced photometry data using the AMPEL unit \texttt{ZTFFPbotForcedPhotometryAlertSupplier}\footnote{\href{https://github.com/AmpelAstro/Ampel-ZTF/blob/master/ampel/ztf/alert/ZTFFPbotForcedPhotometryAlertSupplier.py}{github.com/AmpelAstro/Ampel-ZTF/ ... /ZTFFPbotForcedPhotometryAlertSupplier.py}}. Objects where it was not possible to remove the baseline, for example, due to long-lasting stochastic variability, were removed from the sample. This accounted for approximately 13\% of the sample at this stage. The data were then truncated using the unit \texttt{T2PhaseLimit}\footnote{\href{https://github.com/AmpelAstro/Ampel-HU-astro/blob/main/ampel/contrib/hu/t2/T2PhaseLimit.py}{github.com/AmpelAstro/Ampel-HU-astro/.../T2PhaseLimit.py}}, which selects the portion of the light curve with a significant supernova-like peak. The forced photometry data was utilised in the following two analysis methods of Sect. \ref{subsection:DESI} and \ref{subsection:cuts} to further narrow down our candidates.

We note that the following analysis is based on point source forced photometry, whereas our aim is to detect multiple point sources (e.g. two or four lensed images) that appear unresolved. For glSNe Ia with larger angular separations between the images, this can result in biased light curves. Additionally, it is likely that glSNe Ia with larger image separations (and therefore, displaying blended or multiple images) would be rejected by the ZTF real bogus classifier at the stage where we filter the alerts. Therefore, a different search pipeline (likely image-based) would be required to detect glSNe Ia with larger image separations, which is not the focus of this work.

Executing the archive search pipeline produced 7075 candidate events. This set was analysed further based on whether a DESI spectroscopic redshift was available (Sect. \ref{subsection:DESI}) or not (Sect. \ref{subsection:cuts}). 

\subsection{Selection method 1: Spectroscopic sample with DESI cross-matching (S1)}
\label{subsection:DESI}
    The Dark Energy Spectroscopic Instrument \citep[DESI;][Bailey et al. in prep.]{DESILevi2013,DESI2016a,DESI2016b,DESI2022,DESI2023,DESI2024,DESISilber2023,DESIMiller2023,DESIGuy2023,DESIHahn2023} is a spectroscopic instrument designed to measure the impact of dark energy on the expansion of the Universe. The 5-year DESI survey is currently ongoing and is collecting redshifts for tens of millions of galaxies and quasars (QSOs). The DESI Y1 Key Project publications are \citet{DESI2024VI,DESI2024IV,DESI2024III}.
    As part of our study, collaborators working with the DESI survey contributed redshifts (including data not yet public) for galaxies and quasars that were less than $10\arcsec$ away from our archival transients, and with redshift values greater than 0.15.
    
    We submitted the 7075 candidates that passed the `base cuts' in the pipeline described in Sect. \ref{subsection:pipeline}, for which DESI had lens or host redshifts for 1269 of the candidates. Of these, 519 were labelled as galaxies (and not QSOs) and within $3 \arcsec$ of the ZTF archival transient. We used the DESI redshifts where available to redo the SALT2 fits, as is described in Sect. \ref{subsection:pipeline}. A total of 461 objects had a converging SALT2 fit and were utilised in the next stage of the analysis. The objects that did not pass this stage were likely not supernova-like, with long-duration light curves and stochastic variability.  This long-term variability may not have been present during the alert stage, but is visible in the forced photometry data. After reapplying the base peak $B$-band absolute magnitude cut of $M_B < -19.5\: \mathrm{mag}$, 257 transients remained out of the 461. From this sample, we can assert that approximately 56\% of the photometrically over-luminous candidates in our sample (with a converging SALT2 fit) were also spectroscopically over-luminous (in this case, over-luminous means peak $M_B < -19.5\: \mathrm{mag}$). This is a useful statistic for when we consider the sample with only photometric redshifts in Sect. \ref{subsection:cuts}.

    Next, a stricter peak $B$-band absolute magnitude cut of $M_B < -20\: \mathrm{mag}$ was applied. This was done primarily to select objects that are significantly magnified, and therefore are more likely to be glSNe Ia (as opposed to belonging to over-luminous SNe Ia subclasses such as type Ia-91T, Ia-CSM, or Ia-03fg). From the simulations of SC24, we can see that this cut removes almost all unlensed SNe Ia and approximately 20\% of real glSNe Ia events. We applied this cut after obtaining forced photometry because the additional data points close to the magnitude limit should allow the SALT2 fit to better characterise the light curve shape (including the peak). In other words, we were more confident in our measurement of $M_B$ with the forced photometry data and could apply a stricter cut. 
    
    The remaining 117 candidate light curves were inspected by at least two scanners. Initially, objects were excluded if they did not look supernova-like (i.e. with long-term AGN-like variability). From this inspection, 50 candidates were selected as possible glSNe Ia. At this point, we requested the DESI galaxy spectra for these candidates, to confirm that the redshifts were consistent with the spectral emission lines. The parameters for this sample of 50 are included in Table A.2 (available at the CDS).

     From our spectroscopic sample of 50, we narrowed down our candidates by visual inspection once again. This time, objects were excluded if they did not look Ia-like (e.g. if they looked similar to superluminous supernovae, tidal disruption events, or AGN; refer to Sect. \ref{subsection:contamination}) or if they were previously classified on TNS. Additionally, redshifts were updated as DESI data reprocessing and re-observations occurred throughout this programme, which meant that seven objects were removed from the sample. Consequently, we had 27 final candidates from the spectroscopic sample, which are discussed in Sect. \ref{subsection:candidates} and shown in Table \ref{table:desi_short}.
    
    \subsection{Selection method 2: Photometric sample from selection with additional cuts (S2)}
    \label{subsection:cuts}

\begin{figure}[t!]
   \centering
   \includegraphics[width=\hsize]{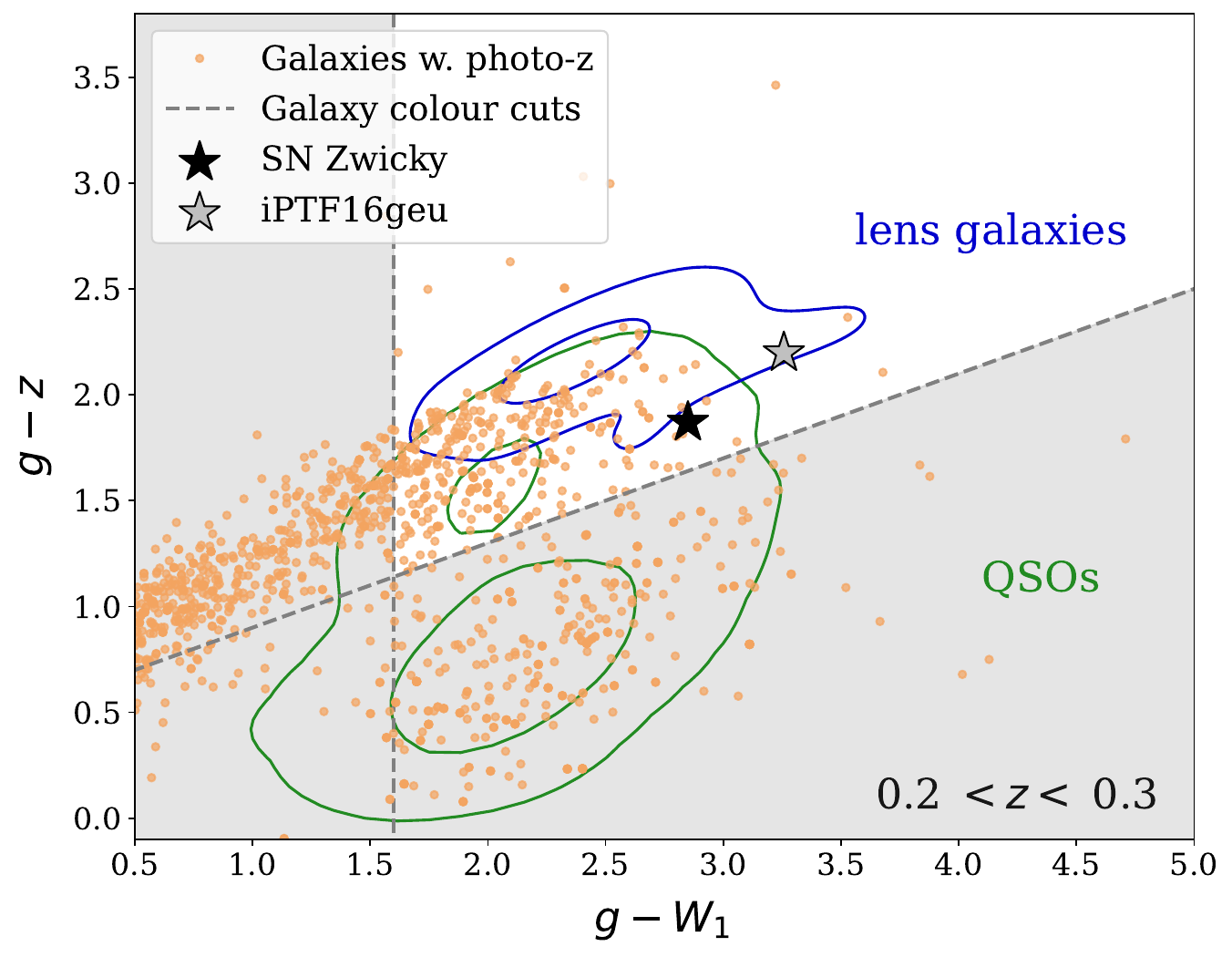}
      \caption{Plot of the LS $g-z$ versus $g-W_1$ galaxy colours for the candidates remaining after the base cuts (orange), compared to contours from confirmed and candidate lens galaxies (blue), and QSOs detected by WISE (green). The data was acquired by cross-matching the AGN and lens catalogues with LS \citep{Duncan2022}. The dashed lines indicate the colour cuts which were devised to select likely elliptical and lens galaxies, shown in Table \ref{table:gal_cuts}. A single redshift bin of $0.2<z<0.3$ is shown for clarity. The two star markers show the lens galaxy colours for iPTF16geu (grey) and SN Zwicky (black), which had lens redshifts of 0.216 and 0.226, respectively. For SN Zwicky, the $g-z$ versus $g-W_1$ colours from Pan-STARRS are shown instead, because the lens galaxy was not present in the LS catalogue).}
         \label{figure:gal_cuts}
   \end{figure}

    The majority of the objects that passed the base cuts described in Sect. \ref{subsection:pipeline} did not have a spectroscopic redshift from DESI. It was not possible to manually inspect these remaining objects, so we applied further cuts motivated by the simulations of SC24 and observational data. For example, Fig. \ref{figure:gal_cuts}, illustrates the LS $g-z$ versus $g-W_1$ lens galaxy colours for the sample of candidates remaining after the base cuts within the redshift bin $0.2 < z < 0.3$. By comparing to catalogues of known lens galaxies\footnote{L. A. Moustakas, J. Brownstein et al., priv. comm. The Master Lens Database of confirmed and probable lenses from various sources. \href{http://admin.masterlens.org}{http://admin.masterlens.org}} and QSOs \citep[from the WISE catalogue of AGN;][]{Assef2018}, it is evident that a large fraction of the candidates' associated galaxies do not have the expected colours of a lensing galaxy.

    By implementing stricter cuts, we aimed to increase the likelihood that a transient was truly a lensed supernova, even though photometric redshifts are less reliable. We required that the transient more closely resembled the expected features of a glSN Ia. These characteristics include the light curve colour evolution, the lens galaxy colours, and the SALT2 fit parameters. We note that the additional cuts were applied to all remaining candidates at the end of Sect. \ref{subsection:pipeline}, not just the objects without a DESI galaxy match in Sect. \ref{subsection:DESI}.
    
    First, we refitted the light curves with the SALT2 template to recalculate the peak absolute magnitude values from the new forced photometry data. If a SALT2 fit could not converge, or the new peak absolute magnitude was dimmer than $-20 \: \mathrm{mag}$, the candidate was removed from the sample. The justification for this stricter absolute magnitude cut is the same as in Sect. \ref{subsection:DESI}.
    
    The remaining candidates were selected from a set of additional cuts:
        \begin{enumerate}[label=(\roman*)]
            \item A selection of $g-r$, $g-i$, and $r-i$ colour cuts at light curve phases of $t_0-7$, $t_0$, and $t_0+7$ days (where $t_0$ is the time at peak $B$ magnitude) were determined based on simulations from SC24. The values for these cuts are summarised in Table \ref{table:lc_cuts}. The motivation for these cuts is that they differentiate normal SN Ia from lensed ones, given that the latter are usually significantly redder (when observed at the same peak apparent magnitude). In SC24, the simulations show that this is the colour space where we expect to find 90\% lensed and 10\% unlensed SNe Ia. Candidates were labelled as passing this cut if they passed at least one of the colour cuts at a certain epoch (i.e. they only needed to pass one out of nine of the cuts). The leniency of this cut is due to some light curves missing bands -- usually the $i$-band -- and/or certain epochs. Additionally, we used light curve information from a window of $\pm3$ days around the phases of $t_0-7$, $t_0$, and $t_0+7$ days, so that we did not penalise objects that were observed with a cadence of two or three days. 
            
\begin{table}[h!]
\caption {\label{table:lc_cuts} Light curve colour cuts for the $g$-, $r$-, and $i$-bands applied at a phase of $t_0-7$, $t_0$, and $t_0+7$ days.}
\centering
\begin{tabular}{l| c c c}
\hline
\hline
&& Light curve colour cut (>) & \\
Phase (days) & $g-r$ & $g-i$ & $r-i$ \\
\hline
$t_0-7$ &   $-0.08$   & $-0.08$    & $-0.34$ \\
$t_0$   &    $0.12$   & $0.06$     & $-0.28$ \\
$t_0+7$ &    $0.33$   & $0.33$     & $-0.23$ \\
\hline
\end{tabular}
\tablefoot{The values are based on simulations from SC24.}
\end{table}

            \item Using galaxy colours from LS, we implemented colour cuts that aimed to exclude AGN and blue star-forming galaxies whilst retaining lens-like galaxies. For this cut, we assumed that the catalogue-matched galaxy is the lens or a blend of the lens and the SN host galaxy (the blended systems, similar to iPTF16geu, will be more in the upper right corner, and so will also be preserved with the cuts). The values are chosen from the empirical observations in Fig. \ref{figure:gal_cuts} and are given in Table \ref{table:gal_cuts}. We note that the population of lenses that host glSNe Ia in surveys such as ZTF may be different that than the population of lenses found in galaxy surveys \citep[as is discussed in][]{Goobar2023}. However, this is not possible to quantify with such a small sample of known glSNe Ia. Additionally, a minimum galaxy brightness of greater than 21.07 in the LS $z$-band (corrected for Milky Way extinction) was required. This was based on simulations of lens galaxies from \citet{Wojtak2019} and \citet{Arendse2024} with K-corrections from Lenspop \citep{Collett2015}\footnote{\href{https://github.com/tcollett/LensPop}{github.com/tcollett/LensPop}}. The simulations show that we expect 90\% of lens galaxies to be brighter than this magnitude limit.
            
\begin{table}[h!]
\caption {\label{table:gal_cuts} Table of galaxy colour cuts for the LS $g$-, $z$- and $W_1$-bands in different redshift bins.}
\centering
\begin{tabular}{l| c }
\hline
\hline
Redshift bin & Galaxy colour cuts\\
\hline
$0.1<z<0.2$ & \makecell{$(g-z)>0.2+0.4\times(g-W_1)$\\ $(g-W_1) > 1.0$} \\
$0.2<z<0.3$ & \makecell{$(g-z)>0.5+0.4\times(g-W_1)$\\ $(g-W_1) > 1.6$} \\
$0.3<z<0.4$ & \makecell{$(g-z)>0.9+0.3\times(g-W_1)$\\ $(g-W_1) > 2.2$} \\
$0.4<z<0.5$ & \makecell{$(g-z)>0.45+0.4\times(g-W_1)$\\ $(g-W_1)> 2.3$} \\
$0.5<z<0.6$ & \makecell{$(g-z)>0.3+0.4\times(g-W_1)$\\ $(g-W_1) > 2.5$} \\
$z>0.6$ & \makecell{$(g-z)>-0.1+0.5\times(g-W_1)$\\ $(g-W_1) > 1.0$} \\
\hline
\end{tabular}
\tablefoot{The cuts are derived from the observations in Fig. \ref{figure:gal_cuts}, to exclude AGN and blue star-forming galaxies but retain lens-like galaxies.}
\end{table}

        \item Based on the simulated distributions shown in SC24, we predict that $90\%$ of glSNe Ia light curves have a SALT2 colour parameter, $c$, of greater than 0 (which is synonymous with being redder). Thus, a cut of $c>0$ was applied to the sample.
        \item As was discussed in Sect. \ref{section:detecting}, we expect that glSNe Ia with narrower light curves will have a greater total magnification (and conversely, wider light curves will have a smaller total magnification). This means that there are likely two populations of glSNe Ia: less magnified objects with larger $x_1$ values that may have poor SALT2 fits, and more magnified objects with smaller $x_1$ values that more closely resemble normal SNe Ia. The transients observed with greater magnification will have a better line-of-sight alignment with the core of the lensing galaxy, meaning that the distance to the catalogue-matched galaxy should be less (the actual value depends on the Einstein radius of the lensing system, but for typical compact systems we would expect a separation of $\theta$ < $1 \arcsec$). This information suggests that we could study the two populations separately so that we could remove contamination by normal SNe Ia more easily. To select the highly magnified glSNe Ia with smaller time delays, we applied a SALT2 stretch cut of $-3<x_1<3$, which would select the objects that look like normal SNe Ia. Additionally, we required that the object was less than $1\arcsec$ away from the catalogue counterpart, to increase the likelihood that the object was a highly magnified glSN Ia. Conversely, to select the less magnified glSNe Ia, we applied a SALT2 stretch cut of $3<x_1<20$, which would exclude the normal SNe Ia. This cut is motivated by the distribution of $x_1$ parameters for simulated lensed and unlensed SNe Ia in SC24. We note that a small angular separation to the lensing galaxy was used as a proxy for magnification, instead of the peak absolute magnitude, to minimise the impact of inaccurate photometric redshifts.
        \end{enumerate}
    The candidates were inspected by at least two scanners. Objects were excluded if they did not look Ia-like (i.e. if they looked similar to other transients, such as the contaminants discussed in Sect. \ref{subsection:contamination}), if they were classified as something else, or if the data was too limited or noisy to determine what they were.

    After this manual inspection stage, 27 photometric candidates were remaining. Seven of the candidates also had a DESI cross-match and were identified in the spectroscopic sample from Sect. \ref{subsection:DESI}. The 27 photometric candidates are discussed in detail in Sect. \ref{subsection:candidates} and shown in Table \ref{table:cuts_short}.

   
     A discussion of the accuracy of the various photometric catalogues utilised in this study is beyond the scope of this paper. However, from our analysis in Sect. \ref{subsection:DESI}, we can assume that at least half of the photometrically over-luminous candidates identified in this selection method would also be spectroscopically over-luminous if a spectrum for the lens or host was obtained.

    The methodology described in Sect. \ref{section:method} is summarised as a flowchart in Fig. \ref{figure:flowchart}, which includes the number of transients that remained after each stage in the analysis (indicated in red).

\subsection{Summary of candidates}
\label{subsection:candidates}
Table \ref{table:desi_short} summarises the 27 candidates from S1 described in Sect. \ref{subsection:DESI}, with spectroscopic redshifts and angular separations from a catalogue match with DESI. This selection of spectroscopically over-luminous candidates is interesting since we expect that the majority of the events are bright enough to be glSNe Ia. However, if they do not pass the cuts specified in Sect. \ref{subsection:cuts}, it is more likely that they are superluminous supernova candidates. Despite this, we cannot exclude the possibility that some of them are glSNe Ia (or other lensed transients) and the reason(s) that they did not pass the cuts are:
\begin{enumerate}
    \item The fit to the SALT2 template fails to model the light curve well enough, which may be the case if the lensed object is not a typical SN Ia or it is a SN II. This could mean that it does not pass the $x_1$, $c$, and light curve colour cuts, or that a SALT2 fit does not converge at all.
    \item The catalogue-matched galaxy information is not present in the LS catalogue, the LS catalogue galaxy is dimmer, or the colours are different than what we would expect from a lens galaxy. This would mean that it does not pass the galaxy cuts.
\end{enumerate}

\begin{table}[h!]
\caption{Spectroscopic sample}
\label{table:desi_short}
\centering
\begin{tabular}{c | c c c c}
\hline\hline
ZTF ID & DESI $z$ & Ang. sep. ($\arcsec$) & Peak $M_B$ \\
\hline
ZTF19aarzfoz & 0.1765   & 0.87                 & $-20.00\pm0.04$       \\
ZTF19aavrvoe & 0.2894   & 0.07                 & $-20.87\pm0.04$      \\
ZTF19abdbskh & 0.2832   & 2.23                 & $-21.15\pm0.06$       \\
ZTF19abpjicm & 0.2383   & 0.92                 & $-20.00\pm0.08$        \\
ZTF20aarvtor & 0.3192   & 1.07                 & $-20.56\pm0.04$       \\
ZTF20aatpwrh & 0.3357   & 1.03                 & $-20.92\pm0.06$       \\
ZTF20aawabry & 0.2930   & 0.94                 & $-20.85\pm0.03$       \\
ZTF20aawlfwk & 0.3138   & 0.17                 & $-20.64\pm0.07$       \\ 
ZTF20abatzio & 0.2872   & 0.63                 & $-20.57\pm0.07$       \\
ZTF20abfhluf & 0.2198   & 0.96                 & $-20.59\pm0.05$       \\
ZTF20abjyrxf & 0.2199   & 0.76                 & $-20.47\pm0.05$       \\
ZTF20acirhoc & 1.3710   & 1.34                 & $-26.58\pm0.05$       \\
ZTF20aclwssg & 0.3309   & 0.31                 & $-21.60\pm0.08$       \\
ZTF20acmvzuo & 0.1762   & 1.14                 & $-20.21\pm0.03$       \\
ZTF20acxtayx & 0.2041   & 0.84                 & $-20.69\pm0.03$      \\
ZTF21aablrfe & 0.3072   & 2.35                 & $-20.34\pm0.09$       \\
ZTF21aacsvko & 0.2287   & 0.12                 & $-20.61\pm0.04$       \\
ZTF21aapreuq & 0.2222   & 0.43                 & $-20.93\pm0.03$       \\
ZTF21abcwuhh & 0.3334   & 1.57                 & $-20.17\pm0.12$       \\
ZTF21abfvefa & 0.3288   & 1.81                 & $-20.93\pm0.07$       \\
ZTF21abwppte & 0.2163   & 1.12                 & $-20.42\pm0.04$       \\
ZTF22aabifrp & 0.1886   & 2.85                 & $-20.94\pm0.06$       \\
ZTF22aadeqlh & 0.3225   & 2.71                 & $-21.61\pm0.04$       \\
ZTF22aahmovu & 0.2429   & 0.65                 & $-20.34\pm0.08$       \\
ZTF22aaqkvvm & 0.3775   & 2.81                 & $-21.35\pm0.06$       \\
ZTF22aatstmd & 0.3120   & 2.90                 & $-21.09\pm0.07$      \\
ZTF22aauihjc & 0.2690   & 0.28                 & $-20.38\pm0.08$        \\      
\hline
\end{tabular}
\tablefoot{The error on the DESI redshift values is 0.0001. The full table including the right ascension (RA), declination, and SALT2 fit parameters is Table A.2 (available at the CDS).}
\end{table}

\begin{table}[h!]
\caption{Photometric sample}
\label{table:cuts_short}
\centering
\begin{tabular}{c | c c c c}
\hline\hline
ZTF ID & Phot. $z$ & Ang. sep. ($\arcsec$) & Peak $M_B$ \\
\hline
ZTF19abamkfs & 0.29    & 0.40            & $-21.22 \pm 0.27$         \\
ZTF19abctwkl & 0.22    & 0.04            & $-20.07 \pm 0.53$         \\
ZTF19abdkdze & 0.18    & 2.39            & $-20.17 \pm 0.28$         \\
ZTF19abheyzp & 0.23    & 0.34            & $-20.80 \pm 1.14$         \\
ZTF19abpfedt & 0.16    & 0.97            & $-20.06 \pm 0.26$         \\
ZTF19abpjicm & 0.24    & 0.87            & $-20.06 \pm 0.36$         \\
ZTF19adbmdsa & 0.18    & 0.53            & $-20.11 \pm 0.25$         \\
ZTF20aahgirj & 0.17    & 0.75            & $-20.13 \pm 0.39$         \\
ZTF20aaiqpgv & 0.29    & 0.69            & $-20.12 \pm 0.45$         \\
ZTF20abjyrxf & 0.19    & 0.78            & $-20.15 \pm 0.34$         \\
ZTF20ablmtxz & 0.26    & 0.18            & $-20.26 \pm 0.84$         \\
ZTF20aceekkd & 0.16    & 0.69            & $-20.35 \pm 0.33$         \\
ZTF20achutix & 0.25    & 2.20            & $-20.55 \pm 0.28$         \\
ZTF20acotfan & 0.23    & 0.25            & $-20.21 \pm 0.86$         \\
ZTF21aablrfe & 0.31    & 2.35            & $-20.35 \pm 0.52$         \\
ZTF21aaxvxbu & 0.24    & 0.51            & $-20.61 \pm 0.20$         \\
ZTF21abawmyd & 0.26    & 0.86            & $-20.31 \pm 0.77$         \\
ZTF21abcwuhh & 0.34    & 1.55            & $-20.21 \pm 0.13$         \\
ZTF21acdypeu & 0.19    & 0.67            & $-20.10 \pm 0.44$         \\
ZTF21acmqvwv & 0.25    & 0.97            & $-20.06 \pm 1.00$         \\
ZTF22aabfojs & 0.19    & 0.21            & $-20.72 \pm 0.62$         \\
ZTF22aabifrp & 0.25    & 0.14            & $-21.45 \pm 0.42$         \\
ZTF22aadeqlh & 0.27    & 2.65            & $-21.11 \pm 0.26$         \\
ZTF22aahmovu & 0.22    & 0.58            & $-20.05 \pm 0.26$         \\
ZTF22aamltry & 0.16    & 1.02            & $-20.21 \pm 0.78$         \\
ZTF22aaoncel & 0.19    & 0.19            & $-20.39 \pm 0.21$         \\
ZTF22abakgnp & 0.25    & 0.62            & $-20.25 \pm 0.33$        \\
\hline
\end{tabular}
\tablefoot{The full table including the right ascension (RA), declination, and SALT2 fit parameters is Table A.2 (available at the CDS). The errors on the photometric redshifts are given in Table~\ref{table:photz_err} in the Appendix.}
\end{table}

To eliminate these possibilities, careful inspection of each of the candidates is necessary. To increase the likelihood that a candidate is a glSN Ia, we could fit the light curves to a combined SALT2 model with multiple images, and check that the resulting fit parameters are reasonable compared to the results from simulation studies.

Table~\ref{table:cuts_short} summarises the 27 candidates from S2 described in Sect. \ref{subsection:cuts}, with the photometric redshifts and angular separations from catalogue matches within AMPEL. The uncertainty on the peak absolute magnitude is dominated by the uncertainty on the photometric redshifts. As a result of the large uncertainties, the majority of the candidates from S2 are consistent with a normal SN Ia within $3 \sigma$, so it is hard to assign a confidence level to the claim that a candidate is truly a lensed supernova. Therefore, the photometric sample in Table \ref{table:cuts_short} will need to be targeted in future spectroscopic follow-up missions to get the lens and host redshifts before we can make any conclusions about them being glSNe Ia. At the time of publication, this was not yet possible, but it is something that we aim to do in future work.

\section{Sample properties}
\label{section:comparison}

\subsection{Sources of contamination}
\label{subsection:contamination}
    Five main sources of contamination may have passed the cuts in our pipeline: AGN, tidal disruption events (TDEs), superluminous supernovae (SLSNe), lensed core-collapse SNe, and unlensed SNe Ia (either due to their peculiar over-luminosity or an incorrect photometric redshift).

   \begin{figure*}
   \begin{center}
   \resizebox{0.8\hsize}{!}{\includegraphics{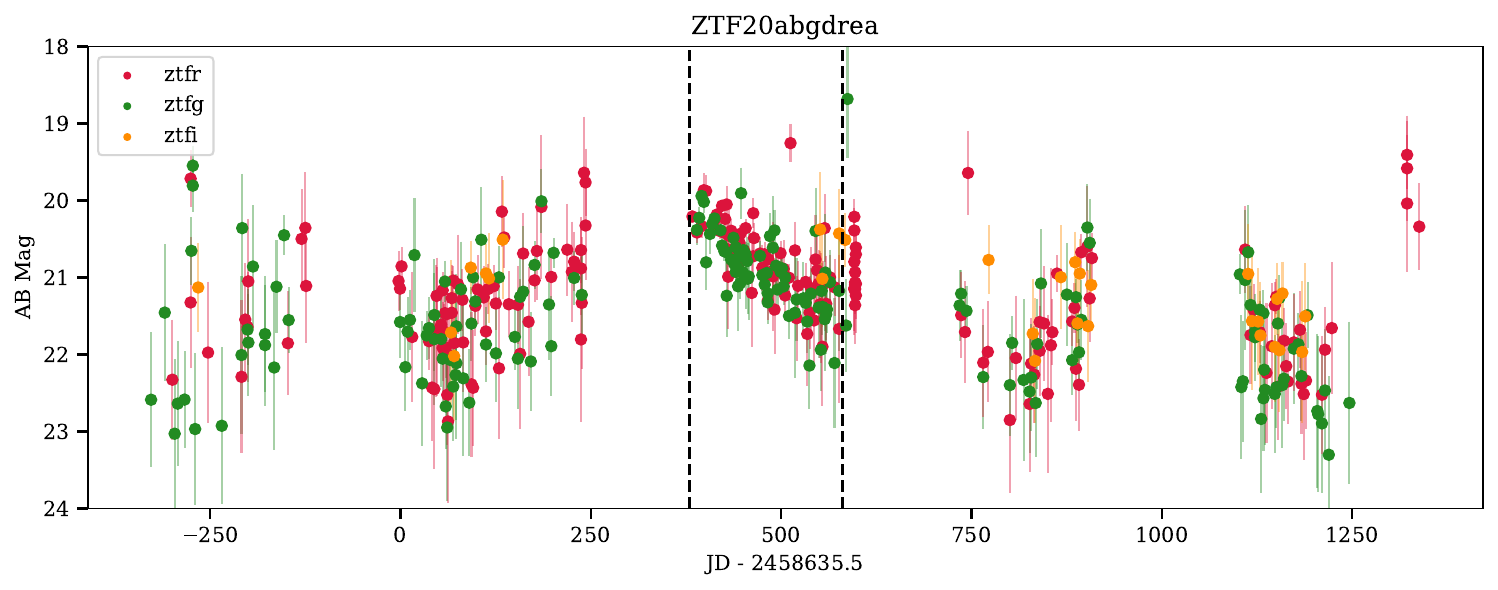}}
      \caption{Light curve of ZTF20abgdrea, a candidate that passed S1 and S2, but was removed during visual inspection due to the long-term variability that suggests it is a likely AGN or QSO. The dotted black lines indicate the section of the light curve that was selected by the pipeline for the SALT2 fit. The baseline was not removed from this forced photometry, to more clearly illustrate the variability.
              }
         \label{figure:ZTF20abgdrea}
    \end{center}
   \end{figure*}

       \begin{figure}[h!]
   \centering
   \includegraphics[width=\hsize]{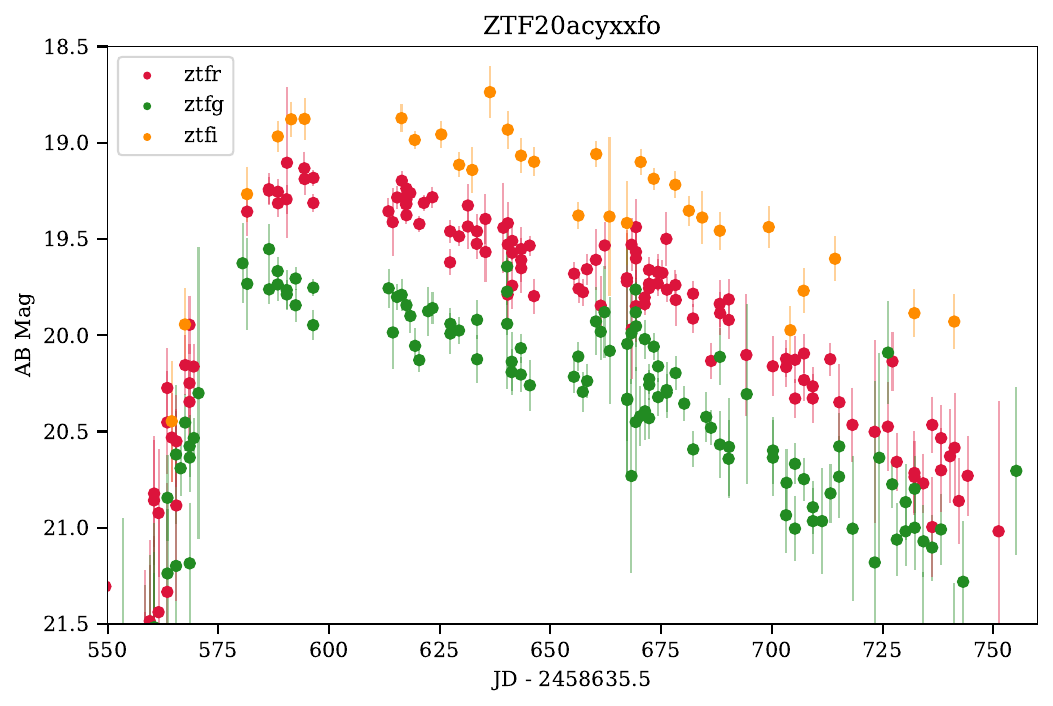}
      \caption{Light curve of ZTF20acyxxfo, a candidate that passed S1 and S2, but was removed during visual inspection because it was identified as a likely TDE due to the approximate power-law decline.
              }
         \label{figure:ZTF20acyxxfo}
   \end{figure}

       \begin{figure}[h!]
   \centering
   \includegraphics[width=\hsize]{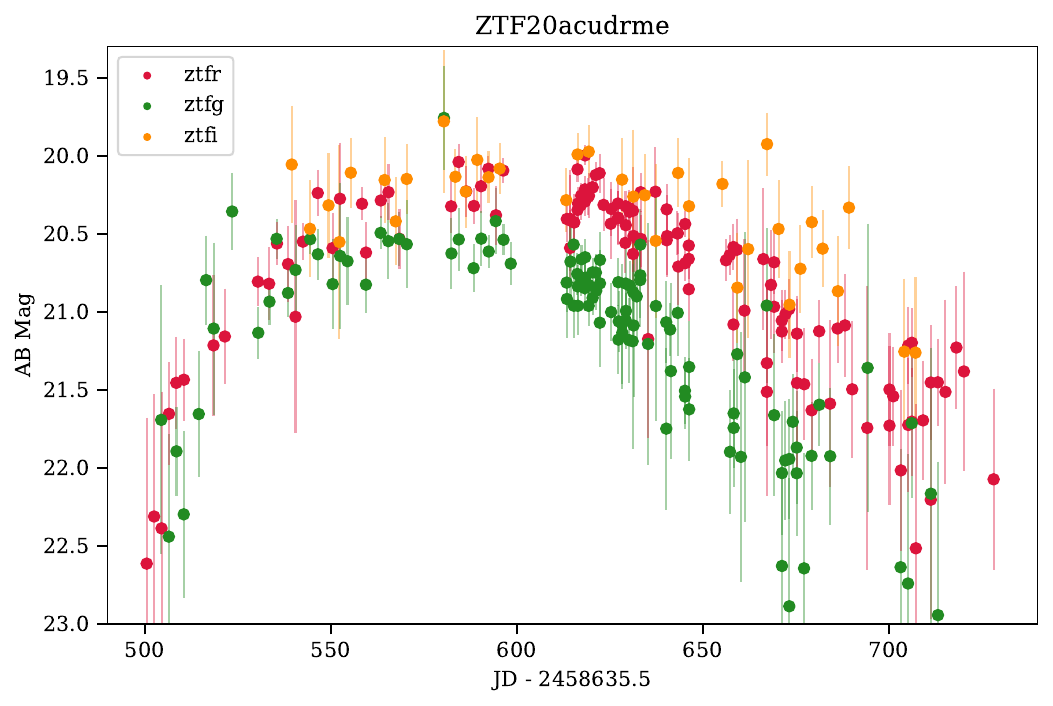}
      \caption{Light curve of ZTF20acudrme, a candidate that passed S2, but was removed during visual inspection because the long duration indicates it is a likely SLSN.
              }
         \label{figure:ZTF20acudrme}
   \end{figure}
        
    Firstly, AGN are luminous enough to look similar to lensed events and constitute a large fraction of the 7075 transients that pass the base cuts with alert photometry. However, they were likely removed by the cuts applied in Sects.\ref{subsection:DESI} and \ref{subsection:cuts} after we obtained forced photometry (because the FP should contain any long-term variations that were slightly below the detection threshold for the alert photometry). Even if there were AGN that passed these additional cuts, the long-term variability in the light curves would be clear enough to remove during the visual inspection stages. An example of this in our study is ZTF20abgdrea, shown in Fig. \ref{figure:ZTF20abgdrea}. This object was present in both selection methods, but was filtered out during the visual inspection stage. The dotted black lines indicate the section of the light curve that was selected by the pipeline for the SALT2 fit, which explains why it passed our stringent cuts from selection method 2 (S2). Furthermore, ZTF20abgdrea is $0.7\arcsec$ away from a candidate lens system in the SuGOHI VI catalogue \citep{Sonnenfeld2020}, which means that it could also be a lensed AGN or QSO. However, the DESI spectrum shows no evidence of AGN activity, so this is not possible to confirm without further study of the lens system.
    
    Secondly, TDEs reside close to galaxy cores and can also be very luminous, so they satisfy most of our criteria. However, they are quite rare events (although not as rare as glSNe Ia) and their light curves have distinct features, such as an approximate power-law decline, and are sometimes accompanied by a dust echo (for example, detected in the WISE infrared bands) \citep{Reusch2022}. This distinctive light curve shape allowed us to identify and remove likely TDEs from our sample. An example of this is ZTF20acyxxfo, shown in Fig. \ref{figure:ZTF20acyxxfo}. This object (with a DESI redshift of $z=0.2$ which implies a peak $M_B<-20\:\,\mathrm{mag}$) was present in both selection methods but was filtered out during the visual inspection stage, as it has a good fit to a TDE template light curve (Reusch et al., in prep.).

    Thirdly, SLSNe are a large contaminant because they are very luminous (with a peak absolute magnitude of approximately $-21 \: \mathrm{mag}$) and some of their light curves closely resemble the light curves of glSNe Ia with longer time delays. However, we can exclude some SLSNe with longer durations that would not be feasible for glSNe Ia (i.e. a rest-frame duration of greater than 100 days). Additionally, the galaxy brightness cut that we describe in Sect. \ref{subsection:cuts} should exclude the hosts of hydrogen-poor superluminous supernovae (SLSNe-I), which are typically bluer and fainter \citep{Lunnan2014,Leloudas2015,Schulze2018}. To fully exclude SLSNe as contaminants, a careful examination of each candidate and all the data available is necessary; we perform this examination in Sect. \ref{section:gold} for our most likely candidates. An example of a possible SLSN that we removed during the visual inspection stage of S2, ZTF20acudrme, is shown in Fig. \ref{figure:ZTF20acudrme}. This object had a photometric redshift of $z=0.25$, which allowed us to estimate a peak $M_B < -20 \: \mathrm{mag}$. However, the long duration and large separation from its galaxy counterpart (at $2 \arcsec$) suggest that ZTF20acudrme is a SLSN candidate.
    
    Additionally, some lensed core-collapse supernovae may be present in this sample. Typically, core-collapse SNe have a peak $B$ absolute magnitude of approximately $-17 \: \mathrm{mag}$ \citep{Perley2020}, which means that they would have to be extremely magnified to pass our final peak cut of $M<-20 \: \mathrm{mag}$. Furthermore, the simulations of SC24 showed that applying a $c>0$ cut should remove almost all lensed core-collapse SNe. This significantly reduces the probability of their detection by our algorithm, although their presence is not ruled out. If we did find something magnified but looked more similar to a core-collapse SN than a SN Ia, we would not exclude it from our analysis (as this would also be an interesting discovery). However, none of the candidates in our sample have the obvious characteristics of a lensed core-collapse supernova.

\begin{figure}[t!]
   \centering
   \includegraphics[width=\hsize]{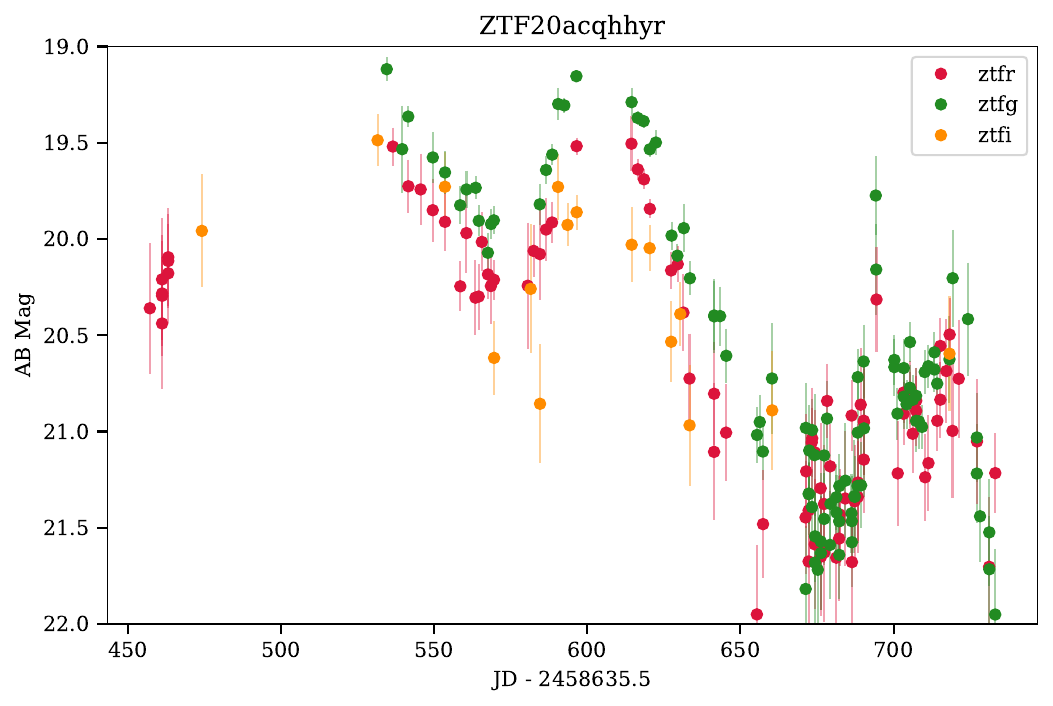}
      \caption{Light curve of ZTF20acqhhyr, a candidate that passed S1, but was removed during visual inspection because of the presence of multiple bumps in the light curve of not-equal width, which suggests that CSM interaction is causing the high luminosity.
              }
         \label{figure:ZTF20acqhhyr}
   \end{figure}

    Transients that interact with the circumstellar material (CSM) are possible contaminants. Supernova classes like type Ia-CSM and type IIn can be more luminous than normal SNe Ia due to the supernova ejecta interacting with the material that surrounds the progenitor, causing an increase in the brightness we observe \citep{Sharma2023}. However, typically the light curve will display signatures of CSM interaction, such as multiple bumps or plateauing phases.  Fig. \ref{figure:ZTF20acqhhyr} illustrates an example candidate that was removed during the visual inspection of S1, due to the multiple peaks of varying widths, which could not be caused by a multiply imaged SN.

   Other peculiar subclasses of SNe Ia can be over-luminous, such as 03fg-like (or super-Chandrasekhar) and 91T-like SNe Ia. SNe Ia-03fg are believed to be the result of white dwarf progenitors that exceed the Chandrasekhar limit before they explode \citep{Howell2006}. These objects can surpass a peak $M_B$ of $-20\: \mathrm{mag}$; however, they are typically found in low-mass galaxies with a lower luminosity \citep[for example,][]{Howell2006,Childress2011,Hsiao2020,Lu2021}. As a result, it is unlikely that they would pass our galaxy cuts in Sect. \ref{subsection:cuts}. Examples of 03fg-like SNe found in more massive galaxies \citep[e.g.][]{Taubenberger2011,Chakradhari2014} are further from the centre of the host and so would not pass our requirement for less than $3\arcsec$ from a catalogue-matched galaxy. Additionally, 03fg-like SNe are typically less red than what we expect from glSNe Ia, so only a small fraction would pass our light curve colour cuts. Similarly, SNe Ia-91T are believed to arise from non-typical progenitor systems that cause the SN explosion to be more luminous. However, SNe Ia-91T are typically 0.2 mag brighter than normal SNe Ia \citep{Yang2022}, so it is unlikely that they would pass our peak magnitude cut.

    Finally, we note that some objects with erroneous photometric redshifts will be present in the photometric sample in Table \ref{table:cuts_short}. These are likely to be normal SNe Ia at redshifts of $z\sim0.1$.

    \subsection{Comparison with the Bright Transient Survey (BTS)}
    \label{subsection:bts}
    BTS is a spectroscopic supernova survey within ZTF that aims to acquire spectra of all transients brighter than 18.5 magnitudes (excluding galactic sources, AGN, or moving objects), using the Spectral Energy Distribution Machine (SEDM) that operates on the Palomar 60-inch telescope \citep{Blagorodnova2018}. The survey aims to create the largest unbiased, brightness-limited sample of supernovae. Additionally, they target objects brighter than 19 magnitudes when the spectroscopic resources are available. The first spectrum of SN Zwicky was obtained as a result of BTS scanners submitting it to the SEDM for spectroscopic classification \citep{Goobar2023}.

\begin{figure}[h!]
   \centering
   \includegraphics[width=0.9\hsize]{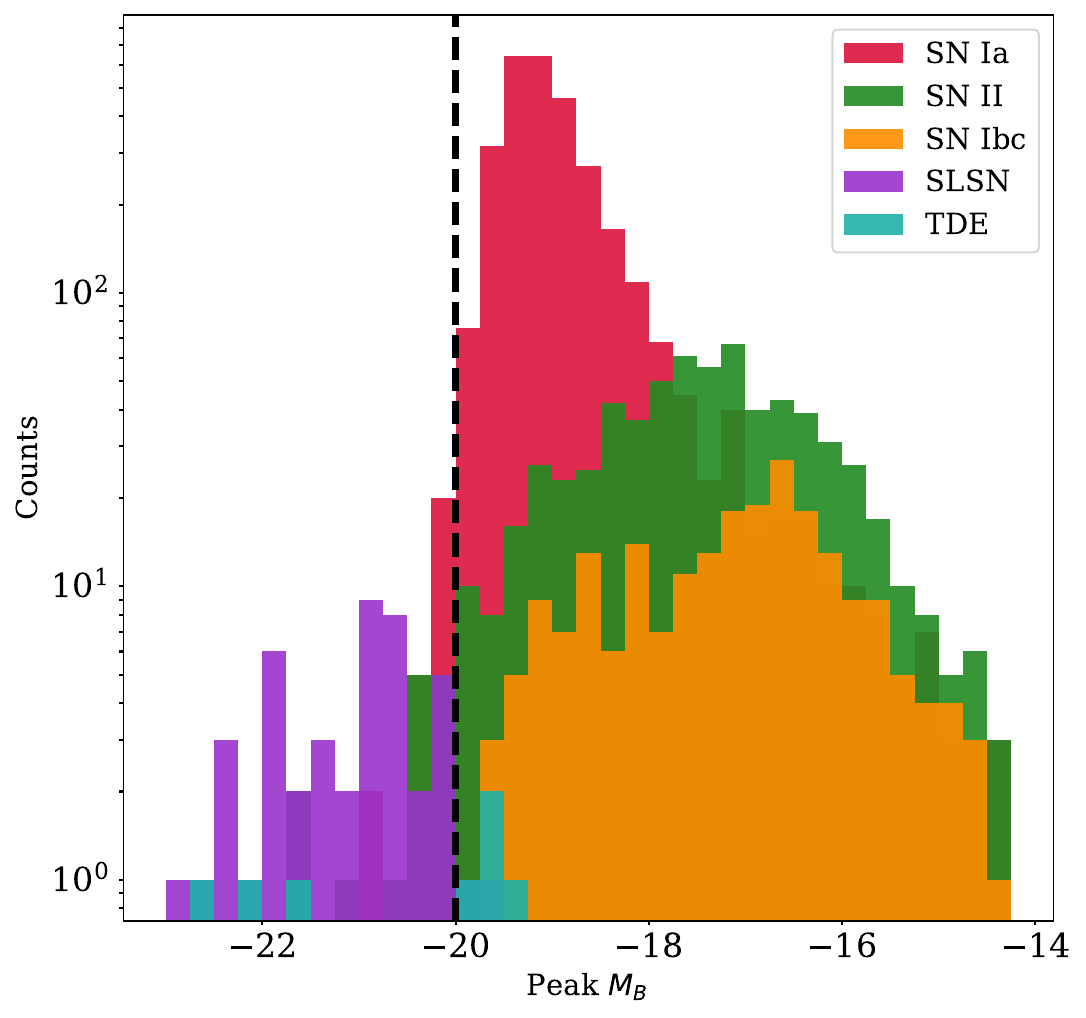}
      \caption{Histogram illustrating the peak B absolute magnitude distribution for the 5 main classes of transients (SN Ia, SN II, SN Ib/c, SLSNe, and TDEs) recorded by the BTS. The dashed black line illustrates the objects with $M_B < -20\: \mathrm{mag}$.
              }
         \label{figure:bts_hist}
   \end{figure}

    To quantify the amount of contamination from unlensed supernova-like sources in this study, a comparison to the BTS was performed. We obtained forced photometry for 4127 transients in the BTS catalogue\footnote{Publicly available here: \href{https://sites.astro.caltech.edu/ztf/bts/explorer.php}{sites.astro.caltech.edu/ztf/bts/explorer.php}} with a redshift (including the classes SLSN, SN Ia, SN Ib/c, TDE, and their subclasses), and processed the data as is described in Sect. \ref{subsection:pipeline}. A SALT2 template fit was performed using the redshift from the BTS catalogue. Figure \ref{figure:bts_hist} shows a histogram with the derived peak $M_B$ from the SALT2 fits for all the objects in the sample.

   First, we applied the base cuts as is described in Sect. \ref{subsection:pipeline} ($M_B<-19.5\: \mathrm{mag}$, $z>0.1$, less than $3\arcsec$ distance from a catalogue-matched galaxy, at least 5 detections around the peak). In the sample remaining after base cuts, we retained 43\% of the original SLSNe, 29\% of the TDEs, 2\% of the SNe II (of which the majority were SN IIn), 2\% of the SNe Ia, and less than 1\% of the SNe Ib/c. After applying the stricter absolute magnitude and redshift cuts ($M_B<20\: \mathrm{mag}$ and $z>0.15$) described in Sects. \ref{subsection:DESI} and \ref{subsection:cuts}, the only contamination in the sample was due to SLSNe (24\% remaining), TDEs (14\% remaining), and a small contribution from SNe II (less than 1\% remaining). After applying all the cuts described in Sect. \ref{subsection:cuts}, only one object remained (ZTF21aahfjrr, classified as a SN IIn, but likely a SLSN-IIn). This study illustrates that we should expect non-negligible contamination from SLSNe and TDEs from the selection method of Sect. \ref{subsection:DESI}, but very little contamination from the method of Sect. \ref{subsection:cuts}. The most effective cuts from the method of Sect. \ref{subsection:cuts} for removing objects were the galaxy colour cuts, as they removed over half of the remaining sample.
   
   However, it is important to note that the brightness range of transients in the BTS does not represent the range expected in this study (as we are considering objects up to the detection limit of ZTF). As a result, we expect more contamination from intrinsically bright objects at larger redshifts.

\subsection{Comparison with known glSNe Ia}
\label{subsection:sample}
\begin{figure}[h!]
   \centering
   \includegraphics[width=\hsize]{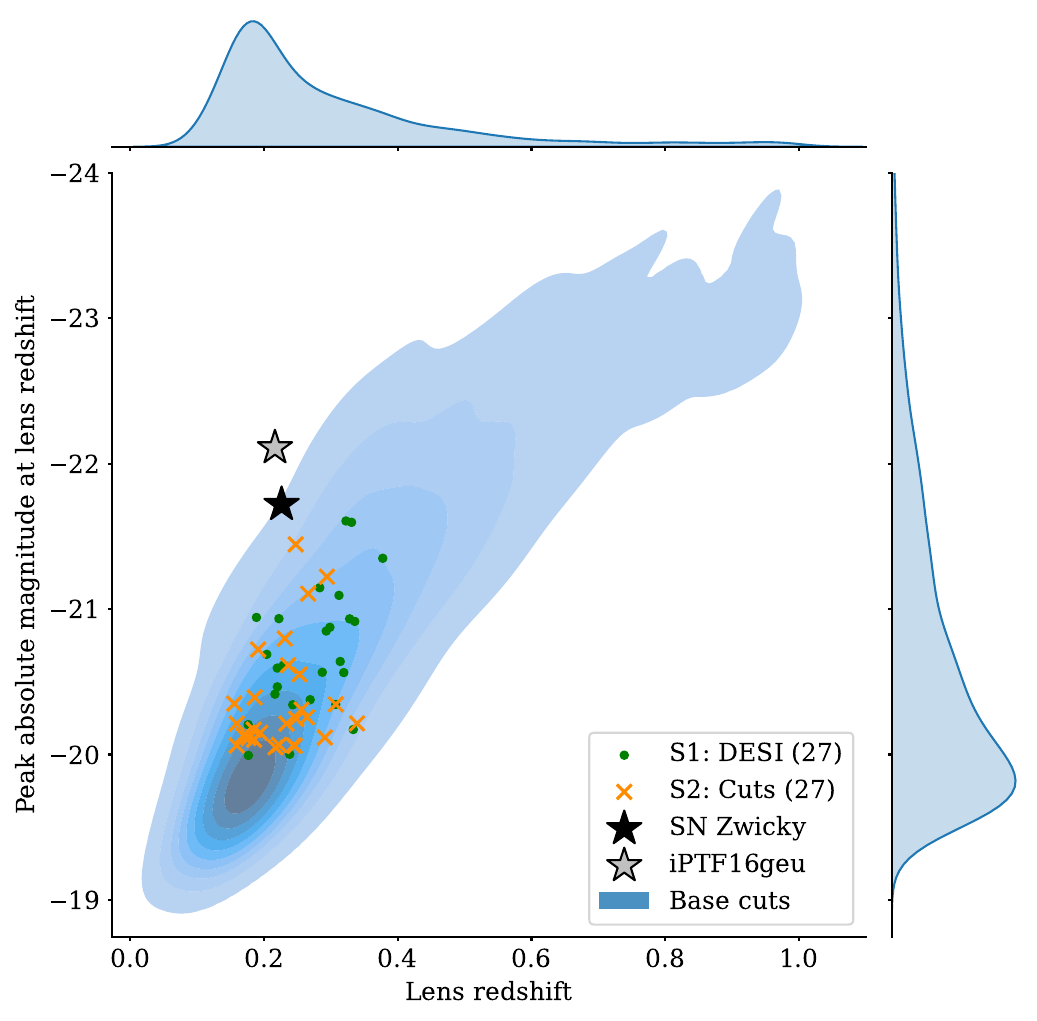}
      \caption{Distribution of the peak $B$-band absolute magnitude at the lens redshift versus the lens redshift for the transients that passed the base cuts described in Sect. \ref{subsection:pipeline} (blue contours), the candidates passing the two selection methods described in Sects. \ref{subsection:DESI} and \ref{subsection:cuts} (green and orange crosses, respectively), and for the already discovered glSNe Ia SN Zwicky and iPTF16geu (black and grey stars, respectively). The upper limit of the $x$ axis has been adjusted for clarity, omitting one higher redshift object from S1 (ZTF20acirhoc).
              }
         \label{figure:magvsz}
   \end{figure}

It is valuable to compare the statistics of our candidates with the glSNe Ia already observed by ZTF and iPTF. Figure \ref{figure:magvsz} shows the distribution of the peak $B$-band absolute magnitude versus the redshift for the candidates in the sample, assuming that the object is at the lens redshift. This peak absolute magnitude would be a lower limit on the true peak if the object were truly lensed. The blue contour displays the transients that passed the base cuts described in Sect. \ref{subsection:pipeline} and the data points illustrate the candidates passing the two selection methods (S1 and S2) described in Sects. \ref{subsection:DESI} and \ref{subsection:cuts}, respectively. The corresponding values for SN Zwicky and iPTF16geu are also displayed. From Fig. \ref{figure:magvsz}, it is apparent that SN Zwicky and iPTF16geu match the brightest magnification tail. This implies that, if our candidates are truly glSNe Ia, they would belong to a less magnified population. Excluding a single high redshift candidate from the S1 sample (ZTF20acirhoc, $z=1.4$), the majority of the candidates are found to have lens galaxies within the range $0.2<z<0.4$.

      \begin{figure}[h!]
   \centering
   \includegraphics[width=\hsize]{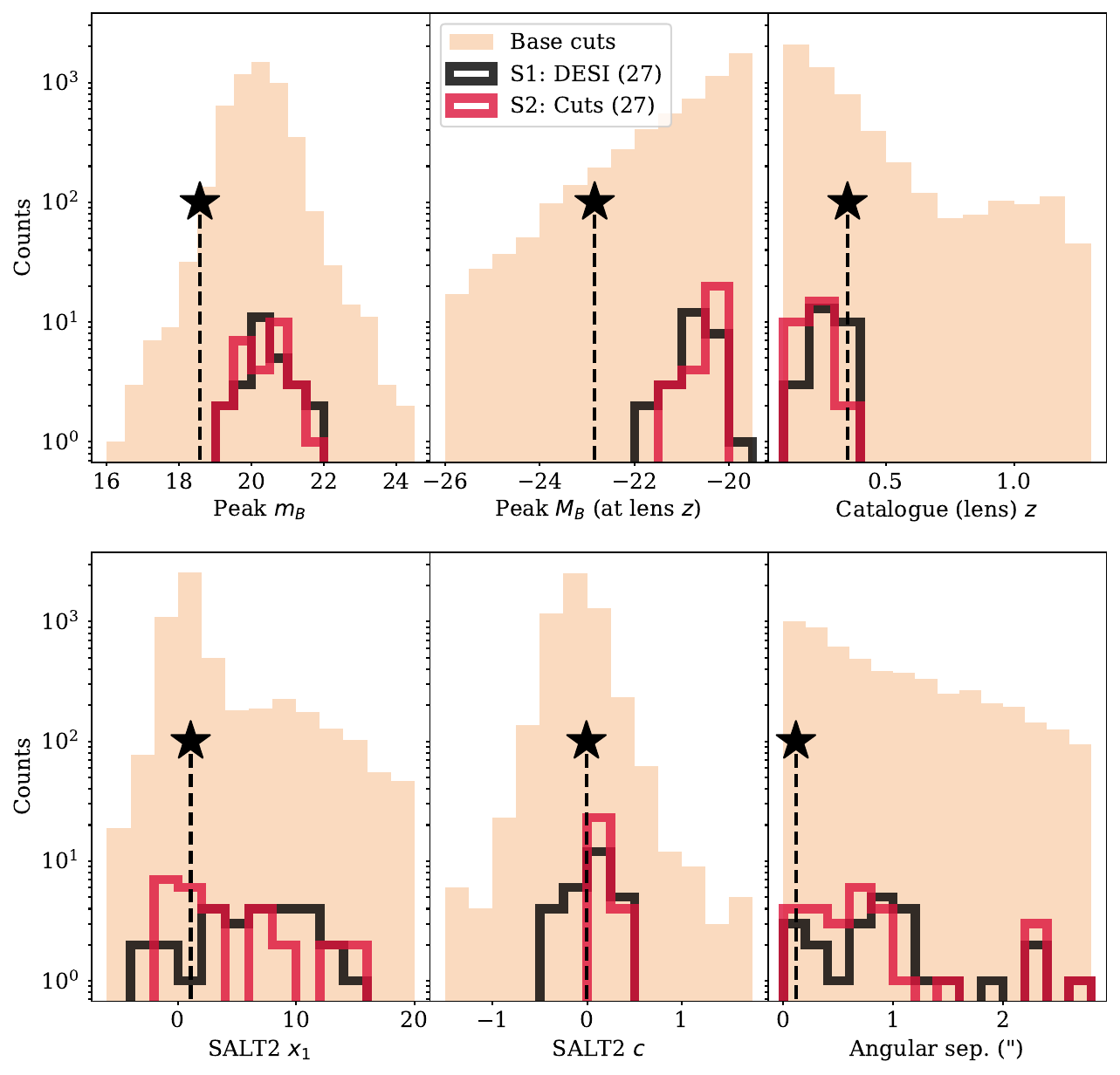}
      \caption{Set of histograms displaying the distributions for peak $m_B$, peak $M_B$, lens $z$, SALT2 $x_1$, SALT2 $c$, and angular separation values for the transients that passed the base cuts described in Sect. \ref{subsection:pipeline} (orange bars), and the candidates passing the two selection methods described in Sects. \ref{subsection:DESI} and \ref{subsection:cuts} (solid black and red lines, respectively). The dotted line with a star marker indicates the corresponding values for SN Zwicky for each distribution parameter.
              }
         \label{figure:hist}
   \end{figure}

   Figure \ref{figure:hist} shows the distributions for peak $B$-band apparent magnitude ($m_B$), peak $B$-band absolute magnitude ($M_B$), lens redshift ($z$), SALT2 $x_1$, SALT2 $c$, and angular separation values for the transients that passed each stage of the cuts (the base cuts, S1 and S2). For comparison, the dotted line with a star marker indicates the corresponding values for SN Zwicky for each distribution parameter, using the values given in \citet{Goobar2023} (with the caveat that the SN Zwicky parameters were fitted using a SALT2 time-delay model with four separate images, while our pipeline is only fitting for one resulting SALT2 light curve).
   
   Additionally, we applied our pipeline to the alert photometry for SN Zwicky and found that it passed all of the base cuts, as well as all of the additional cuts described in Sect. \ref{subsection:cuts}, except for the galaxy colour cuts (as they rely on Legacy Survey data, which is not available for this area of sky). However, as is shown in Fig. \ref{figure:gal_cuts}, a catalogue counterpart for SN Zwicky observed by Pan-STARRS would have passed our galaxy colour cuts.

\section{Gold sample of glSN Ia candidates}
\label{section:gold}

\begin{figure*}
        \centering
        \begin{subfigure}[b]{0.32\textwidth}
            \centering
            \includegraphics[width=\textwidth]{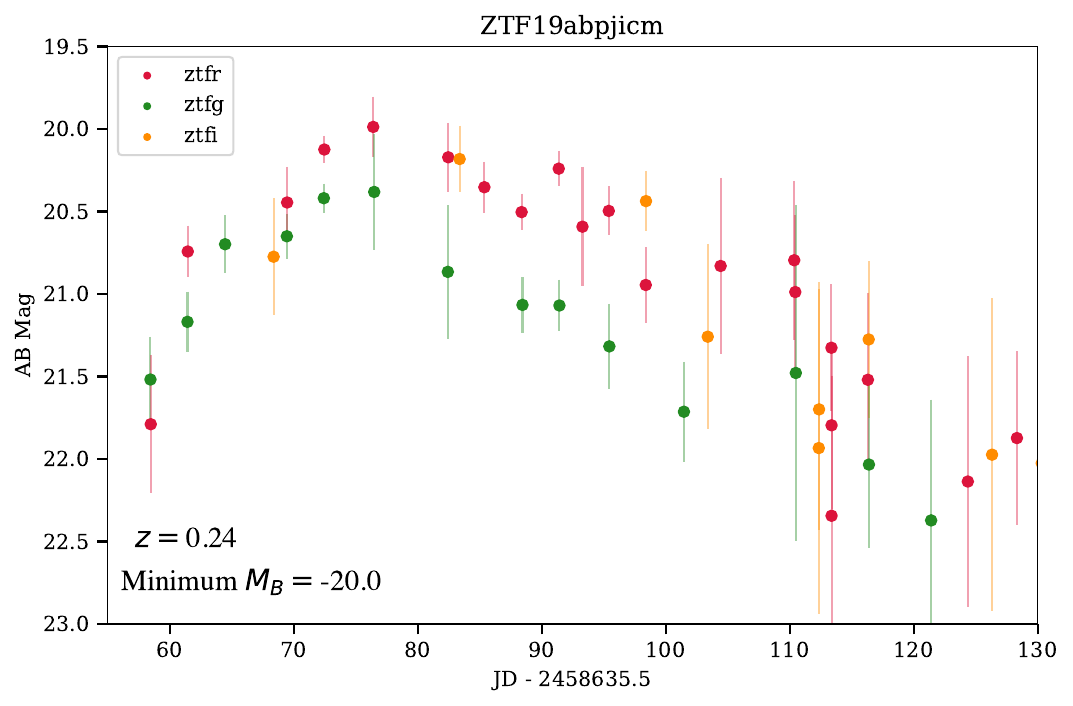}   
        \end{subfigure}
        \begin{subfigure}[b]{0.32\textwidth}  
            \centering 
            \includegraphics[width=\textwidth]{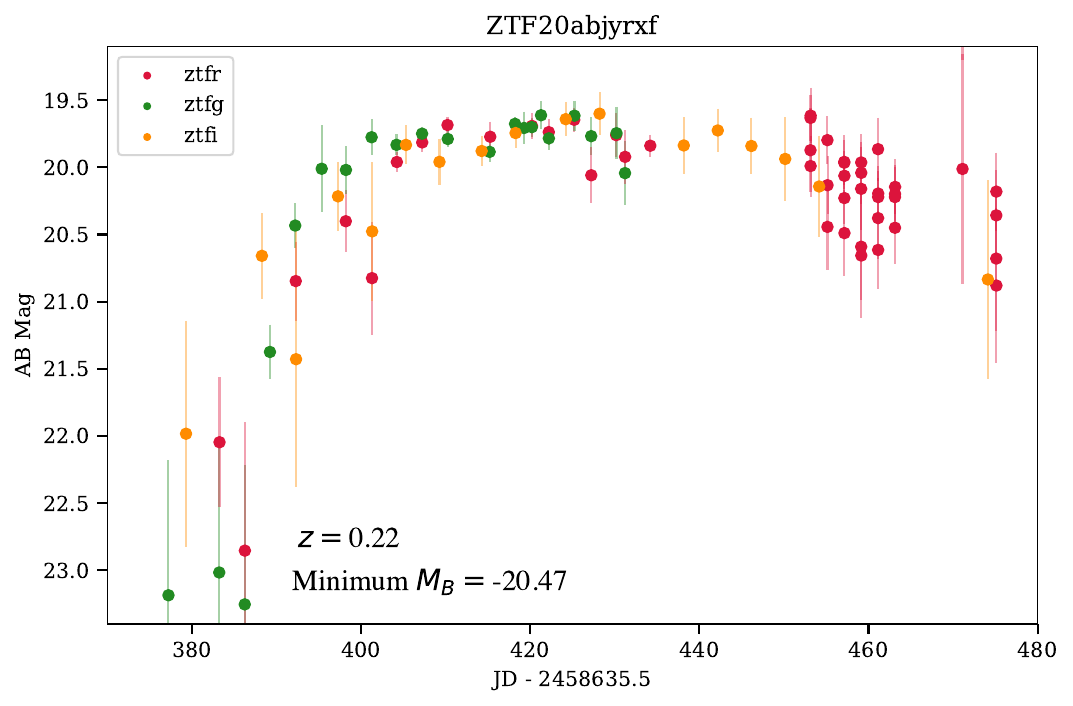}   
        \end{subfigure}
        \begin{subfigure}[b]{0.32\textwidth}   
            \centering 
            \includegraphics[width=\textwidth]{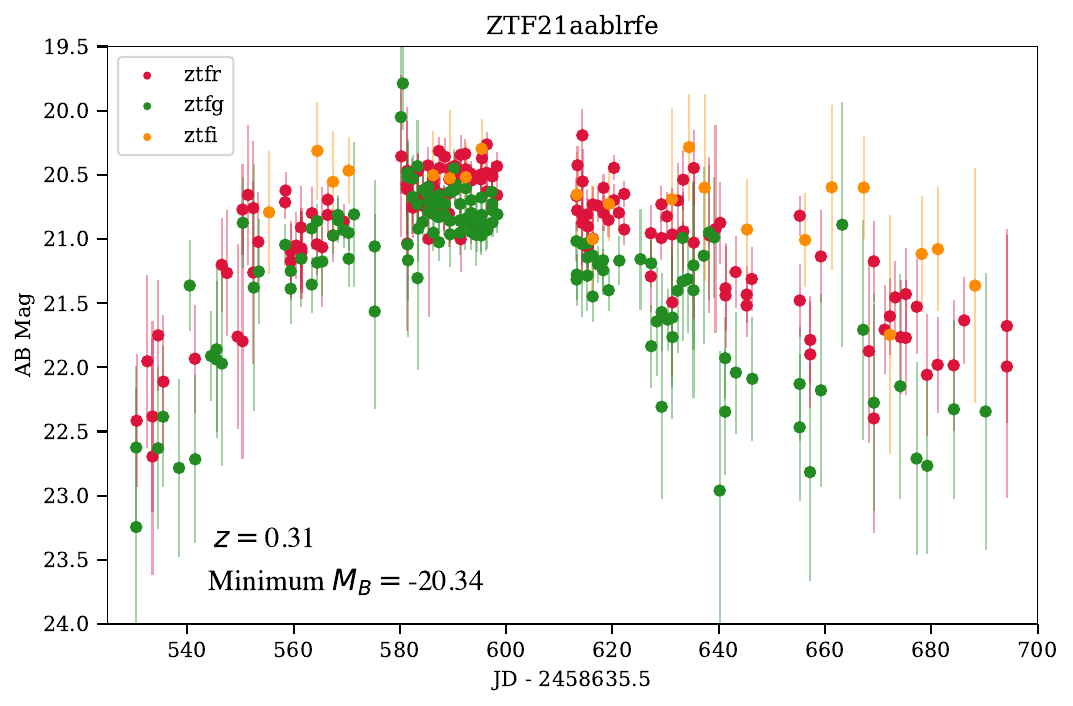}   
        \end{subfigure}
        \begin{subfigure}[b]{0.32\textwidth}   
            \centering 
            \includegraphics[width=\textwidth]{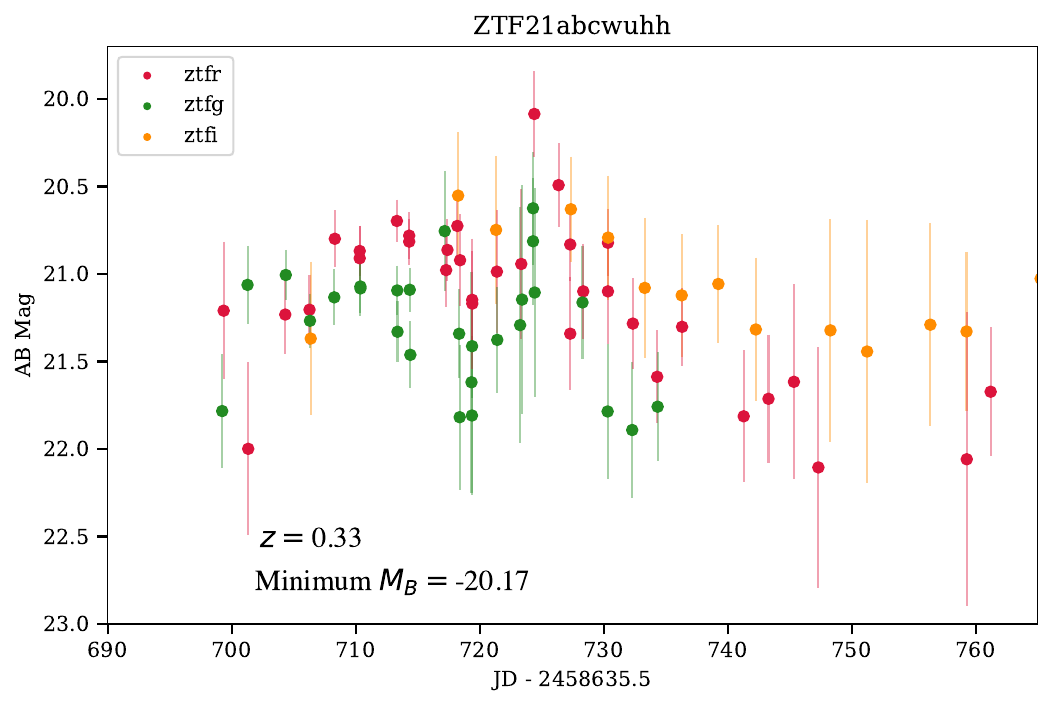} 
        \end{subfigure}
        \begin{subfigure}[b]{0.32\textwidth}   
            \centering 
            \includegraphics[width=\textwidth]{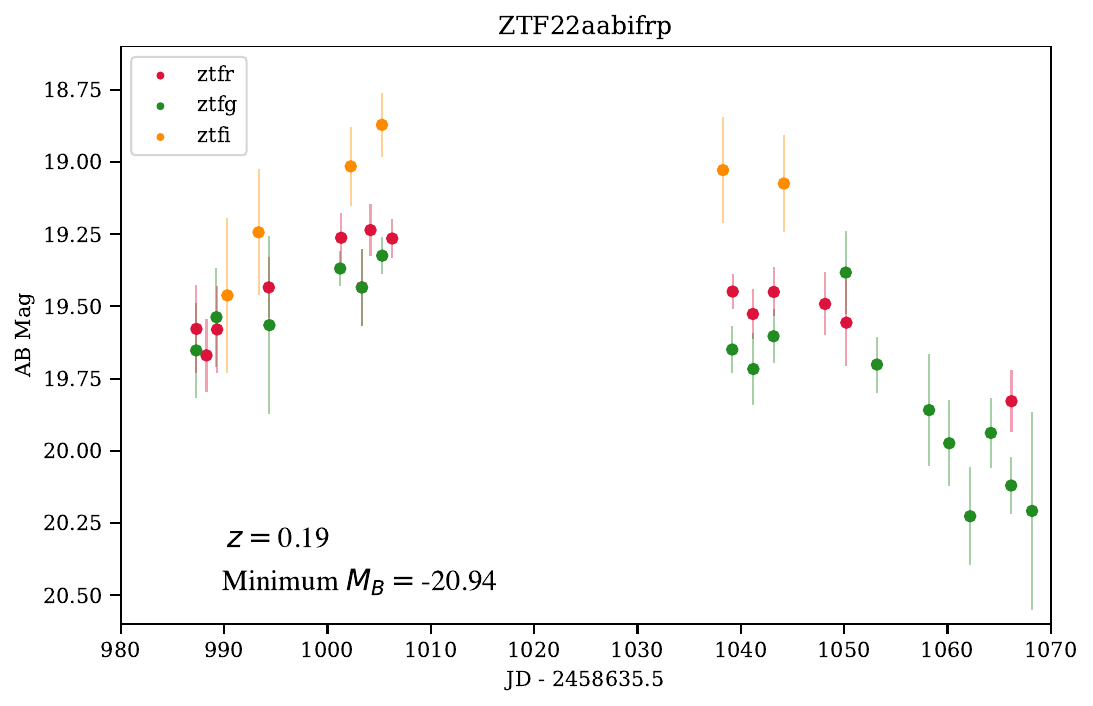}   
        \end{subfigure}
        \begin{subfigure}[b]{0.32\textwidth}   
            \centering 
            \includegraphics[width=\textwidth]{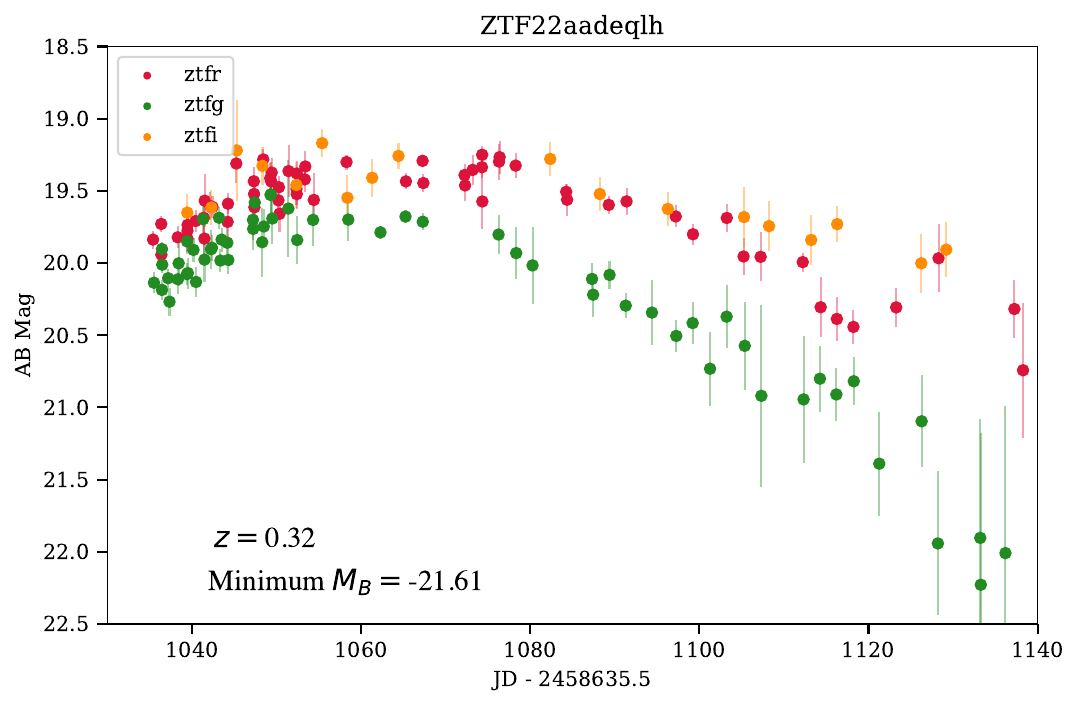}  
        \end{subfigure}
        \begin{subfigure}[b]{0.001\textwidth}   
            \centering
            \color{white}
            \rule{0.5cm}{0.5cm}
        \end{subfigure}
        \begin{subfigure}[b]{0.32\textwidth}   
            \centering 
            \includegraphics[width=\textwidth]{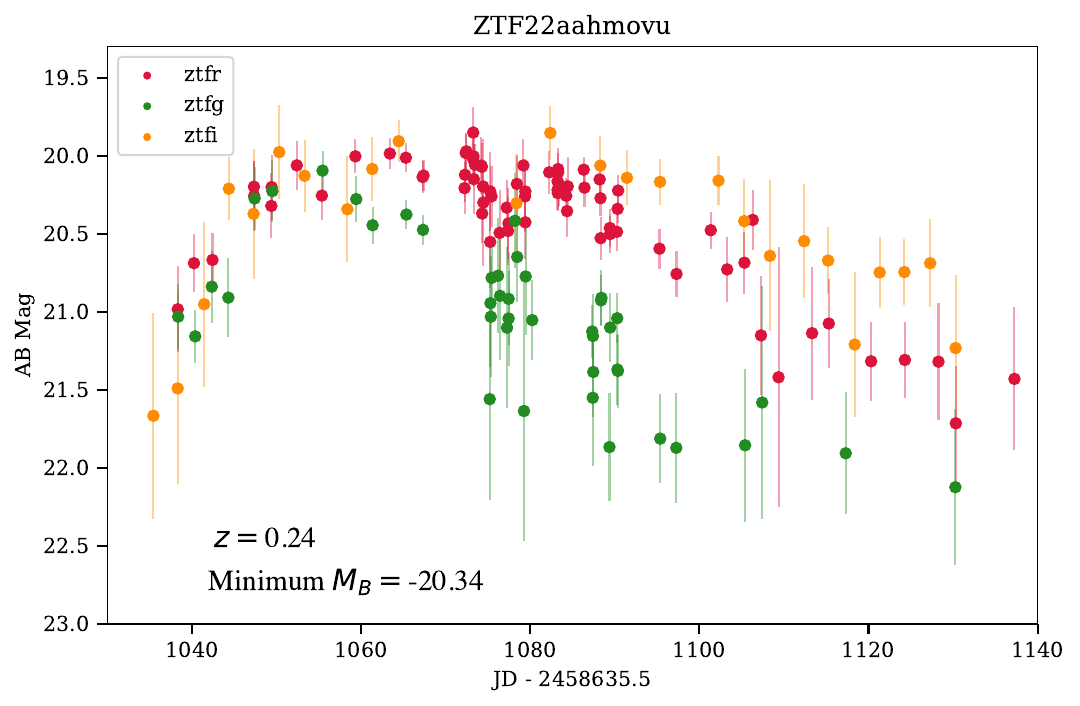}   
        \end{subfigure}
        \caption
        {Light curve plots for the seven candidates in the gold sample that had a spectroscopic redshift from the DESI catalogue and passed all the cuts described in Sect. \ref{subsection:cuts}. Each plot shows the light curve, the lens redshift, and peak absolute magnitude at the lens redshift (from the DESI cross-match).
        } 
        \label{figure:gold_lcs}
    \end{figure*}

\begin{figure*}
        \centering
        \begin{subfigure}[b]{0.245\textwidth}
            \centering
            \includegraphics[width=\textwidth]{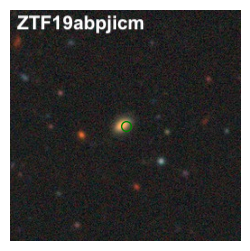}   
        \end{subfigure}
        \begin{subfigure}[b]{0.245\textwidth}  
            \centering 
            \includegraphics[width=\textwidth]{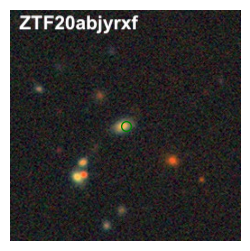}   
        \end{subfigure}
        \begin{subfigure}[b]{0.245\textwidth}   
            \centering 
            \includegraphics[width=\textwidth]{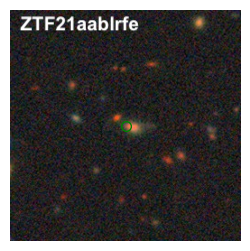}   
        \end{subfigure}
        \begin{subfigure}[b]{0.245\textwidth}   
            \centering 
            \includegraphics[width=\textwidth]{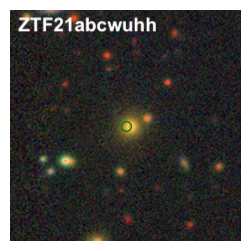} 
        \end{subfigure}
        \begin{subfigure}[b]{0.245\textwidth}   
            \centering 
            \includegraphics[width=\textwidth]{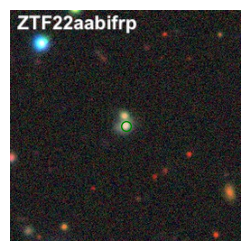}   
        \end{subfigure}
        \begin{subfigure}[b]{0.245\textwidth}   
            \centering 
            \includegraphics[width=\textwidth]{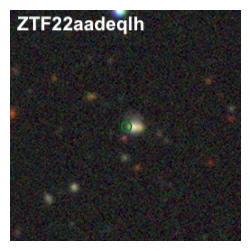}  
        \end{subfigure}
        \begin{subfigure}[b]{0.245\textwidth}   
            \centering 
            \includegraphics[width=\textwidth]{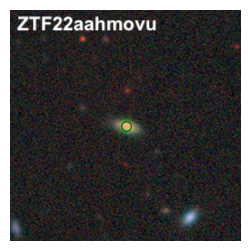}   
        \end{subfigure}
        \caption
        {Image cutouts of the LS field at the location of the transient (indicated by a green circle) for the seven gold sample candidates (the ZTF name for each transient is shown in the top left of the cutout). The galaxies we see are likely the lensing galaxies, and the host is more distant and faint.
        } 
        \label{figure:gold_LS_cutout}
    \end{figure*}

There is a subsample of candidates that pass all the cuts described in Sect. \ref{subsection:cuts} and also have a cross-match with a galaxy counterpart observed by DESI. Table \ref{table:all_short} summarises the candidates that were found by both selection methods. Hereafter we shall refer to this sample of candidates as the `gold sample’. The light curves for the seven candidates in the gold sample are displayed in Fig. \ref{figure:gold_lcs}, which includes their DESI galaxy redshift and peak absolute magnitude (assuming the object is located at the catalogue redshift). Additionally, the Legacy Survey (LS) field cutouts for the seven candidates are shown in Fig. \ref{figure:gold_LS_cutout}, where the location of the ZTF transient is indicated by a green circle. This sample represents our most likely glSN Ia candidates because we are confident that they are more luminous than normal SNe Ia and they have the expected light curve and lens galaxy characteristics of glSNe Ia.

\begin{table}[h!]
\caption{Gold sample}
\label{table:all_short}
\centering
\begin{tabular}{c | c c c c}
\hline\hline
ZTF ID & Phot. $z$ & DESI $z$ & Peak $M_B$ (DESI) \\
\hline
ZTF19abpjicm & 0.24     & 0.24   & $-20.00\pm0.08$         \\
ZTF20abjyrxf & 0.19     & 0.22   & $-20.47\pm0.05$         \\
ZTF21aablrfe & 0.31     & 0.31   & $-20.34\pm0.09$         \\
ZTF21abcwuhh & 0.34     & 0.33   & $-20.17\pm0.12$         \\
ZTF22aabifrp & 0.25     & 0.19   & $-20.94\pm0.06$         \\
ZTF22aadeqlh & 0.27     & 0.32   & $-21.61\pm0.04$         \\
ZTF22aahmovu & 0.22     & 0.24   & $-20.34\pm0.08$         \\
\hline
\end{tabular}
\tablefoot{The full table including the right ascension (RA), declination, and SALT2 fit parameters is Table A.2 (available at the CDS).}
\end{table}

In the following subsections, we examined the seven gold sample candidates further by fitting to a simple two-image SN Ia model (Sect. \ref{subsection:salt2}), calculating the rise, decline, and duration of the light curves (Sect. \ref{subsection:rise_decline}), and by obtaining the galaxy photometry (Sect. \ref{subsection:host_obs}). Each object is also discussed in depth, starting with the most likely candidates (Sect. \ref{subsection:indiv_cand}).

\subsection{Two-image combined SALT2 fit}
\label{subsection:salt2}

We fitted the light curves of the gold sample to a combined SALT2 model with two images. While simulations from \citet{Goldstein2019} predict that most events detected by ZTF would have four images (at a percentage of 62\%, versus 22\% for two images), constraining the parameters for four images simultaneously is challenging without additional information about the system (e.g. the flux ratios of the images to constrain each magnification) and also not possible with the limited number of detections we have for some candidates. By including a second component to the SALT2 light curve, we can show that the transient differs from a typical unlensed SN, which strengthens the argument that the object is multiply imaged. The combined model, in flux space, is given as
\begin{equation}
    f_{\mathrm{model}}(t)= F_{1} \times f_{\mathrm{SALT2}}(t) + F_{2} \times f_{\mathrm{SALT2}}(t- \Delta t),
\end{equation}
where $F_1$, $F_2$ are the relative image magnifications, $\Delta t$ is the second image time delay (in the observer-frame), $t$ is the phase of the light curve, and $f_{\mathrm{SALT2}}$ is the flux model for a single SALT2 template. The SALT2 model also depends on the following parameters: the time of the light curve peak ($t_0$), the amplitude ($x_0$), colour ($c$), and stretch ($x_1$) of the light curve, and the redshift of the transient ($z$). We note that these image magnifications are relative, and not absolute values. If we assume that the lenses can be modelled as a singular isothermal sphere or halo (SIS), then the absolute total image magnification (which is the sum of the individual image magnifications) for the two-image system is given by
\begin{equation}
    \mu_\mathrm{tot} = \frac{2(r + 1)}{{r-1}},
\end{equation}
where we define the flux ratio as $r = F_{1}/F_{2}$.\footnote{For completeness, the individual image magnifications are given as
\begin{equation}
    \mu_\mathrm{1} = \frac{2r}{{r-1}}, \; \mu_\mathrm{2} = \frac{2}{{r-1}}.
\end{equation}}

We allowed the redshift parameter to vary between a lower limit of the DESI $z$, and two times this value. This is because we expect the DESI $z$ to belong to the lens, and the actual supernova to belong to a more distant background galaxy. To avoid degeneracy with the image magnifications, the intrinsic amplitude of the SNe Ia light curve ($x_0$) is fixed to an arbitrary value. We implemented bounds of $[-2,2]$ on $x_1$ and $[-0.5,0.5]$ on $c$, to match the parameters of a normal SN Ia. We limited the time delay for the second image to values between zero and 60 days \citep[this value is motivated by the simulations of SC24 and ][and also implied by the absence of a distinct secondary peak]{Goldstein2019}.

\begin{table}[h!]
\caption{Two-image combined SALT2 model parameters for the gold sample}
\label{table:gold_salt}
\centering
\begin{tabular}{p{1.4cm} | m{1.9cm} m{1.9cm} m{1.9cm} }
\hline\hline
Parameter & ZTF19abpjicm & ZTF20abjyrxf & ZTF22aahmovu \\
\hline
$z$ & $0.44 \pm 0.05$  & $0.24 \pm 0.02$ & $0.35 \pm 0.01$ \\
$t_0$ (MJD) & $58710 \pm 2$ & $59069 \pm 1$ & $59692 \pm 1$ \\
$x_0$ & $(2\pm1)\times10^{-4}$ & $(6\pm1)\times10^{-4}$ & $(2\pm1)\times10^{-4}$ \\
$x_1$ & $0.13 \pm 0.97$ & $2.00 \pm 0.05$ & $2.00 \pm 0.06$ \\
$c$ & $-0.23 \pm 0.05$ & $-0.11 \pm 0.04$ & $0.00 \pm 0.02$  \\
$\Delta t$ (days) & $22 \pm 3$ &  $27 \pm 1$ & $34 \pm 1$ \\
$F_1$ & $0.46 \pm 0.16$ & $0.28 \pm 0.06$ & $0.72 \pm 0.08$ \\
$F_2$ & $0.13 \pm 0.06$ & $0.23 \pm 0.05$ & $0.26 \pm 0.03$ \\
$\mu_\mathrm{tot}$ & $3.6\pm1.3$ & $20.4\pm6.2$ & $4.3\pm0.6$ \\
$\chi^2_r$ & $0.86$ & $2.53$ & $1.38$ \\
\hline
\end{tabular}
\end{table}

\begin{figure*}
        \centering
        \begin{subfigure}[b]{\textwidth}
            \centering
            \includegraphics[width=\textwidth]{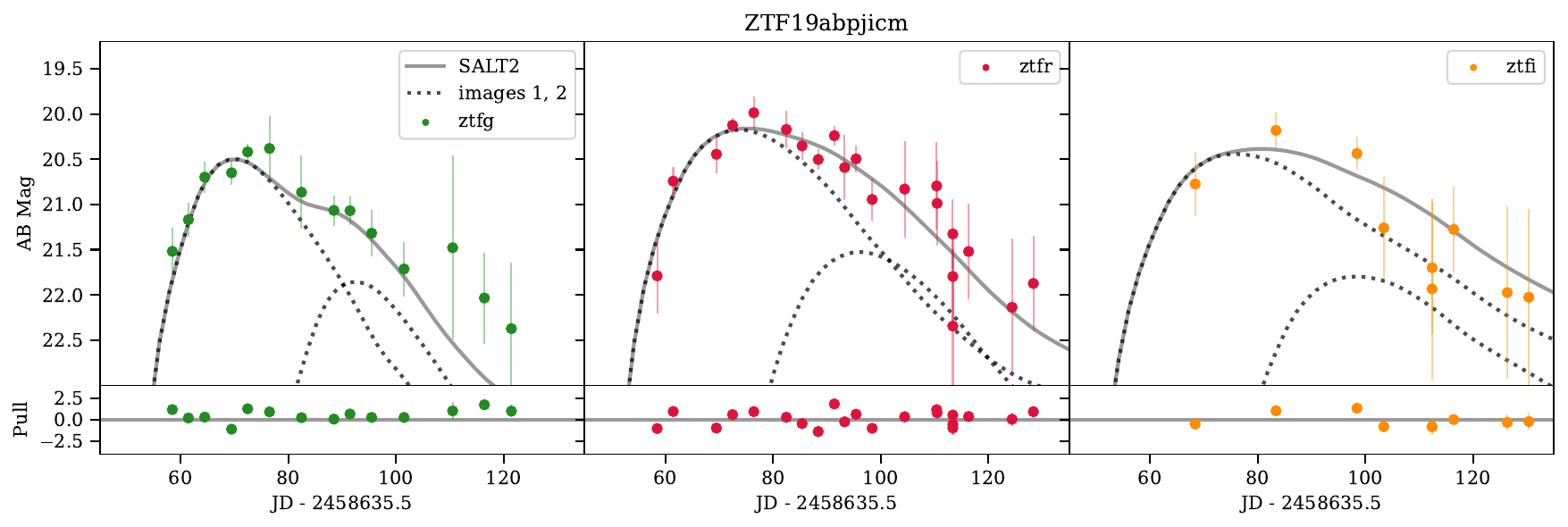}   
        \end{subfigure}
        \begin{subfigure}[b]{\textwidth}
            \centering
            \includegraphics[width=\textwidth]{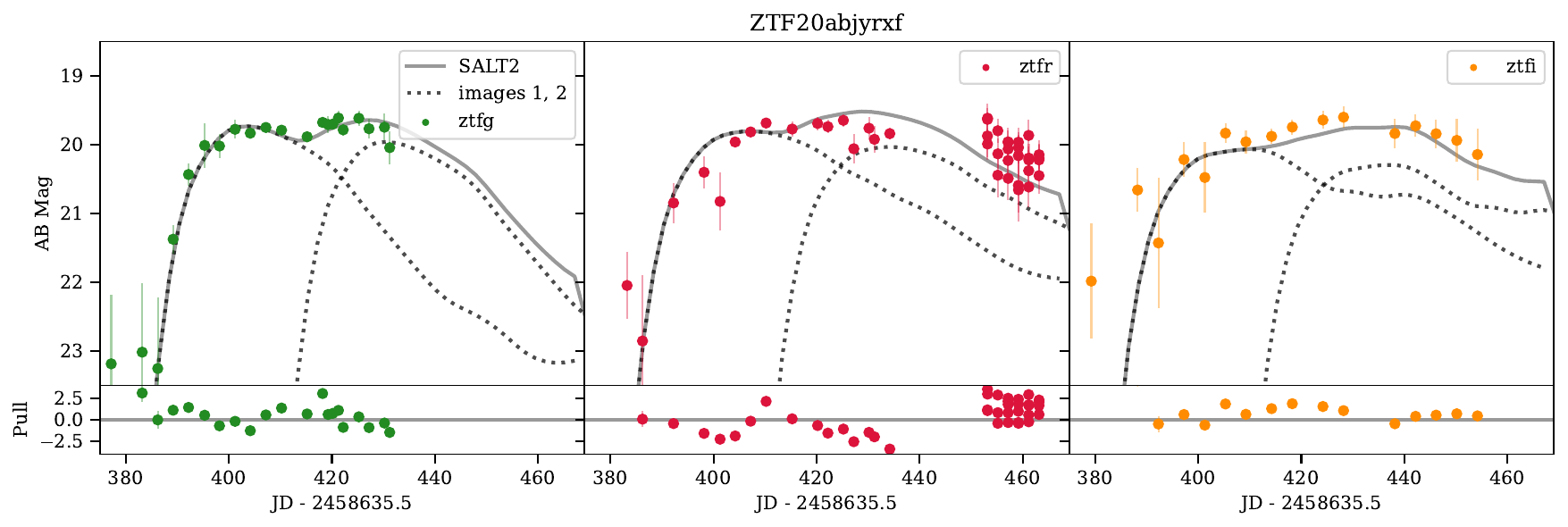}   
        \end{subfigure}
        \begin{subfigure}[b]{\textwidth}
            \centering
            \includegraphics[width=\textwidth]{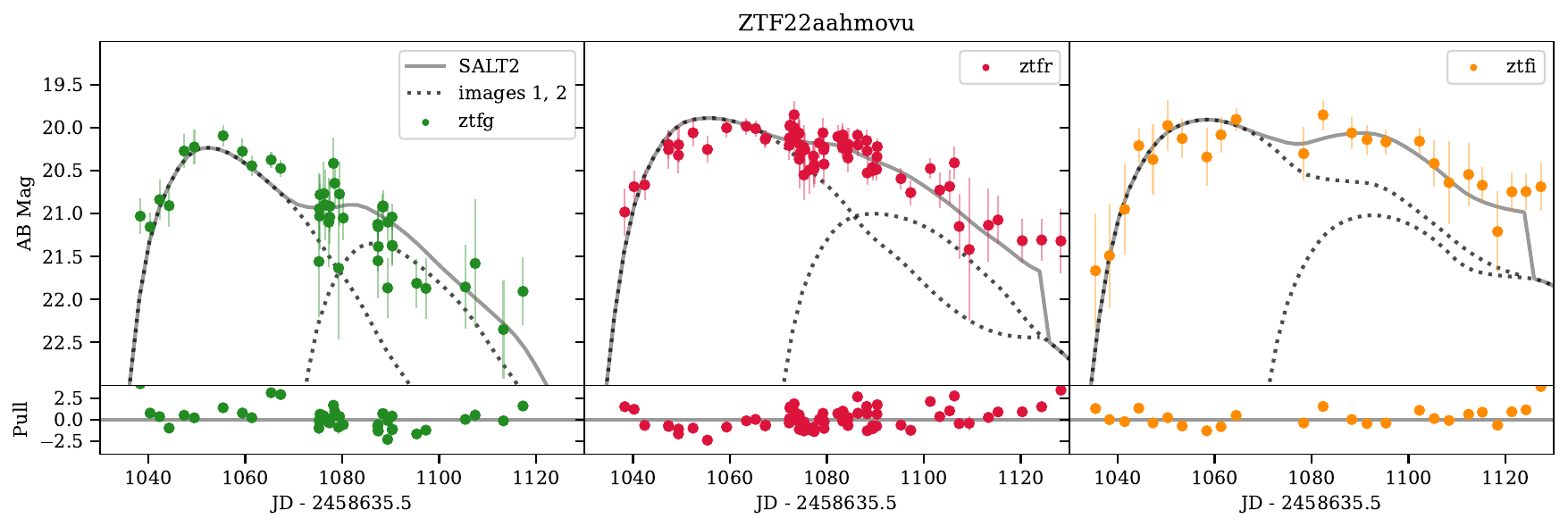}   
        \end{subfigure}
        \caption
        {Two-image combined SALT2 template fits for the three candidates in the gold sample that had a reasonable fit; ZTF19abpjicm, ZTF20abjyrxf, and ZTF22aahmovu. Each plot shows the light curve in $g$-, $r$-, and $i$-bands, the best fit combined SALT2 model (solid grey line), and the model for each lensed image (dashed grey lines). The fit parameters are given in Table \ref{table:gold_salt}.
        } 
        \label{figure:gold_salt}
    \end{figure*}

Of the seven candidates in the gold sample, only three of them converged to reasonable two-image combined SALT2 fits: ZTF19abpjicm, ZTF20abjyrxf, and ZTF22aahmovu. Figure \ref{figure:gold_salt} illustrates the best fit two-image model for each candidate and Table \ref{table:gold_salt} displays the fit parameters (as well as the derived parameters $\mu_\mathrm{tot}$, $\mu_1$, and $\mu_2$) for each model. The best fit time delays are larger than what we observed for SN Zwicky (where the time delays for all the images were less than a day) but are not unexpected, according to the simulations of SC24. They find that the median maximum time delay for glSNe Ia with a peak apparent magnitude of $m>19$ is 8.9 days in the rest-frame (which would correspond to 12.5 days in the observer-frame, assuming a glSN at $z=0.4$). Additionally, they predict that approximately 20\% of glSNe Ia in this magnitude regime would have maximum time delays of greater than 25 days in the rest-frame.

The remaining four objects in the gold sample have poor combined two-image SALT2 fits because they are too wide to only display a single peak (ZTF21aablrfe, ZTF22aabifrp, and ZTF22aadeqlh) or they are too noisy to provide a convincing fit (ZTF21abcwuhh). The light curves and corresponding fits are shown in Fig.~\ref{figure:badgold_salt2} of the Appendix. The two-image fits for the wide objects produced larger time delays, for which we would have expected two distinct light curve peaks. However, this is not evident from any of the light curves. We note that a model with three or four images could explain the wider light curves and the absence of bumps, but this is not something that we explicitly modelled (due to the difficulty in constraining the extra model parameters).

\begin{table}[h!]
\caption{Two-image combined SALT2 model parameters for the gold sample}
\label{table:gold_img}
\centering
\begin{tabular}{p{1.4cm} |p{1.9cm} p{1.9cm} p{1.9cm}}
\hline\hline
Parameter & ZTF19abpjicm & ZTF20abjyrxf & ZTF22aahmovu \\
\hline
$\Delta \theta$ ($\arcsec$) & $1.1$  & $6.7$ & $1.7$ \\
$v$ ($\mathrm{km\,s^{-1}}$) & $190$ & $480$ & $240$ \\
\hline
\end{tabular}
\tablefoot{These values are heavily reliant on the source redshift and the time delays, which are unknown and only estimated by the SALT2 two-image fits.}
\end{table}

Additionally, we calculated the expected image separation ($\Delta \theta$) from our best fit two-image SALT2 parameters. Once again assuming the lens to be an SIS, we have the following equation for image separation, \citep[derived from e.g.][]{Mortsell2020}:
\begin{equation}
\label{equation:deltatheta}
    \Delta \theta = 0.27\arcsec \sqrt{ \left(\frac{\Delta t}{\mathrm{days}} \right) \left( \frac{500 \: \mathrm{Mpc}}{2 \: D_l} \right) \left( \frac{D_s}{D_{ls}} \right) \frac{(r+1)}{(r-1)} \frac{1}{(1+z_l)} },
\end{equation}
where $z_l$ is the lens redshift, and $D_l$, $D_s$, $D_{ls}$ are the angular diameter distances to the lens, source, and between the lens and the source, respectively. Furthermore, the image separation scales with the square of the velocity dispersion, $v$, of the lensing galaxy, according to the following equation:
\begin{equation}
\label{equation:v}
    \Delta \theta = 1.15\arcsec \left( \frac{v}{200 \: \mathrm{km\,s^{-1}}}  \right)^2.
\end{equation}
We combined Equation \ref{equation:deltatheta} and \ref{equation:v} to calculate $v$ in terms of our fitted parameters:
\begin{equation}
    v = 47 \: \mathrm{km\,s^{-1}} \left[ \left(\frac{\Delta t}{\mathrm{days}} \right) \left( \frac{500 \: \mathrm{Mpc}}{2 \: D_l} \right) \left( \frac{D_s}{D_{ls}} \right) \frac{(r+1)}{(r-1)} \frac{1}{(1+z_l)} \right]^{1/4}
\end{equation}

The calculated values for $\Delta \theta$ and $v$ are displayed in Table \ref{table:gold_img} for the three objects with reasonable two-image combined SALT2 fits, derived from the parameters in Table \ref{table:gold_salt}.

\subsubsection{Fitting a two-image SALT2 model to superluminous supernovae and tidal disruption events}
\label{subsubsection:twoimage_SLSN}
To illustrate the difference between the light curves of our candidates and those of our typical contaminants, we also performed fits of the two-image combined SALT2 model to the forced photometry light curves of SLSNe and TDEs from BTS.

For a sample of 55 SLSNe (34 SLSNe-I and 21 SLSNe-II), we found that only two (3.6\%) objects could be fit with a reduced $\chi^2$ of less than five. The ZTF names of the SLSNe are ZTF19aavouyw and ZTF21aaarmti, with reduced $\chi^2$ values of 3.6 and 1.4, respectively. Both of these objects were SLSNe-I, and we observed a trend that the SLSNe-II fits had higher reduced $\chi^2$ values and poorer fits by eye. The light curves and corresponding fits are shown in Fig.~\ref{figure:btsslsn_salt2} of the Appendix. It should be noted that these fits do not look convincing, and the pull values are larger than we would expect for lower reduced $\chi^2$ values. This is because there are large uncertainties in the resulting SALT2 model (particularly in the $g$-band) that are included in the $\chi^2$ calculation. As a result, we conclude that both SLSN-I and SLSN-II are not well modelled by the two-image SALT2 model. 

For a sample of 18 TDEs,  we found that only one (5.6\%) object could be fit with a reduced $\chi^2$ of less than five. The ZTF name of this TDE is ZTF22aagvrlq, with a reduced $\chi^2$ value of 2.8. The light curve and corresponding fit are shown in Fig.~\ref{figure:btstde_salt2} of the Appendix. Once again, it is notable that the fit is very poor by eye. This is not surprising, given the characteristic light curve shape of TDEs. Therefore, we conclude that it is unlikely that a typical TDE would be well modelled by the two-image SALT2 model.

This adds further evidence in favour of a glSN Ia interpretation for our candidates ZTF19abpjicm, ZTF20abjyrxf, and ZTF22aahmovu.

\subsubsection{Comparison with SLSN and TDE model fit}
\label{subsubsection:csm_tde_model_fit}
To strengthen the argument that our gold candidates are glSNe Ia, we also performed fits to SLSN and TDE models as a comparison for the three objects that are well fit by the two-image SALT2 template. To do this, we used the software package \texttt{redback}\footnote{\href{https://github.com/nikhil-sarin/redback}{https://github.com/nikhil-sarin/redback}} \citep{Sarin2024}, which relies on the Bayesian parameter inference tool \texttt{Bilby}\footnote{\href{https://git.ligo.org/lscsoft/bilby}{https://git.ligo.org/lscsoft/bilby}} \citep{Ashton2019}. Within \texttt{redback}, it is possible to fit a wide range of transient models with user-defined priors and use Bayesian inference to determine the best fit parameters for the model.

Although the exact powering mechanism behind SLSNe is unknown, several models exist to explain the high luminosities that we observe. One possible power source is strong circumstellar interaction (CSI) between the ejecta and the material surrounding the transient, which was likely ejected by the progenitor before the explosion. It is possible that the high luminosities of hydrodogen-rich SLSN-II are caused by CSI, which is usually evidenced by strong hydrogen emission lines similar to the narrow H$\alpha$ lines shown in the spectra of SN IIn \citep[e.g.][]{Smith2007}. Another possible powering mechanism is a central engine, such as the spin-down of a magnetar. There is evidence that the spin-down magnetar model \citep{Kasen2010,Woosley2010}, where energy is transferred from a strongly magnetised neutron star to the ejecta, is the cause of the high lumninosities of hydrogen-poor SLSN-I. However, since SLSN-II are the only SLSN subclass that have been found in luminous, Milky Way-like galaxies \citep{GalYam2012}, this subtype is the most likely contaminant in our sample. Conversely, SLSN-I are found in lower mass, lower luminosity, metal-poor hosts \citep{Lunnan2014,Perley2016,Chen2017}, which would not pass our galaxy cuts described in Sect.~\ref{subsection:cuts}.

Therefore, we performed a \texttt{redback} fit to an SLSN model that is consistent with the characteristics of a hydrogen-rich subtype. Our model combines CSI and nickel-cobalt decay \citep[the \texttt{csmni} model described in MOSFiT;][]{Guillochon2018}. We used the nested sampler \texttt{nessai}\footnote{\href{https://github.com/mj-will/nessai}{https://github.com/mj-will/nessai}} \citep{Williamsnessai,Williams2021,Williams2023} and implemented broad uniform priors. Figure~\ref{figure:gold_slsn} shows the resulting models for the three promising candidates.

For the TDE case, we performed a \texttt{redback} fit to the \texttt{tde\_fallback} model \citep[which is based on the \texttt{tde} model described in MOSFiT;][]{Guillochon2013,Guillochon2018,Mockler2019}. We used the nested sampler \texttt{nestle}\footnote{\href{https://github.com/kbarbary/nestle}{https://github.com/kbarbary/nestle}} \citep{Barbary2021} and implemented broad uniform priors. Figure~\ref{figure:gold_tde} shows the resulting models for the three promising candidates.

To compare models, we calculated the Bayesian information criterion \citep[BIC;][]{Schwarz1978} for the light curve fits to the individual SALT2 template, the two-image SALT2 template, SLSN-II model, and the TDE model, which is shown in Table~\ref{table:bic}.

\begin{table}[h!]
\caption{BIC comparison for several model fits to our best glSN candidates from the gold sample}
\label{table:bic}
\centering
\begin{tabular}{p{1.5cm} | m{1.9cm} m{1.9cm} m{1.9cm} }
\hline\hline
Model & ZTF19abpjicm & ZTF20abjyrxf & ZTF22aahmovu \\
\hline
Single SALT2& 61.8 & 361.3 & 331.9 \\
Two-img. SALT2& 67.3 & 263.8 & 217.2 \\
SLSN-II (CSI+Ni)& 90.7 & 52.4 & 40.9 \\
TDE& 111.6 & 40.1 & 52.9 \\
\hline
\end{tabular}
\end{table}

For the case of ZTF19abpjicm, we see that a single SALT2 fit is the preferred model, closely followed by a two-image SALT2 template. This is likely due to the BIC penalising the two-image model for the additional fit parameters. Additionally, the brightness of ZTF19abpjicm excludes a typical SN Ia explanation, as we shall discuss in Sect~\ref{subsubsection:ZTF19abpjicm}. The larger BIC values for the SLSNe and TDE models support our hypothesis that this object is more likely to be a glSN Ia. Additionally, we can see in Figs.~\ref{figure:gold_slsn} and \ref{figure:gold_tde} that the SLSNe and TDE models fail to capture the behaviour of the light curve.

For ZTF20abjyrxf, the BIC values suggest that a TDE or SLSN interpretation is preferred. The posterior resulting from the TDE model predicts a blackhole mass ($M_{\mathrm{BH}}$) of $9.75^{+0.89}_{-0.99} \times10^{6}~\mathrm{M_{\odot}}$ and a disrupted star mass ($M_{\mathrm{star}}$) of $12.62^{+1.57}_{-2.50} \:~\mathrm{M_{\odot}}$, which are on the upper limits of the mass ranges of known TDEs \citep{Wevers2017,Ryu2020,Chang2024} but are still consistent with expectations. However, without any additional evidence supporting a TDE interpretation (such as the presence of a dust echo), this conclusion is less likely. Additionally, we can see in Fig.~\ref{figure:gold_tde} that the model fails to capture the behaviour of the light curve at the pre-peak epoch. Considering the SLSN interpretation, the posterior predicts a high ejecta mass ($M_{\mathrm{ej}}$) of $5.42^{+1.13}_{-0.78} \:~\mathrm{M_{\odot}}$ and a high nickel fraction ($f_{\mathrm{Ni}}$) of $0.81^{+0.13}_{-0.19}$. This would correspond to a nickel mass of greater than a solar mass, which would suggest a pair instability supernova \citep{GalYam2009}. This is also a very rare transient event, with a small number of possible candidates \citep[e.g.][]{Schulze2024}, and is therefore very unlikely. As a result, we expect that the two-image SALT2 model for ZTF20abjyrxf is the most likely in this comparison.

Finally, the BIC values for ZTF22aahmovu also suggest that a TDE or SLSN interpretation is preferred. Once again, the posterior for the TDE model predicts a high blackhole mass ($M_{\mathrm{BH}}$) of $18.99^{+0.68}_{-1.16} \times10^{6}~\mathrm{M_{\odot}}$ and a disrupted star mass ($M_{\mathrm{star}}$) of $11.29^{+1.50}_{-1.64} \:~\mathrm{M_{\odot}}$ which are physically reasonable. However, the TDE model once again fails to capture the behaviour of the light curve at the pre-peak epoch in Fig.~\ref{figure:gold_tde}. The posterior parameters of the SLSN model produce unrealistic results; the model predicts a very low $M_{\mathrm{ej}}$ of $3.53^{+1.43}_{-0.06} \times10^{-4}~\mathrm{M_{\odot}}$ which would suggest a failed supernova or a different type of stellar event. As this is not compatible with the observed luminosity, this SLSN-II model does not accurately describe ZTF22aahmovu, and we propose that the two-image SALT2 model is the most likely in this comparison.

\subsection{Rise, decline, and duration calculations}
\label{subsection:rise_decline}

We calculated the time of rise, decline, and duration in the $r$-band for each of the seven gold sample candidates, shown in Table \ref{table:rise_decline}. This is a useful characteristic for comparison with contaminants like SLSNe, which typically have a longer rise time and duration. However, we note that glSNe Ia with a brighter second bump would display a longer rise and a shorter decline (such as in the case of ZTF20abjyrxf in Fig. \ref{figure:gold_salt}).

We defined the rise time as the time it takes the transient to rise from $50\%$ of the peak flux value to the peak ($t_{\mathrm{peak}}-t_{50\%,\: \mathrm{r}}$), and conversely for the decline time ($t_{50\%,\: \mathrm{d}} - t_{\mathrm{peak}}$). Hence, the duration was defined as the sum of the rise and decline times. The $r$-band was chosen as this band usually had the most detections, due to the regular cadence and redness of the objects. For simplicity, and to avoid introducing any biases from models, we assumed that the largest flux value was the peak of the light curve.  The values $t_{50\%,\: \mathrm{r}}$ and $t_{50\%,\: \mathrm{d}}$ were determined by averaging the time of detection before and after the flux values crossed the $50\%$ threshold. If there was no detection before (or after) the $50\%$ peak flux threshold, then the first (or last) detection date was chosen. From our crude estimations, we expect uncertainties on the rise or decline times of approximately $\pm1$ day (half the typical survey cadence).

\begin{table}[h!]
\caption{Duration, rise and decline times for the gold sample}
\label{table:rise_decline}
\centering
\begin{tabular}{l | lll}
\hline\hline
\multicolumn{1}{l}{} & \multicolumn{3}{c}{Time in observer-frame (days)}  \\
ZTF ID           & Duration & Rise & Decline \\
\hline
ZTF19abpjicm & 31.4            & 10.9             & 20.5                \\
ZTF20abjyrxf & 58.4            & 50.4             & 8.0                 \\
ZTF21aablrfe & 72.8            & 63.4             & 9.4                 \\
ZTF21abcwuhh & 18.5            & 12.6             & 5.9                 \\
ZTF22aabifrp & 78.9            & 14.0             & 64.9                \\
ZTF22aadeqlh & 92.4            & 53.4             & 39.0                \\
ZTF22aahmovu & 57.5            & 28.4             & 29.1                \\
\hline
\end{tabular}
\end{table}

\begin{figure}[h!]
   \centering
   \includegraphics[width=\hsize]{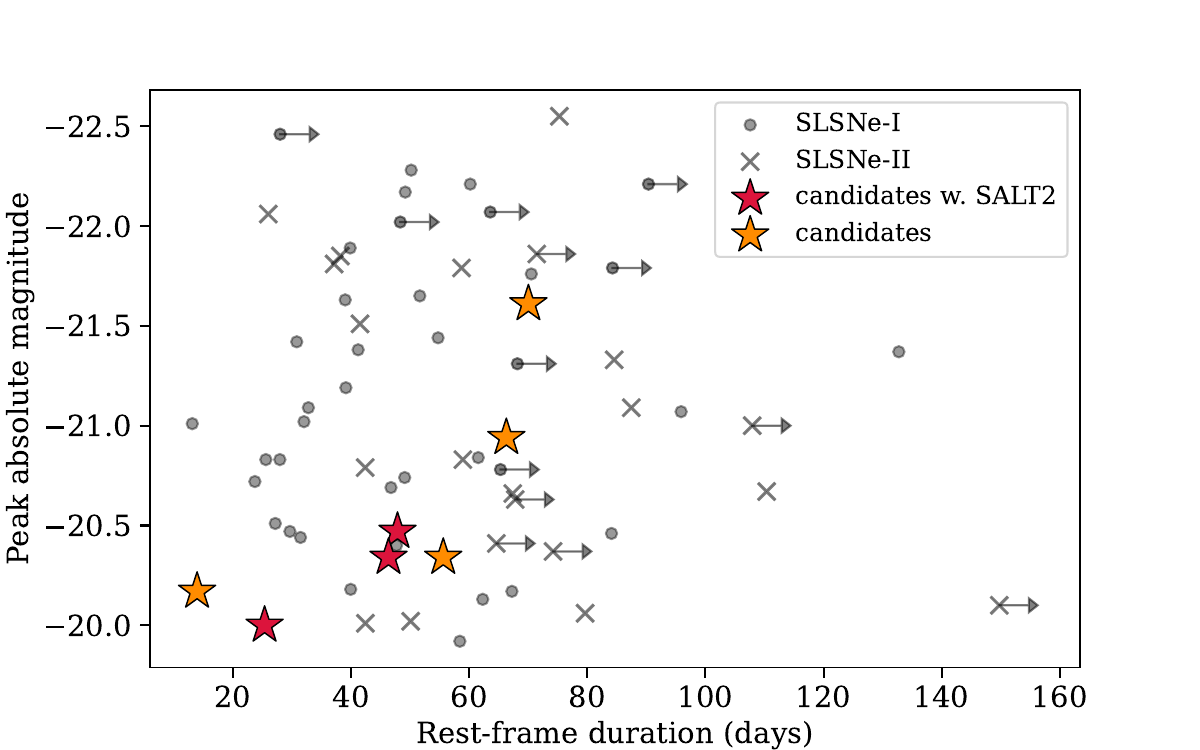}
      \caption{Plot comparing the rest-frame duration of the seven gold candidates (red stars for candidates with a reasonable two-image SALT2 fit and gold stars for the remaining objects) with SLSNe-I and SLSNe-II from the ZTF Bright Transient Survey. In this plot, we assume that the catalogue-matched redshift for our candidates belongs to the host (which would be the case if we treated the candidates like potential SLSNe).
              }
         \label{figure:duration_slsn}
   \end{figure}

A comparison of the duration of the glSN Ia candidates in the gold sample with the duration of SLSNe-I and SLSNe-II from the Bright Transient Survey \citep{Perley2020} is shown in Fig. \ref{figure:duration_slsn}. We do not see a clear distinction between our glSN candidates and the SLSNe, but we note that the selection of candidates that had a reasonable two-image SALT2 fit displayed a shorter duration.

\subsection{Candidate lens galaxy properties}
\label{subsection:host_obs}

Next, we estimated the stellar masses of the gold sample candidate's galaxies to determine whether they are consistent with SLSN host galaxies or lens galaxies. For each candidate in the gold sample, we retrieved science-ready co-added images from the Galaxy Evolution Explorer general release 6/7 \citep[GALEX;][]{Martin2005a}, the Sloan Digital Sky Survey DR 9 \citep[SDSS;][]{Ahn2012a}, the Panoramic Survey Telescope and Rapid Response System (Pan-STARRS, PS1) DR1 \citep{Chambers2016a}, DESI Legacy Imaging Surveys (Legacy Surveys, LS) \cite{Dey2018a}, the Two Micron All Sky Survey \citep[2MASS;][]{Skrutskie2006a}, and images from the Wide-Field Infrared Survey Explorer \citep[WISE;]{Wright2010a} processed by \citet{Lang2014a}. We measured the brightness of the potential lensing galaxy using LAMBDAR\footnote{\href{https://github.com/AngusWright/LAMBDAR}{https://github.com/AngusWright/LAMBDAR}} \citep[Lambda Adaptive Multi-Band Deblending Algorithm in R;][]{Wright2016a} and the methods described in \citet{Schulze2021a}. Tables \ref{table:host_phot1} to \ref{table:host_phot7} in the Appendix display the measurements in the different bands. 

\begin{table}
  \caption{Candidate lens galaxy properties of the glSN Ia candidates in the gold sample\label{table:host_prop}}
  \centering
  \begin{tabular}{l|cc}
  \hline\hline
    ZTF ID & $\log (M_\star/M_\odot)$ & SFR \\
      &                   & $(M_\odot\,\rm yr^{-1})$\\
  \hline
ZTF19abpjicm &$ 10.88^{+0.10}_{-0.30} $&$ 13.34^{+8.10 }_{- 3.86}$\\
ZTF20abjyrxf &$ 10.40^{+0.19}_{-0.39} $&$  0.71^{+5.34 }_{- 0.69}$\\
ZTF21aablrfe &$ 10.76^{+0.13}_{-0.14} $&$ 19.08^{+7.66 }_{- 4.73}$\\
ZTF21abcwuhh &$ 11.61^{+0.11}_{-0.24} $&$  1.57^{+3.36 }_{- 1.57}$\\
ZTF22aabifrp &$  9.90^{+0.21}_{-0.38} $&$  7.56^{+13.13}_{- 4.66}$\\
ZTF22aadeqlh &$ 10.65^{+0.16}_{-0.18} $&$ 31.45^{+17.48}_{-12.65}$\\
ZTF22aahmovu &$ 10.71^{+0.17}_{-0.49} $&$  4.48^{+11.43}_{- 3.17}$\\
\hline
\end{tabular}
\end{table}

\begin{figure}[h!]
   \centering
   \includegraphics[width=\hsize]{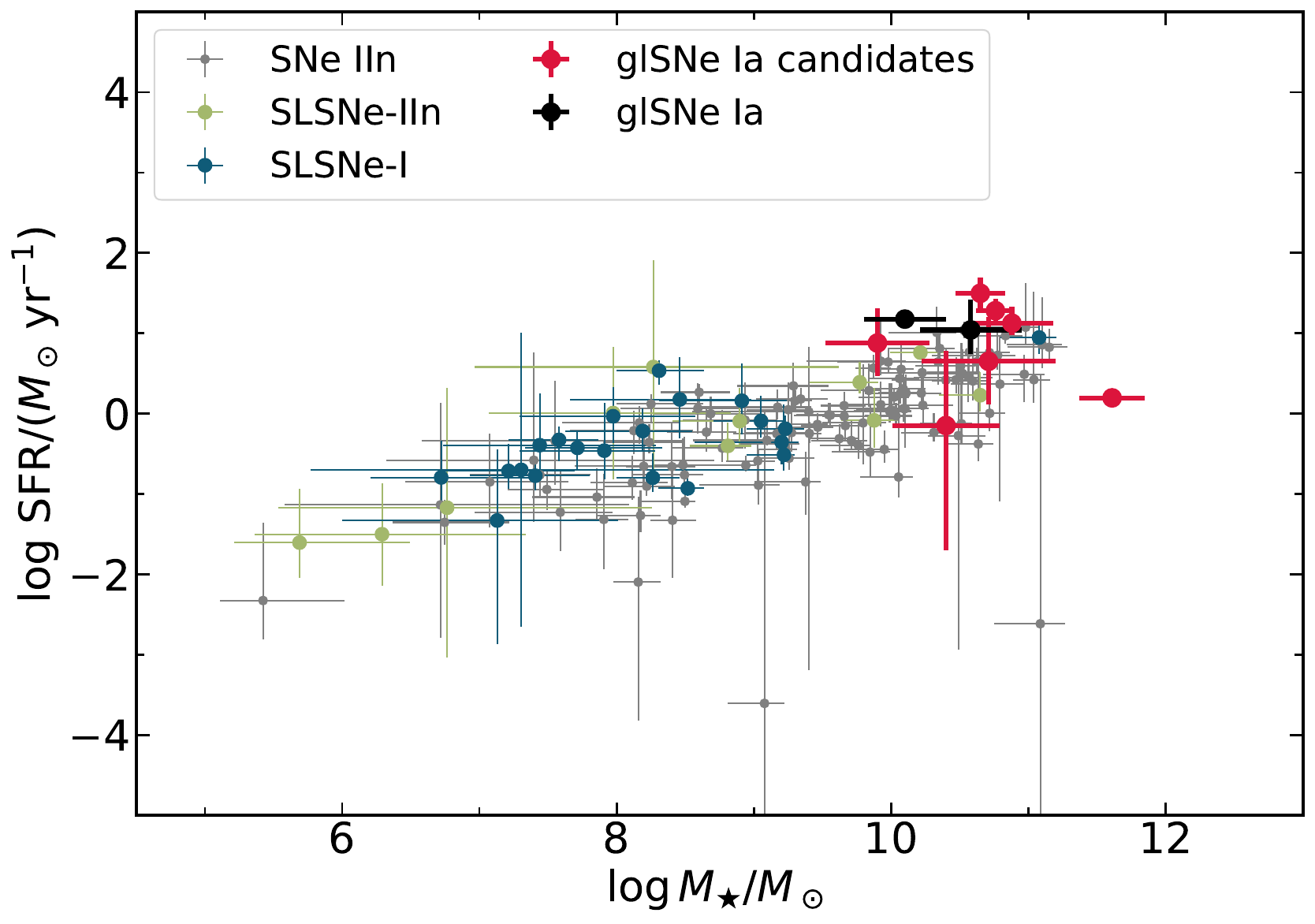}
      \caption{Star formation rate and stellar mass of the glSN Ia candidate lens galaxies (black), known glSNe Ia lens galaxies (red), and SLSNe-I (green), SLSNe-II (blue) and SNe IIn (grey) host galaxies. The lens galaxy of the glSN Ia candidates are similar to hosts of SLSNe-IIn and SNe IIn, suggesting that some glSN Ia candidates could be SLSNe-IIn. However, the masses are also consistent with previously observed galaxy lenses.
              }
         \label{figure:host_mass_sfr}
   \end{figure}

We modelled the observed spectral energy distribution with the software package \texttt{Prospector} version 1.1 \citep{Johnson2021a}\footnote{\texttt{Prospector} uses the Flexible Stellar Population Synthesis (FSPS) code \citep{Conroy2009a} to generate the underlying physical model and \texttt{python-fsps} \citep{ForemanMackey2014a} to interface with FSPS in Python. The FSPS code also accounts for the contribution from the diffuse gas based on the \texttt{Cloudy} models from \citet{Byler2017a}. We use the dynamic nested sampling package \texttt{dynesty} \citep{Speagle2020a} to sample the posterior probability.}. We assume a Chabrier IMF \citep{Chabrier2003a} and approximate the star formation history (SFH) by a linearly increasing SFH at early times followed by an exponential decline at late times [functional form $t \times \exp\left(-t/\tau\right)$, where $t$ is the age of the SFH episode and $\tau$ is the $e$-folding timescale]. The model is attenuated with the \citet{Calzetti2000a} model. The priors of the model parameters are set identically to those used by \citet{Schulze2021a}. Table \ref{table:host_prop} summarises the properties of the galaxy.

To put these measurements in context of hosts of other SN classes, we show in Fig. \ref{figure:host_mass_sfr} the mass versus star formation rate of the candidate glSNe Ia lens galaxies and the hosts of hydrogen-poor superluminous supernovae (SLSNe-I), interaction-powered supernovae (SNe IIn) and superluminous supernovae (SLSNe-IIn) from the PTF survey \citep{Schulze2021a}. The lens galaxies of the known glSNe Ia, SN Zwicky and iPTF16geu, are also plotted for comparison. The potential lens galaxies of the glSN Ia candidates are significantly more massive than the hosts of H-poor SLSNe, but they are consistent with the properties of SLSNe-IIn and SNe IIn hosts. However, they are also consistent with previously observed galaxy lenses, so we cannot exclude them as potential lensing galaxies.


\subsection{Individual candidate discussion}
\label{subsection:indiv_cand}

\subsubsection{ZTF19abpjicm}
\label{subsubsection:ZTF19abpjicm}
ZTF19abpjicm has a good two-image combined SALT2 fit (with a reduced $\chi^2$ of 0.86), and a relatively short duration of 31.4 days. From our BIC comparison to SLSN and TDE models in Sect.~\ref{subsubsection:csm_tde_model_fit}, the two-image combined SALT2 model is the preferred interpretation. Therefore, we state that the light curve is not typical for contaminants such as SLSNe or TDEs.

Fitting the light curve with a single SALT2 template also produces a reasonable fit, with a larger $x_1$ of $2.40\pm0.62$ and a reduced $\chi^2$ of 1.04, which indicates the object could be an unlensed SN Ia. The photometric and spectroscopic redshift values agree, which suggests that we can trust that this object is over-luminous. This means that it cannot be a normal SN Ia. The only SN Ia subclasses known to surpass a peak $M_B$ of $-20\: \mathrm{mag}$ are Ia-CSM and 03fg-like. As was discussed in Sect.~\ref{subsection:contamination}, 03fg-like SNe are commonly found in low-mass, star-forming galaxies, which is not consistent with the galaxy properties for ZTF19abpjicm given in Table~\ref{table:host_prop}.  Also, we do not see obvious signs of CSM interaction in the light curve (there are no plateau phases, late-time bumps, or rebrightening in the forced photometry) for a SN Ia-CSM. However, SNe Ia-CSM are a diverse subtype, and it is possible that we do not see CSM signatures such as bumps and plateaus due to the SN being close to the detection limit.


From the relative magnification of the two fitted images alone, we calculated a total magnification of $\mu_\mathrm{tot} = 3.6\pm1.3$ in Sect. \ref{subsection:salt2}. We can also estimate the total magnification by comparing the peak apparent magnitude to the apparent magnitude we would expect from a normal SN Ia (with $M_B=-19.4 \: \mathrm{mag}$) at a redshift of 0.44 (as predicted by the combined SALT2 model). Although this requires the assumption of a cosmological model \citep[here we assume a flat $\Lambda \mathrm{CDM}$ model with $H_0=67.4 \: \mathrm{km\,s^{-1}Mpc^{-1}}$ and $\Omega_M=0.315$; ][]{Planck2020}, and relies on the fitted redshift, it is independent of the previous method that relies solely on the flux ratios of the images. From this method, we calculate a total magnification of $\mu_\mathrm{tot} = 4.7$. This is consistent with the $\mu_\mathrm{tot}$ estimated from the parameters of the combined SALT2 model, where the deviation is within one standard deviation of the predicted value.

Also in Sect. \ref{subsection:salt2}, we calculated an image separation ($\Delta \theta$) of $1.1 \arcsec$ for the case of ZTF19abpjicm, from the estimate of the source redshift and the time delays. This is within the range of values we would expect for typical lensing systems and agrees with our observation that the separate images were unresolved by ZTF. We also calculated a lens galaxy velocity dispersion ($v$) of $190 \: \mathrm{km\,s^{-1}}$, which is typical for strong lens galaxies with the stellar mass quoted in Table \ref{table:host_prop}.

As a result, we have justified that our two-image model fit parameters are physical and consistent with our photometric measurements of the light curve and the suspected lens.

\subsubsection{ZTF22aahmovu}
\label{subsubsection:ZTF22aahmovu}
ZTF22aahmovu has a good two-image combined SALT2 fit with a reduced $\chi^2$ of $1.38$. From our comparison to SLSN and TDE models in Sect.~\ref{subsubsection:csm_tde_model_fit}, the TDE model failed to accurately model the light curve in the pre-peak epoch and the SLSN model predicted an unphysical ejecta mass. As a result, we claim that the light curve is not typical for contaminants such as SLSNe or TDEs. Additionally, the light curve does not resemble that of a normal SN Ia, so we can also exclude over-luminous Ia subclasses as being a contaminant.

From the relative magnification of the two fitted images alone, we calculated a total magnification of $\mu_\mathrm{tot}=4.3\pm0.6$ in Sect. \ref{subsection:salt2}. As with ZTF19abpjicm, we also estimated the total magnification by comparing the peak apparent magnitude to the apparent magnitude that we would expect from a normal SN Ia at a redshift of $0.35$ (assuming the same flat $\Lambda \mathrm{CDM}$ cosmology). From this method, we calculate a total magnification of $\mu_\mathrm{tot}=2.8$. These values are consistent within $3\sigma$, which is reasonable when we consider that there are additional uncertainties we have not accounted for (such as microlensing by intervening stars, which could increase or decrease the magnification from the expected value).

The calculated image separation from our best fit two-image SALT2 parameters was $1.7 \arcsec$. We also determined a lens galaxy velocity dispersion of $240 \: \mathrm{km\,s^{-1}}$ for this system. These values are typical for strong lensing galaxies with the stellar mass quoted in Table \ref{table:host_prop}.

Therefore, like with ZTF19abpjicm, we have justified that our two-image model fit parameters are physical and consistent with our photometric measurements of the light curve and the suspected lens.

\subsubsection{ZTF20abjyrxf}
ZTF20abjyrxf also has a two-image combined SALT2 fit (with a reduced $\chi^2$ of 2.53). The model displays a slight decrease in the $g$-band brightness as the fit transitions from being dominated by the first image to the second, which is consistent with what we observe in the light curve. Additionally, the light curve is broad with a slower rise time (approximately 58.4 days). This is attributed to the relatively equal flux contributions from the two images, which causes the peak of the model to occur later (at the peak of the second image). However, this slower rise or plateauing feature could be characteristic of an SLSN light curve. From Fig. \ref{figure:host_mass_sfr} in Sect. \ref{subsection:host_obs}, the galaxy associated with ZTF20abjyrxf could be consistent with the host of a SLSN-II(n). However, from our comparison to an SLSN model in Sect.~\ref{subsubsection:csm_tde_model_fit}, the best fit model predicts a nickel mass of greater than a solar mass, which would suggest a pair-instability supernova (an extremely rare and therefore unlikely event). The TDE model fit from Sect.~\ref{subsubsection:csm_tde_model_fit} also failed to accurately model the light curve in the pre-peak epoch, and so we assert that this object is not a typical TDE.

In Sect. \ref{subsection:salt2}, we estimated a total magnification of $\mu_\mathrm{tot}=20.4\pm6.2$. This is quite large, similar to the magnification we observed from SN Zwicky (with $\mu_\mathrm{tot} = 24.3 \pm 2.7$). For this to be correct, we would expect the source redshift to be $\sim 2.8$ times the value of the lens redshift (assuming the same flat $\Lambda \mathrm{CDM}$ cosmology as for ZTF19abpjicm), which is possible, although not consistent with the best fit redshift parameter from the SALT2 two-image fit. We would also expect larger image separation for ZTF20abjyrxf, although this is highly dependent on the source redshift (it could vary from $2.5 \arcsec$ for $z_s = 2.3 \: z_l$, to $6.7 \arcsec$ for $z_s=0.24$ as was determined from the SALT2 fit). Considering that the spatial resolution of ZTF is approximately $1 \arcsec$ (combined with the $2 \arcsec$ seeing at Palomar Observatory), it should be possible to distinguish the separate images in the case of a $6.7 \arcsec$ separation. However, we did not find any evidence of spatial variability in the ZTF alert images.

It is also possible that our simplified lens model does not accurately resemble the system of ZTF20abjyrxf. This has been observed before; in the case of SN Zwicky, the observed relative flux contributions for two out of the four images did not agree with the model predictions. Furthermore, we cannot ignore the possibility that four images could be present in our case, which would change the relative image magnifications and the resulting flux ratios. As a result, we conclude that, if ZTF20abjyrxf is a glSN Ia, it is not well described by a simple two-image system with a SIS lens, and further study of the potential lens galaxy is required.
    
\subsubsection{ZTF21aablrfe}
It was not possible to fit ZTF21aablrfe with a reasonable two-image SALT2 model, due to the width of the light curve and the longer rise time (approximately $60$ days). It also had a poor fit to the single-image SALT2 model (with a reduced $\chi^2$ of $4.7$). Additionally, it is further away from the DESI galaxy, at a separation of $2.4\arcsec$. The combination of these characteristics suggests that this object could be a SLSN-II(n) candidate.

Despite this, the light curve is sufficiently red to pass the colour cuts and the galaxy counterpart is not consistent with the hosts of SLSNe-I (as is shown in Fig. \ref{figure:host_mass_sfr}). A four-image SALT2 model might be able to more accurately match the light curve we observe, although this would require us to know more about the lens system.

\subsubsection{ZTF22aabifrp}
ZTF22aabifrp has an interesting LS cutout, which shows two galaxies close together (less than three arcseconds apart). The DESI catalogue-matched redshift ($z=0.1886$) belongs to the object above the green circle plotted in Fig. \ref{figure:gold_lcs}. The photometric redshift for the lower galaxy, which is closer to ZTF22aabifrp at a separation of $0.1\arcsec$, is given as $z=0.25 \pm 0.04$ in the LS DR9 catalogue. Therefore, the $z=0.19$ object could be a lens for the host at $z=0.25$. However, the redshifts of the two galaxies are consistent within 1.5 standard deviations, and so this claim is not statistically significant.

Similarly to ZTF21aablrfe, it was not possible to fit ZTF22aabifrp with a reasonable two-image SALT2 model, due to the width of the light curve. From our fit, we would expect to see two distinct peaks for the separate images; however, there is no evidence for this in the data (unfortunately, there is a large period of no observations where the valley between the two peaks would be). This is another case where a four-image SALT2 model might be able to more accurately match the light curve we observe. Also like ZTF21aablrfe, the light curve is sufficiently red to pass the colour cuts and the closest galaxy is not consistent with the hosts of SLSNe-I. Despite this, we cannot exclude the possibility that this object is a SLSN-II(n).

\subsubsection{ZTF22aadeqlh}
The LS cutout for ZTF22aadeqlh in Fig. \ref{figure:gold_LS_cutout} displays two possible galaxy lenses or hosts (a larger blob to the left and a yellower blob to the right). It is not certain which galaxy is the lens or the host, although it appears that the transient is more aligned with the object on the left. The two galaxies are at approximately the same redshift according to DESI ($z=0.32$), so they may be a combined two-galaxy lens.

ZTF22aadeqlh was reported to be a candidate SLSN by \citet{Gkini2022} in an AstroNote. This is possible, because of the broad light curve and lack of a reasonable two-image SALT2 fit. Similarly to ZTF21aablrfe and ZTF22aabifrp, the light curve is sufficiently red to pass the colour cuts and the closest galaxy is not consistent with the hosts of SLSNe-I. However, once again, we cannot exclude the possibility that this object is a SLSN-II(n).

\subsubsection{ZTF21abcwuhh}
ZTF21abcwuhh has a single-image SALT2 fit with a reduced $\chi^2$ of $1.0$. However, the light curve is too faint (and, therefore, too noisy) to provide a convincing two-image combined SALT2 fit. The photometric and spectroscopic redshift values agree, which suggests that we can trust that this object is over-luminous. ZTF21abcwuhh may be a glSN Ia candidate, but the limited data means that we cannot further deduce the true nature of this faint transient.

\subsection{Summary of gold sample candidates}
\label{subsubsection:summarygold}
We present two candidates, ZTF19abpjicm and ZTF22aahmovu, as likely glSNe Ia. The observed characteristics of the two objects are not typical for contaminants such as SLSNe, TDEs, or over-luminous SN Ia subclasses. They can also be well modelled by a two-image combined SALT2 fit, and the resulting parameters are consistent with our photometric measurements and adhere to the expected physical characteristics of lensing systems.

An additional candidate, ZTF20abjyrxf, was also well modelled by a two-image combined SALT2 fit. However, the resulting parameters suggest that we should have observed a larger image separation (and therefore should have observed the two images as separate sources in ZTF). As this was not the case, we conclude that ZTF20abjyrxf is not well described by a simple two-image system with an SIS lens. Further study of the potential lens galaxy is required for us to present ZTF20abjyrxf as a likely glSN Ia.

ZTF21aablrfe, ZTF22aabifrp, and ZTF22aadeqlh were not well modelled by a two-image combined SALT2 fit due to the width of their light curves (we only see evidence for a single peak in the data; however, the model predicts that we should see two distinct peaks). Despite this, we see evidence of two galaxies at different redshifts (a potential lens and host) in the LS cutout for ZTF22aabifrp. We also observe two galaxies at approximately the same redshift for ZTF22aadeqlh, which could be a combined two-galaxy lens. As a result, we conclude that these candidates are not well described by a simple two-image system with an SIS lens, but a more complex model (e.g. with four images) could potentially provide a more accurate fit to the observed data. As with ZTF20abjyrxf, follow-up of the potential lens and/or host galaxies is required.


\section{Conclusions}
\label{section:conc}
In this work, we presented a systematic search of ZTF archival data between June 1 2019 and September 1 2022 for strongly lensed type SNe Ia. We created an analysis pipeline using the AMPEL processing platform, which allowed us to query the ZTF alert photometry, apply filters for SN-like objects, and perform SALT2 template fits. Then, we applied some initial cuts and acquired forced photometry for the remaining candidates. Finally, we narrowed down our most likely candidates through two selection methods: by cross-matching to the DESI spectroscopic redshift catalogues, or applying stricter cuts motivated by simulations and previous observations. Additionally, we performed a similar analysis on the ZTF Bright Transient Survey sample of supernovae, to estimate the amount of contamination we should expect from each class.

The conclusions of our analysis and the plans for future work are as follows:

   \begin{enumerate}
      \item We have shown that it is possible to filter the alert photometry of a large optical survey to find transients that match the expected characteristics of glSNe Ia. We devised an initial filtering algorithm to reduce 31930 SN-like alerts to 7075 candidates that passed some preliminary cuts. We reduced the 7075 initial candidates to 50 final glSN Ia candidates by applying additional cuts and matching to the DESI spectroscopic catalogue.
      \item Of these final candidates, 27 were the result of a DESI cross-match (spectroscopic sample) and 27 were the result of applying stricter cuts based on simulations (photometric sample). A careful inspection of each of the spectroscopic candidates is necessary to exclude the possibility that they are SLSNe. Both the spectroscopic and photometric samples will be targeted in future spectroscopic follow-up missions to determine the lens and/or host redshifts.
      \item The vast majority of the final candidates were found within the lens redshift range of $0.2<z<0.4$. Furthermore, they are all dimmer than SN Zwicky and iPTF16geu, which suggests that, if they are truly glSNe Ia, they belong to a less magnified population.
      \item Seven candidates were present in both the spectroscopic and photometric samples, and these were discussed in more depth in this work. We present two candidates, ZTF19abpjicm and ZTF22aahmovu, as likely glSNe Ia. They are well modelled by a combined two-image SALT2 template and the model fit parameters are consistent with the physical parameters we would expect for strong lensing systems. From this two-image modelling, we estimate time delays of $22\pm3$ and $34\pm1$ days for the two events, respectively, which suggests that we have uncovered a population of glSNe Ia with longer time delays. The remaining five objects are still promising candidate glSNe Ia, but we would require more information about their lensing systems to model them more accurately.
      \item The two most promising candidates, ZTF19abpjicm and ZTF22aahmovu, should be targeted in a follow-up programme to confirm that they occurred in a strong lensing system. To do this, we suggest acquiring high-resolution photometry of the system, to search for evidence of a strongly lensed background host galaxy. Alternatively, if the host galaxy is not lensed, we suggest acquiring deeper spectra of the potential lens galaxy, to search for additional lines from a background host galaxy.

   \end{enumerate}

In this paper, we have highlighted the importance of lensed transients (particularly standardisable candles like SNe Ia) for cosmology and presented a pipeline that is both efficient and sensitive enough to parse full alert streams from optical observatories to search for glSNe Ia. The methodology presented in this systematic search is currently applied as a live search in ZTF and could be the foundation for future live searches in LSST and beyond. To fully remove the sources of contamination, spectroscopic follow-up while the transient is live is necessary. Despite this, we have shown that it is still possible to eliminate contaminants based on light curve modelling and photometric observations of the galaxy counterpart, and, as a result, we present two likely glSNe Ia.


\section{Data availability}
\label{section:datavail}
Table A.2 is only available in electronic form at the CDS via anonymous ftp to \href{cdsarc.u-strasbg.fr}{cdsarc.u-strasbg.fr} (130.79.128.5) or via \href{http://cdsweb.u-strasbg.fr/cgi-bin/qcat?J/A+A/}{http://cdsweb.u-strasbg.fr/cgi-bin/qcat?J/A+A/}.
All ZTF light curve data is publicly available at \href{https://www.ztf.caltech.edu/ztf-public-releases.html}{https://www.ztf.caltech.edu/ztf-public-releases.html}. The instructions for how to install and run the AMPEL platform can be found at \href{https://github.com/AmpelAstro/Ampel-HU-astro/}{https://github.com/AmpelAstro/Ampel-HU-astro/}. The specific AMPEL workflow we used to obtain the alert photometry, cross-match to galaxy catalogues, and perform initial filtering is found at \href{https://github.com/AmpelAstro/Ampel-HU-astro/blob/main/notebooks/run_lensing_query.ipynb}{https://github.com/AmpelAstro/Ampel-HU-astro/blob/main/notebooks/run\_lensing\_query.ipynb}. The forced photometry pipeline for ZTF data that we utilised can be found at \href{https://github.com/simeonreusch/fpbot}{https://github.com/simeonreusch/fpbot}. The BTS sample, which we utilised to estimate the contaminants present in our sample, is publicly accessible at \href{https://sites.astro.caltech.edu/ztf/bts/explorer.php}{https://sites.astro.caltech.edu/ztf/bts/explorer.php}. All external data sources are referenced in the main text. Supplemental data are accessible at \href{https://zenodo.org/records/11105340}{https://zenodo.org/records/11105340}.

\begin{acknowledgements}
      Based on observations obtained with the Samuel Oschin Telescope 48-inch and the 60-inch Telescope at the Palomar Observatory as part of the Zwicky Transient Facility project. ZTF is supported by the National Science Foundation under Grants No. AST-1440341 and AST-2034437 and a collaboration including current partners Caltech, IPAC, the Weizmann Institute of Science, the Oskar Klein Center at Stockholm University, the University of Maryland, Deutsches Elektronen-Synchrotron and Humboldt University, the TANGO Consortium of Taiwan, the University of Wisconsin at Milwaukee, Trinity College Dublin, Lawrence Livermore National Laboratories, IN2P3, University of Warwick, Ruhr University Bochum, Northwestern University and former partners the University of Washington, Los Alamos National Laboratories, and Lawrence Berkeley National Laboratories. Operations are conducted by COO, IPAC, and UW. \\

      This work has been supported by the research project grant “Understanding the Dynamic Universe” funded by the Knut and Alice Wallenberg Foundation under Dnr KAW 2018.0067,  {\em Vetenskapsr\aa det}, the Swedish Research Council, project 2020-03444, and the G.R.E.A.T research environment, project number 2016-06012. \\
      
      This material is based upon work supported by the U.S. Department of Energy (DOE), Office of Science, Office of High-Energy Physics, under Contract No. DE–AC02–05CH11231, and by the National Energy Research Scientific Computing Center, a DOE Office of Science User Facility under the same contract. Additional support for DESI was provided by the U.S. National Science Foundation (NSF), Division of Astronomical Sciences under Contract No. AST-0950945 to the NSF’s National Optical-Infrared Astronomy Research Laboratory; the Science and Technology Facilities Council of the United Kingdom; the Gordon and Betty Moore Foundation; the Heising-Simons Foundation; the French Alternative Energies and Atomic Energy Commission (CEA); the National Council of Science and Technology of Mexico (CONACYT); the Ministry of Science and Innovation of Spain (MICINN), and by the DESI Member Institutions: \url{https://www.desi.lbl.gov/collaborating-institutions}. Any opinions, findings, and conclusions or recommendations expressed in this material are those of the author(s) and do not necessarily reflect the views of the U. S. National Science Foundation, the U. S. Department of Energy, or any of the listed funding agencies.\\

      The authors are honored to be permitted to conduct scientific research on Iolkam Du’ag (Kitt Peak), a mountain with particular significance to the Tohono O’odham Nation.\\

      S. Schulze is partially supported by LBNL Subcontract NO.\ 7707915.\\

      M. W. Coughlin acknowledges support from the National Science Foundation with grant numbers PHY-2308862 and PHY-2117997.
\end{acknowledgements}

\bibliographystyle{aa}
\bibliography{lens.bib}

\begin{thebibliography}{129}
\expandafter\ifx\csname natexlab\endcsname\relax\def\natexlab#1{#1}\fi

\bibitem[{{Ahn} {et~al.}(2012){Ahn}, {Alexandroff}, {Allende Prieto}, {Anderson}, {Anderton}, {Andrews}, {Aubourg}, {Bailey}, {Balbinot}, {Barnes}, \& et~al.}]{Ahn2012a}
{Ahn}, C.~P., {Alexandroff}, R., {Allende Prieto}, C., {et~al.} 2012, \apjs, 203, 21

\bibitem[{{Arendse} {et~al.}(2024){Arendse}, {Dhawan}, {Sagu{\'e}s Carracedo}, {Peiris}, {Goobar}, {Wojtak}, {Alves}, {Biswas}, {Huber}, {Birrer}, \& {The LSST Dark Energy Science Collaboration}}]{Arendse2024}
{Arendse}, N., {Dhawan}, S., {Sagu{\'e}s Carracedo}, A., {et~al.} 2024, \mnras, 531, 3509

\bibitem[{{Ashton} {et~al.}(2019){Ashton}, {H{\"u}bner}, {Lasky}, {Talbot}, {Ackley}, {Biscoveanu}, {Chu}, {Divakarla}, {Easter}, {Goncharov}, {Hernandez Vivanco}, {Harms}, {Lower}, {Meadors}, {Melchor}, {Payne}, {Pitkin}, {Powell}, {Sarin}, {Smith}, \& {Thrane}}]{Ashton2019}
{Ashton}, G., {H{\"u}bner}, M., {Lasky}, P.~D., {et~al.} 2019, \apjs, 241, 27

\bibitem[{{Assef} {et~al.}(2018){Assef}, {Stern}, {Noirot}, {Jun}, {Cutri}, \& {Eisenhardt}}]{Assef2018}
{Assef}, R.~J., {Stern}, D., {Noirot}, G., {et~al.} 2018, \apjs, 234, 23

\bibitem[{{Bag} {et~al.}(2021){Bag}, {Kim}, {Linder}, \& {Shafieloo}}]{Bag2021}
{Bag}, S., {Kim}, A.~G., {Linder}, E.~V., \& {Shafieloo}, A. 2021, \apj, 910, 65

\bibitem[{{Barbary}(2021)}]{Barbary2021}
{Barbary}, K. 2021, {nestle: Nested sampling algorithms for evaluating Bayesian evidence}, Astrophysics Source Code Library, record ascl:2103.022

\bibitem[{Barbary {et~al.}(2022)Barbary, Bailey, Barentsen, Barclay, Biswas, Boone, Craig, Feindt, Friesen, Goldstein, Jha, Jones, Mondon, Papadogiannakis, Perrefort, Pierel, Rodney, Rose, Saunders, Sipőcz, Sofiatti, Thomas, van Santen, Vincenzi, Wang, \& Wood-Vasey}]{sncosmo}
Barbary, K., Bailey, S., Barentsen, G., {et~al.} 2022, SNCosmo

\bibitem[{{Beck} {et~al.}(2021){Beck}, {Szapudi}, {Flewelling}, {Holmberg}, {Magnier}, \& {Chambers}}]{Beck2021}
{Beck}, R., {Szapudi}, I., {Flewelling}, H., {et~al.} 2021, \mnras, 500, 1633

\bibitem[{{Bellm} {et~al.}(2019{\natexlab{a}}){Bellm}, {Kulkarni}, {Barlow}, {Feindt}, {Graham}, {Goobar}, {Kupfer}, {Ngeow}, {Nugent}, {Ofek}, {Prince}, {Riddle}, {Walters}, \& {Ye}}]{Bellm2019b}
{Bellm}, E.~C., {Kulkarni}, S.~R., {Barlow}, T., {et~al.} 2019{\natexlab{a}}, \pasp, 131, 068003

\bibitem[{{Bellm} {et~al.}(2019{\natexlab{b}}){Bellm}, {Kulkarni}, {Graham}, {Dekany}, {Smith}, {Riddle}, {Masci}, {Helou}, {Prince}, {Adams}, {Barbarino}, {Barlow}, {Bauer}, {Beck}, {Belicki}, {Biswas}, {Blagorodnova}, {Bodewits}, {Bolin}, {Brinnel}, {Brooke}, {Bue}, {Bulla}, {Burruss}, {Cenko}, {Chang}, {Connolly}, {Coughlin}, {Cromer}, {Cunningham}, {De}, {Delacroix}, {Desai}, {Duev}, {Eadie}, {Farnham}, {Feeney}, {Feindt}, {Flynn}, {Franckowiak}, {Frederick}, {Fremling}, {Gal-Yam}, {Gezari}, {Giomi}, {Goldstein}, {Golkhou}, {Goobar}, {Groom}, {Hacopians}, {Hale}, {Henning}, {Ho}, {Hover}, {Howell}, {Hung}, {Huppenkothen}, {Imel}, {Ip}, {Ivezi{\'c}}, {Jackson}, {Jones}, {Juric}, {Kasliwal}, {Kaspi}, {Kaye}, {Kelley}, {Kowalski}, {Kramer}, {Kupfer}, {Landry}, {Laher}, {Lee}, {Lin}, {Lin}, {Lunnan}, {Giomi}, {Mahabal}, {Mao}, {Miller}, {Monkewitz}, {Murphy}, {Ngeow}, {Nordin}, {Nugent}, {Ofek}, {Patterson}, {Penprase}, {Porter}, {Rauch}, {Rebbapragada}, {Reiley}, {Rigault}, {Rodriguez}, {van Roestel},
  {Rusholme}, {van Santen}, {Schulze}, {Shupe}, {Singer}, {Soumagnac}, {Stein}, {Surace}, {Sollerman}, {Szkody}, {Taddia}, {Terek}, {Van Sistine}, {van Velzen}, {Vestrand}, {Walters}, {Ward}, {Ye}, {Yu}, {Yan}, \& {Zolkower}}]{Bellm2019a}
{Bellm}, E.~C., {Kulkarni}, S.~R., {Graham}, M.~J., {et~al.} 2019{\natexlab{b}}, \pasp, 131, 018002

\bibitem[{{Betoule} {et~al.}(2014){Betoule}, {Kessler}, {Guy}, {Mosher}, {Hardin}, {Biswas}, {Astier}, {El-Hage}, {Konig}, {Kuhlmann}, {Marriner}, {Pain}, {Regnault}, {Balland}, {Bassett}, {Brown}, {Campbell}, {Carlberg}, {Cellier-Holzem}, {Cinabro}, {Conley}, {D'Andrea}, {DePoy}, {Doi}, {Ellis}, {Fabbro}, {Filippenko}, {Foley}, {Frieman}, {Fouchez}, {Galbany}, {Goobar}, {Gupta}, {Hill}, {Hlozek}, {Hogan}, {Hook}, {Howell}, {Jha}, {Le Guillou}, {Leloudas}, {Lidman}, {Marshall}, {M{\"o}ller}, {Mour{\~a}o}, {Neveu}, {Nichol}, {Olmstead}, {Palanque-Delabrouille}, {Perlmutter}, {Prieto}, {Pritchet}, {Richmond}, {Riess}, {Ruhlmann-Kleider}, {Sako}, {Schahmaneche}, {Schneider}, {Smith}, {Sollerman}, {Sullivan}, {Walton}, \& {Wheeler}}]{Betoule2014}
{Betoule}, M., {Kessler}, R., {Guy}, J., {et~al.} 2014, \aap, 568, A22

\bibitem[{{Bilicki} {et~al.}(2014){Bilicki}, {Jarrett}, {Peacock}, {Cluver}, \& {Steward}}]{Bilicki2014}
{Bilicki}, M., {Jarrett}, T.~H., {Peacock}, J.~A., {Cluver}, M.~E., \& {Steward}, L. 2014, \apjs, 210, 9

\bibitem[{{Bilicki} {et~al.}(2016){Bilicki}, {Peacock}, {Jarrett}, {Cluver}, {Maddox}, {Brown}, {Taylor}, {Hambly}, {Solarz}, {Holwerda}, {Baldry}, {Loveday}, {Moffett}, {Hopkins}, {Driver}, {Alpaslan}, \& {Bland-Hawthorn}}]{Bilicki2016}
{Bilicki}, M., {Peacock}, J.~A., {Jarrett}, T.~H., {et~al.} 2016, \apjs, 225, 5

\bibitem[{{Birrer} {et~al.}(2022){Birrer}, {Dhawan}, \& {Shajib}}]{Birrer2022}
{Birrer}, S., {Dhawan}, S., \& {Shajib}, A.~J. 2022, \apj, 924, 2

\bibitem[{{Blagorodnova} {et~al.}(2018){Blagorodnova}, {Neill}, {Walters}, {Kulkarni}, {Fremling}, {Ben-Ami}, {Dekany}, {Fucik}, {Konidaris}, {Nash}, {Ngeow}, {Ofek}, {O' Sullivan}, {Quimby}, {Ritter}, \& {Vyhmeister}}]{Blagorodnova2018}
{Blagorodnova}, N., {Neill}, J.~D., {Walters}, R., {et~al.} 2018, \pasp, 130, 035003

\bibitem[{{Brescia} {et~al.}(2015){Brescia}, {Cavuoti}, \& {Longo}}]{Brescia2015}
{Brescia}, M., {Cavuoti}, S., \& {Longo}, G. 2015, \mnras, 450, 3893

\bibitem[{{Byler} {et~al.}(2017){Byler}, {Dalcanton}, {Conroy}, \& {Johnson}}]{Byler2017a}
{Byler}, N., {Dalcanton}, J.~J., {Conroy}, C., \& {Johnson}, B.~D. 2017, \apj, 840, 44

\bibitem[{{Calzetti} {et~al.}(2000){Calzetti}, {Armus}, {Bohlin}, {Kinney}, {Koornneef}, \& {Storchi-Bergmann}}]{Calzetti2000a}
{Calzetti}, D., {Armus}, L., {Bohlin}, R.~C., {et~al.} 2000, \apj, 533, 682

\bibitem[{{Chabrier}(2003)}]{Chabrier2003a}
{Chabrier}, G. 2003, \pasp, 115, 763

\bibitem[{{Chakradhari} {et~al.}(2014){Chakradhari}, {Sahu}, {Srivastav}, \& {Anupama}}]{Chakradhari2014}
{Chakradhari}, N.~K., {Sahu}, D.~K., {Srivastav}, S., \& {Anupama}, G.~C. 2014, \mnras, 443, 1663

\bibitem[{{Chambers} {et~al.}(2016){Chambers}, {Magnier}, {Metcalfe}, {Flewelling}, {Huber}, {Waters}, {Denneau}, {Draper}, {Farrow}, {Finkbeiner}, {Holmberg}, {Koppenhoefer}, {Price}, {Rest}, {Saglia}, {Schlafly}, {Smartt}, {Sweeney}, {Wainscoat}, {Burgett}, {Chastel}, {Grav}, {Heasley}, {Hodapp}, {Jedicke}, {Kaiser}, {Kudritzki}, {Luppino}, {Lupton}, {Monet}, {Morgan}, {Onaka}, {Shiao}, {Stubbs}, {Tonry}, {White}, {Ba{\~n}ados}, {Bell}, {Bender}, {Bernard}, {Boegner}, {Boffi}, {Botticella}, {Calamida}, {Casertano}, {Chen}, {Chen}, {Cole}, {Deacon}, {Frenk}, {Fitzsimmons}, {Gezari}, {Gibbs}, {Goessl}, {Goggia}, {Gourgue}, {Goldman}, {Grant}, {Grebel}, {Hambly}, {Hasinger}, {Heavens}, {Heckman}, {Henderson}, {Henning}, {Holman}, {Hopp}, {Ip}, {Isani}, {Jackson}, {Keyes}, {Koekemoer}, {Kotak}, {Le}, {Liska}, {Long}, {Lucey}, {Liu}, {Martin}, {Masci}, {McLean}, {Mindel}, {Misra}, {Morganson}, {Murphy}, {Obaika}, {Narayan}, {Nieto-Santisteban}, {Norberg}, {Peacock}, {Pier}, {Postman}, {Primak}, {Rae}, {Rai},
  {Riess}, {Riffeser}, {Rix}, {R{\"o}ser}, {Russel}, {Rutz}, {Schilbach}, {Schultz}, {Scolnic}, {Strolger}, {Szalay}, {Seitz}, {Small}, {Smith}, {Soderblom}, {Taylor}, {Thomson}, {Taylor}, {Thakar}, {Thiel}, {Thilker}, {Unger}, {Urata}, {Valenti}, {Wagner}, {Walder}, {Walter}, {Watters}, {Werner}, {Wood-Vasey}, \& {Wyse}}]{Chambers2016a}
{Chambers}, K.~C., {Magnier}, E.~A., {Metcalfe}, N., {et~al.} 2016, arXiv e-prints, arXiv:1612.05560

\bibitem[{{Chang} {et~al.}(2024){Chang}, {Dai}, {Pfister}, {Chowdhury}, \& {Natarajan}}]{Chang2024}
{Chang}, J. N.~Y., {Dai}, L., {Pfister}, H., {Chowdhury}, R.~K., \& {Natarajan}, P. 2024, arXiv e-prints, arXiv:2407.09339

\bibitem[{{Chen} {et~al.}(2017){Chen}, {Smartt}, {Yates}, {Nicholl}, {Kr{\"u}hler}, {Schady}, {Dennefeld}, \& {Inserra}}]{Chen2017}
{Chen}, T.-W., {Smartt}, S.~J., {Yates}, R.~M., {et~al.} 2017, \mnras, 470, 3566

\bibitem[{{Childress} {et~al.}(2011){Childress}, {Aldering}, {Aragon}, {Antilogus}, {Bailey}, {Baltay}, {Bongard}, {Buton}, {Canto}, {Chotard}, {Copin}, {Fakhouri}, {Gangler}, {Kerschhaggl}, {Kowalski}, {Hsiao}, {Loken}, {Nugent}, {Paech}, {Pain}, {Pecontal}, {Pereira}, {Perlmutter}, {Rabinowitz}, {Runge}, {Scalzo}, {Thomas}, {Smadja}, {Tao}, {Weaver}, \& {Wu}}]{Childress2011}
{Childress}, M., {Aldering}, G., {Aragon}, C., {et~al.} 2011, \apj, 733, 3

\bibitem[{{Chornock} {et~al.}(2013){Chornock}, {Berger}, {Rest}, {Milisavljevic}, {Lunnan}, {Foley}, {Soderberg}, {Smartt}, {Burgasser}, {Challis}, {Chomiuk}, {Czekala}, {Drout}, {Fong}, {Huber}, {Kirshner}, {Leibler}, {McLeod}, {Marion}, {Narayan}, {Riess}, {Roth}, {Sanders}, {Scolnic}, {Smith}, {Stubbs}, {Tonry}, {Valenti}, {Burgett}, {Chambers}, {Hodapp}, {Kaiser}, {Kudritzki}, {Magnier}, \& {Price}}]{Chornock2013}
{Chornock}, R., {Berger}, E., {Rest}, A., {et~al.} 2013, \apj, 767, 162

\bibitem[{{Collett}(2015)}]{Collett2015}
{Collett}, T.~E. 2015, \apj, 811, 20

\bibitem[{{Conroy} {et~al.}(2009){Conroy}, {Gunn}, \& {White}}]{Conroy2009a}
{Conroy}, C., {Gunn}, J.~E., \& {White}, M. 2009, \apj, 699, 486

\bibitem[{{Craig} {et~al.}(2024){Craig}, {O'Connor}, {Chakrabarti}, {Rodney}, {Pierel}, {McCully}, \& {Perez-Fournon}}]{Craig2024}
{Craig}, P., {O'Connor}, K., {Chakrabarti}, S., {et~al.} 2024, \mnras, 534, 1077

\bibitem[{{D{\'a}lya} {et~al.}(2018){D{\'a}lya}, {Galg{\'o}czi}, {Dobos}, {Frei}, {Heng}, {Macas}, {Messenger}, {Raffai}, \& {de Souza}}]{Dalya2018}
{D{\'a}lya}, G., {Galg{\'o}czi}, G., {Dobos}, L., {et~al.} 2018, \mnras, 479, 2374

\bibitem[{{Dekany} {et~al.}(2020){Dekany}, {Smith}, {Riddle}, {Feeney}, {Porter}, {Hale}, {Zolkower}, {Belicki}, {Kaye}, {Henning}, {Walters}, {Cromer}, {Delacroix}, {Rodriguez}, {Reiley}, {Mao}, {Hover}, {Murphy}, {Burruss}, {Baker}, {Kowalski}, {Reif}, {Mueller}, {Bellm}, {Graham}, \& {Kulkarni}}]{Dekany2020}
{Dekany}, R., {Smith}, R.~M., {Riddle}, R., {et~al.} 2020, \pasp, 132, 038001

\bibitem[{{Denissenya} {et~al.}(2022){Denissenya}, {Bag}, {Kim}, {Linder}, \& {Shafieloo}}]{Denissenya2022}
{Denissenya}, M., {Bag}, S., {Kim}, A.~G., {Linder}, E.~V., \& {Shafieloo}, A. 2022, \mnras, 511, 1210

\bibitem[{{DESI Collaboration} {et~al.}(2022){DESI Collaboration}, {Abareshi}, {Aguilar}, {Ahlen}, {Alam}, {Alexander}, {Alfarsy}, {Allen}, {Allende Prieto}, {Alves}, {Ameel}, {Armengaud}, {Asorey}, {Aviles}, {Bailey}, {Balaguera-Antol{\'\i}nez}, {Ballester}, {Baltay}, {Bault}, {Beltran}, {Benavides}, {BenZvi}, {Berti}, {Besuner}, {Beutler}, {Bianchi}, {Blake}, {Blanc}, {Blum}, {Bolton}, {Bose}, {Bramall}, {Brieden}, {Brodzeller}, {Brooks}, {Brownewell}, {Buckley-Geer}, {Cahn}, {Cai}, {Canning}, {Capasso}, {Carnero Rosell}, {Carton}, {Casas}, {Castander}, {Cervantes-Cota}, {Chabanier}, {Chaussidon}, {Chuang}, {Circosta}, {Cole}, {Cooper}, {da Costa}, {Cousinou}, {Cuceu}, {Davis}, {Dawson}, {de la Cruz-Noriega}, {de la Macorra}, {de Mattia}, {Della Costa}, {Demmer}, {Derwent}, {Dey}, {Dey}, {Dhungana}, {Ding}, {Dobson}, {Doel}, {Donald-McCann}, {Donaldson}, {Douglass}, {Duan}, {Dunlop}, {Edelstein}, {Eftekharzadeh}, {Eisenstein}, {Enriquez-Vargas}, {Escoffier}, {Evatt}, {Fagrelius}, {Fan}, {Fanning},
  {Fawcett}, {Ferraro}, {Ereza}, {Flaugher}, {Font-Ribera}, {Forero-Romero}, {Frenk}, {Fromenteau}, {G{\"a}nsicke}, {Garcia-Quintero}, {Garrison}, {Gazta{\~n}aga}, {Gerardi}, {Gil-Mar{\'\i}n}, {Gontcho a Gontcho}, {Gonzalez-Morales}, {Gonzalez-de-Rivera}, {Gonzalez-Perez}, {Gordon}, {Graur}, {Green}, {Grove}, {Gruen}, {Gutierrez}, {Guy}, {Hahn}, {Harris}, {Herrera}, {Herrera-Alcantar}, {Honscheid}, {Howlett}, {Huterer}, {Ir{\v{s}}i{\v{c}}}, {Ishak}, {Jelinsky}, {Jiang}, {Jimenez}, {Jing}, {Joyce}, {Jullo}, {Juneau}, {Kara{\c{c}}ayl{\i}}, {Karamanis}, {Karcher}, {Karim}, {Kehoe}, {Kent}, {Kirkby}, {Kisner}, {Kitaura}, {Koposov}, {Kov{\'a}cs}, {Kremin}, {Krolewski}, {L'Huillier}, {Lahav}, {Lambert}, {Lamman}, {Lan}, {Landriau}, {Lane}, {Lang}, {Lange}, {Lasker}, {Le Guillou}, {Leauthaud}, {Le Van Suu}, {Levi}, {Li}, {Magneville}, {Manera}, {Manser}, {Marshall}, {Martini}, {McCollam}, {McDonald}, {Meisner}, {Mena-Fern{\'a}ndez}, {Meneses-Rizo}, {Mezcua}, {Miller}, {Miquel}, {Montero-Camacho}, {Moon},
  {Moustakas}, {Mueller}, {Mu{\~n}oz-Guti{\'e}rrez}, {Myers}, {Nadathur}, {Najita}, {Napolitano}, {Neilsen}, {Newman}, {Nie}, {Ning}, {Niz}, {Norberg}, {Noriega}, {O'Brien}, {Obuljen}, {Palanque-Delabrouille}, {Palmese}, {Zhiwei}, {Pappalardo}, {PENG}, {Percival}, {Perruchot}, {Pogge}, {Poppett}, {Porredon}, {Prada}, {Prochaska}, {Pucha}, {P{\'e}rez-Fern{\'a}ndez}, {P{\'e}rez-R{\`a}fols}, {Rabinowitz}, {Raichoor}, {Ramirez-Solano}, {Ram{\'\i}rez-P{\'e}rez}, {Ravoux}, {Reil}, {Rezaie}, {Rocher}, {Rockosi}, {Roe}, {Roodman}, {Ross}, {Rossi}, {Ruggeri}, {Ruhlmann-Kleider}, {Sabiu}, {Safonova}, {Said}, {Saintonge}, {Salas Catonga}, {Samushia}, {Sanchez}, {Saulder}, {Schaan}, {Schlafly}, {Schlegel}, {Schmoll}, {Scholte}, {Schubnell}, {Secroun}, {Seo}, {Serrano}, {Sharples}, {Sholl}, {Silber}, {Silva}, {Sirk}, {Siudek}, {Smith}, {Sprayberry}, {Staten}, {Stupak}, {Tan}, {Tarl{\'e}}, {Tie}, {Tojeiro}, {Ure{\~n}a-L{\'o}pez}, {Valdes}, {Valenzuela}, {Valluri}, {Vargas-Maga{\~n}a}, {Verde}, {Walther}, {Wang}, {Wang},
  {Weaver}, {Weaverdyck}, {Wechsler}, {Wilson}, {Yang}, {Yu}, {Yuan}, {Y{\`e}che}, {Zhang}, {Zhang}, {Zhao}, {Zhou}, {Zhou}, {Zou}, {Zou}, {Zou}, {Zu}, \& {DESI Collaboration}}]{DESI2022}
{DESI Collaboration}, {Abareshi}, B., {Aguilar}, J., {et~al.} 2022, \aj, 164, 207

\bibitem[{{DESI Collaboration} {et~al.}(2024{\natexlab{a}}){DESI Collaboration}, {Adame}, {Aguilar}, {Ahlen}, {Alam}, {Aldering}, {Alexander}, {Alfarsy}, {Allende Prieto}, {Alvarez}, {Alves}, {Anand}, {Andrade-Oliveira}, {Armengaud}, {Asorey}, {Avila}, {Aviles}, {Bailey}, {Balaguera-Antol{\'\i}nez}, {Ballester}, {Baltay}, {Bault}, {Bautista}, {Behera}, {Beltran}, {BenZvi}, {Beraldo e Silva}, {Bermejo-Climent}, {Berti}, {Besuner}, {Beutler}, {Bianchi}, {Blake}, {Blum}, {Bolton}, {Brieden}, {Brodzeller}, {Brooks}, {Brown}, {Buckley-Geer}, {Burtin}, {Cabayol-Garcia}, {Cai}, {Canning}, {Cardiel-Sas}, {Carnero Rosell}, {Castander}, {Cervantes-Cota}, {Chabanier}, {Chaussidon}, {Chaves-Montero}, {Chen}, {Chen}, {Chuang}, {Claybaugh}, {Cole}, {Cooper}, {Cuceu}, {Davis}, {Dawson}, {de Belsunce}, {de la Cruz}, {de la Macorra}, {de Mattia}, {Demina}, {Demirbozan}, {DeRose}, {Dey}, {Dey}, {Dhungana}, {Ding}, {Ding}, {Doel}, {Doshi}, {Douglass}, {Edge}, {Eftekharzadeh}, {Eisenstein}, {Elliott}, {Escoffier}, {Fagrelius},
  {Fan}, {Fanning}, {Fawcett}, {Ferraro}, {Ereza}, {Flaugher}, {Font-Ribera}, {Forero-S{\'a}nchez}, {Forero-Romero}, {Frenk}, {G{\"a}nsicke}, {Garc{\'\i}a}, {Garc{\'\i}a-Bellido}, {Garcia-Quintero}, {Garrison}, {Gil-Mar{\'\i}n}, {Golden-Marx}, {Gontcho A Gontcho}, {Gonzalez-Morales}, {Gonzalez-Perez}, {Gordon}, {Graur}, {Green}, {Gruen}, {Guy}, {Hadzhiyska}, {Hahn}, {Han}, {Hanif}, {Herrera-Alcantar}, {Honscheid}, {Hou}, {Howlett}, {Huterer}, {Ir{\v{s}}i{\v{c}}}, {Ishak}, {Jana}, {Jiang}, {Jimenez}, {Jing}, {Joudaki}, {Jullo}, {Joyce}, {Juneau}, {Kizhuprakkat}, {Kara{\c{c}}ayl{\i}}, {Karim}, {Kehoe}, {Kent}, {Khederlarian}, {Kim}, {Kirkby}, {Kisner}, {Kitaura}, {Kneib}, {Koposov}, {Kov{\'a}cs}, {Kremin}, {Krolewski}, {L'Huillier}, {Lahav}, {Lambert}, {Lamman}, {Lan}, {Landriau}, {Lang}, {Lange}, {Lasker}, {Le Guillou}, {Leauthaud}, {Levi}, {Li}, {Linder}, {Lyons}, {Magneville}, {Manera}, {Manser}, {Margala}, {Martini}, {McDonald}, {Medina}, {Medina-Varela}, {Meisner}, {Mena-Fern{\'a}ndez}, {Meneses-Rizo},
  {Mezcua}, {Miquel}, {Montero-Camacho}, {Moon}, {Moore}, {Moustakas}, {Mueller}, {Mundet}, {Mu{\~n}oz-Guti{\'e}rrez}, {Myers}, {Nadathur}, {Napolitano}, {Neveux}, {Newman}, {Nie}, {Niz}, {Norberg}, {Noriega}, {Paillas}, {Palanque-Delabrouille}, {Palmese}, {Zhiwei}, {Parkinson}, {Penmetsa}, {Percival}, {P{\'e}rez-Fern{\'a}ndez}, {P{\'e}rez-R{\`a}fols}, {Pieri}, {Poppett}, {Porredon}, {Prada}, {Pucha}, {Raichoor}, {Ram{\'\i}rez-P{\'e}rez}, {Ramirez-Solano}, {Rashkovetskyi}, {Ravoux}, {Rocher}, {Rockosi}, {Ross}, {Rossi}, {Ruggeri}, {Ruhlmann-Kleider}, {Sabiu}, {Said}, {Saintonge}, {Samushia}, {Sanchez}, {Saulder}, {Schaan}, {Schlafly}, {Schlegel}, {Scholte}, {Schubnell}, {Seo}, {Shafieloo}, {Sharples}, {Sheu}, {Silber}, {Sinigaglia}, {Siudek}, {Slepian}, {Smith}, {Sprayberry}, {Stephey}, {Su{\'a}rez-P{\'e}rez}, {Sun}, {Tan}, {Tarl{\'e}}, {Tojeiro}, {Ure{\~n}a-L{\'o}pez}, {Vaisakh}, {Valcin}, {Valdes}, {Valluri}, {Vargas-Maga{\~n}a}, {Variu}, {Verde}, {Walther}, {Wang}, {Wang}, {Weaver}, {Weaverdyck},
  {Wechsler}, {White}, {Xie}, {Yang}, {Y{\`e}che}, {Yu}, {Yuan}, {Zhang}, {Zhang}, {Zhao}, {Zheng}, {Zhou}, {Zhou}, {Zou}, {Zou}, {Zu}, \& {DESI Collaboration}}]{DESI2024}
{DESI Collaboration}, {Adame}, A.~G., {Aguilar}, J., {et~al.} 2024{\natexlab{a}}, \aj, 167, 62

\bibitem[{{DESI Collaboration} {et~al.}(2024{\natexlab{b}}){DESI Collaboration}, {Adame}, {Aguilar}, {Ahlen}, {Alam}, {Aldering}, {Alexander}, {Alfarsy}, {Allende Prieto}, {Alvarez}, \& et~al.}]{DESI2023}
{DESI Collaboration}, {Adame}, A.~G., {Aguilar}, J., {et~al.} 2024{\natexlab{b}}, \aj, 168, 58

\bibitem[{{DESI Collaboration} {et~al.}(2024{\natexlab{c}}){DESI Collaboration}, {Adame}, {Aguilar}, {Ahlen}, {Alam}, {Alexander}, {Alvarez}, {Alves}, {Anand}, {Andrade}, {Armengaud}, {Avila}, {Aviles}, {Awan}, {Bahr-Kalus}, {Bailey}, {Baltay}, {Bault}, {Behera}, {BenZvi}, {Bera}, {Beutler}, {Bianchi}, {Blake}, {Blum}, {Brieden}, {Brodzeller}, {Brooks}, {Buckley-Geer}, {Burtin}, {Calderon}, {Canning}, {Carnero Rosell}, {Cereskaite}, {Cervantes-Cota}, {Chabanier}, {Chaussidon}, {Chaves-Montero}, {Chen}, {Chen}, {Claybaugh}, {Cole}, {Cuceu}, {Davis}, {Dawson}, {de la Macorra}, {de Mattia}, {Deiosso}, {Dey}, {Dey}, {Ding}, {Doel}, {Edelstein}, {Eftekharzadeh}, {Eisenstein}, {Elliott}, {Fagrelius}, {Fanning}, {Ferraro}, {Ereza}, {Findlay}, {Flaugher}, {Font-Ribera}, {Forero-S{\'a}nchez}, {Forero-Romero}, {Frenk}, {Garcia-Quintero}, {Gazta{\~n}aga}, {Gil-Mar{\'\i}n}, {Gontcho}, {Gonzalez-Morales}, {Gonzalez-Perez}, {Gordon}, {Green}, {Gruen}, {Gsponer}, {Gutierrez}, {Guy}, {Hadzhiyska}, {Hahn}, {Hanif},
  {Herrera-Alcantar}, {Honscheid}, {Howlett}, {Huterer}, {Ir{\v{s}}i{\v{c}}}, {Ishak}, {Juneau}, {Kara{\c{c}}ayl{\i}}, {Kehoe}, {Kent}, {Kirkby}, {Kremin}, {Krolewski}, {Lai}, {Lan}, {Landriau}, {Lang}, {Lasker}, {Le Goff}, {Le Guillou}, {Leauthaud}, {Levi}, {Li}, {Linder}, {Lodha}, {Magneville}, {Manera}, {Margala}, {Martini}, {Maus}, {McDonald}, {Medina-Varela}, {Meisner}, {Mena-Fern{\'a}ndez}, {Miquel}, {Moon}, {Moore}, {Moustakas}, {Mudur}, {Mueller}, {Mu{\~n}oz-Guti{\'e}rrez}, {Myers}, {Nadathur}, {Napolitano}, {Neveux}, {Newman}, {Nguyen}, {Nie}, {Niz}, {Noriega}, {Padmanabhan}, {Paillas}, {Palanque-Delabrouille}, {Pan}, {Penmetsa}, {Percival}, {Pieri}, {Pinon}, {Poppett}, {Porredon}, {Prada}, {P{\'e}rez-Fern{\'a}ndez}, {P{\'e}rez-R{\`a}fols}, {Rabinowitz}, {Raichoor}, {Ram{\'\i}rez-P{\'e}rez}, {Ramirez-Solano}, {Ravoux}, {Rashkovetskyi}, {Rezaie}, {Rich}, {Rocher}, {Rockosi}, {Roe}, {Rosado-Marin}, {Ross}, {Rossi}, {Ruggeri}, {Ruhlmann-Kleider}, {Samushia}, {Sanchez}, {Saulder}, {Schlafly}, {Schlegel},
  {Schubnell}, {Seo}, {Shafieloo}, {Sharples}, {Silber}, {Slosar}, {Smith}, {Sprayberry}, {Tan}, {Tarl{\'e}}, {Taylor}, {Trusov}, {Ure{\~n}a-L{\'o}pez}, {Vaisakh}, {Valcin}, {Valdes}, {Vargas-Maga{\~n}a}, {Verde}, {Walther}, {Wang}, {Wang}, {Weaver}, {Weaverdyck}, {Wechsler}, {Weinberg}, {White}, {Yu}, {Yu}, {Yuan}, {Y{\`e}che}, {Zaborowski}, {Zarrouk}, {Zhang}, {Zhao}, {Zhao}, {Zhou}, {Zhuang}, \& {Zou}}]{DESI2024VI}
{DESI Collaboration}, {Adame}, A.~G., {Aguilar}, J., {et~al.} 2024{\natexlab{c}}, arXiv e-prints. Accepted for publication in JCAP, arXiv:2404.03002

\bibitem[{{DESI Collaboration} {et~al.}(2024{\natexlab{d}}){DESI Collaboration}, {Adame}, {Aguilar}, {Ahlen}, {Alam}, {Alexander}, {Alvarez}, {Alves}, {Anand}, {Andrade}, {Armengaud}, {Avila}, {Aviles}, {Awan}, {Bailey}, {Baltay}, {Bault}, {Bautista}, {Behera}, {BenZvi}, {Beutler}, {Bianchi}, {Blake}, {Blum}, {Brieden}, {Brodzeller}, {Brooks}, {Buckley-Geer}, {Burtin}, {Calderon}, {Canning}, {Carnero Rosell}, {Cereskaite}, {Cervantes-Cota}, {Chabanier}, {Chaussidon}, {Chaves-Montero}, {Chen}, {Chen}, {Claybaugh}, {Cole}, {Cuceu}, {Davis}, {Dawson}, {de la Cruz}, {de la Macorra}, {de Mattia}, {Deiosso}, {Dey}, {Dey}, {Ding}, {Ding}, {Doel}, {Edelstein}, {Eftekharzadeh}, {Eisenstein}, {Elliott}, {Fagrelius}, {Fanning}, {Ferraro}, {Ereza}, {Findlay}, {Flaugher}, {Font-Ribera}, {Forero-S{\'a}nchez}, {Forero-Romero}, {Garcia-Quintero}, {Gazta{\~n}aga}, {Gil-Mar{\'\i}n}, {Gontcho}, {Gonzalez-Morales}, {Gonzalez-Perez}, {Gordon}, {Green}, {Gruen}, {Gsponer}, {Gutierrez}, {Guy}, {Hadzhiyska}, {Hahn}, {Hanif},
  {Herrera-Alcantar}, {Honscheid}, {Howlett}, {Huterer}, {Ir{\v{s}}i{\v{c}}}, {Ishak}, {Juneau}, {Kara{\c{c}}ayli}, {Kehoe}, {Kent}, {Kirkby}, {Kremin}, {Krolewski}, {Lai}, {Lan}, {Landriau}, {Lang}, {Lasker}, {Le Goff}, {Le Guillou}, {Leauthaud}, {Levi}, {Li}, {Linder}, {Lodha}, {Magneville}, {Manera}, {Margala}, {Martini}, {Maus}, {McDonald}, {Medina-Varela}, {Meisner}, {Mena-Fern{\'a}ndez}, {Miquel}, {Moon}, {Moore}, {Moustakas}, {Mueller}, {Mu{\~n}oz-Guti{\'e}rrez}, {Myers}, {Nadathur}, {Napolitano}, {Neveux}, {Newman}, {Nguyen}, {Nie}, {Niz}, {Noriega}, {Padmanabhan}, {Paillas}, {Palanque-Delabrouille}, {Pan}, {Penmetsa}, {Percival}, {Pieri}, {Pinon}, {Poppett}, {Porredon}, {Prada}, {P{\'e}rez-Fern{\'a}ndez}, {P{\'e}rez-R{\`a}fols}, {Rabinowitz}, {Raichoor}, {Ram{\'\i}rez-P{\'e}rez}, {Ramirez-Solano}, {Rashkovetskyi}, {Ravoux}, {Rezaie}, {Rich}, {Rocher}, {Rockosi}, {Roe}, {Rosado-Marin}, {Ross}, {Rossi}, {Ruggeri}, {Ruhlmann-Kleider}, {Samushia}, {Sanchez}, {Saulder}, {Schlafly}, {Schlegel},
  {Schubnell}, {Seo}, {Sharples}, {Silber}, {Sinigaglia}, {Slosar}, {Smith}, {Sprayberry}, {Tan}, {Tarl{\'e}}, {Trusov}, {Vaisakh}, {Valcin}, {Valdes}, {Vargas-Maga{\~n}a}, {Verde}, {Walther}, {Wang}, {Wang}, {Weaver}, {Weaverdyck}, {Wechsler}, {Weinberg}, {White}, {Yu}, {Yu}, {Yuan}, {Y{\`e}che}, {Zaborowski}, {Zarrouk}, {Zhang}, {Zhao}, {Zhao}, {Zhou}, \& {Zou}}]{DESI2024IV}
{DESI Collaboration}, {Adame}, A.~G., {Aguilar}, J., {et~al.} 2024{\natexlab{d}}, arXiv e-prints. Accepted for publication in JCAP, arXiv:2404.03001

\bibitem[{{DESI Collaboration} {et~al.}(2024{\natexlab{e}}){DESI Collaboration}, {Adame}, {Aguilar}, {Ahlen}, {Alam}, {Alexander}, {Alvarez}, {Alves}, {Anand}, {Andrade}, {Armengaud}, {Avila}, {Aviles}, {Awan}, {Bailey}, {Baltay}, {Bault}, {Behera}, {BenZvi}, {Beutler}, {Bianchi}, {Blake}, {Blum}, {Brieden}, {Brodzeller}, {Brooks}, {Buckley-Geer}, {Burtin}, {Calderon}, {Canning}, {Carnero Rosell}, {Cereskaite}, {Cervantes-Cota}, {Chabanier}, {Chaussidon}, {Chaves-Montero}, {Chen}, {Chen}, {Claybaugh}, {Cole}, {Cuceu}, {Davis}, {Dawson}, {de la Macorra}, {de Mattia}, {Deiosso}, {Dey}, {Dey}, {Ding}, {Doel}, {Edelstein}, {Eftekharzadeh}, {Eisenstein}, {Elliott}, {Fagrelius}, {Fanning}, {Ferraro}, {Ereza}, {Findlay}, {Flaugher}, {Font-Ribera}, {Forero-S{\'a}nchez}, {Forero-Romero}, {Garcia-Quintero}, {Gazta{\~n}aga}, {Gil-Mar{\'\i}n}, {Gontcho}, {Gonzalez-Morales}, {Gonzalez-Perez}, {Gordon}, {Green}, {Gruen}, {Gsponer}, {Gutierrez}, {Guy}, {Hadzhiyska}, {Hahn}, {Hanif}, {Herrera-Alcantar}, {Honscheid},
  {Howlett}, {Huterer}, {Ir{\v{s}}i{\v{c}}}, {Ishak}, {Juneau}, {Kara{\c{c}}ayl{\i}}, {Kehoe}, {Kent}, {Kirkby}, {Kremin}, {Krolewski}, {Lai}, {Lan}, {Landriau}, {Lang}, {Lasker}, {Le Goff}, {Le Guillou}, {Leauthaud}, {Levi}, {Li}, {Linder}, {Lodha}, {Magneville}, {Manera}, {Margala}, {Martini}, {Maus}, {McDonald}, {Medina-Varela}, {Meisner}, {Mena-Fern{\'a}ndez}, {Miquel}, {Moon}, {Moore}, {Moustakas}, {Mudur}, {Mueller}, {Mu{\~n}oz-Guti{\'e}rrez}, {Myers}, {Nadathur}, {Napolitano}, {Neveux}, {Newman}, {Nguyen}, {Nie}, {Niz}, {Noriega}, {Padmanabhan}, {Paillas}, {Palanque-Delabrouille}, {Pan}, {Penmetsa}, {Percival}, {Pieri}, {Pinon}, {Poppett}, {Porredon}, {Prada}, {P{\'e}rez-Fern{\'a}ndez}, {P{\'e}rez-R{\`a}fols}, {Rabinowitz}, {Raichoor}, {Ram{\'\i}rez-P{\'e}rez}, {Ramirez-Solano}, {Rashkovetskyi}, {Rezaie}, {Rich}, {Rocher}, {Rockosi}, {Roe}, {Rosado-Marin}, {Ross}, {Rossi}, {Ruggeri}, {Ruhlmann-Kleider}, {Samushia}, {Sanchez}, {Saulder}, {Schlafly}, {Schlegel}, {Schubnell}, {Seo}, {Sharples}, {Silber},
  {Slosar}, {Smith}, {Sprayberry}, {Swanson}, {Tan}, {Tarl{\'e}}, {Trusov}, {Vaisakh}, {Valcin}, {Valdes}, {Vargas-Maga{\~n}a}, {Verde}, {Walther}, {Wang}, {Wang}, {Weaver}, {Weaverdyck}, {Wechsler}, {Weinberg}, {White}, {Yu}, {Yu}, {Yuan}, {Y{\`e}che}, {Zaborowski}, {Zarrouk}, {Zhang}, {Zhao}, {Zhao}, {Zhou}, \& {Zou}}]{DESI2024III}
{DESI Collaboration}, {Adame}, A.~G., {Aguilar}, J., {et~al.} 2024{\natexlab{e}}, arXiv e-prints, arXiv:2404.03000

\bibitem[{{DESI Collaboration} {et~al.}(2016{\natexlab{a}}){DESI Collaboration}, {Aghamousa}, {Aguilar}, {Ahlen}, {Alam}, {Allen}, {Allende Prieto}, {Annis}, {Bailey}, {Balland}, {Ballester}, {Baltay}, {Beaufore}, {Bebek}, {Beers}, {Bell}, {Bernal}, {Besuner}, {Beutler}, {Blake}, {Bleuler}, {Blomqvist}, {Blum}, {Bolton}, {Briceno}, {Brooks}, {Brownstein}, {Buckley-Geer}, {Burden}, {Burtin}, {Busca}, {Cahn}, {Cai}, {Cardiel-Sas}, {Carlberg}, {Carton}, {Casas}, {Castander}, {Cervantes-Cota}, {Claybaugh}, {Close}, {Coker}, {Cole}, {Comparat}, {Cooper}, {Cousinou}, {Crocce}, {Cuby}, {Cunningham}, {Davis}, {Dawson}, {de la Macorra}, {De Vicente}, {Delubac}, {Derwent}, {Dey}, {Dhungana}, {Ding}, {Doel}, {Duan}, {Ealet}, {Edelstein}, {Eftekharzadeh}, {Eisenstein}, {Elliott}, {Escoffier}, {Evatt}, {Fagrelius}, {Fan}, {Fanning}, {Farahi}, {Farihi}, {Favole}, {Feng}, {Fernandez}, {Findlay}, {Finkbeiner}, {Fitzpatrick}, {Flaugher}, {Flender}, {Font-Ribera}, {Forero-Romero}, {Fosalba}, {Frenk}, {Fumagalli}, {Gaensicke},
  {Gallo}, {Garcia-Bellido}, {Gaztanaga}, {Pietro Gentile Fusillo}, {Gerard}, {Gershkovich}, {Giannantonio}, {Gillet}, {Gonzalez-de-Rivera}, {Gonzalez-Perez}, {Gott}, {Graur}, {Gutierrez}, {Guy}, {Habib}, {Heetderks}, {Heetderks}, {Heitmann}, {Hellwing}, {Herrera}, {Ho}, {Holland}, {Honscheid}, {Huff}, {Hutchinson}, {Huterer}, {Hwang}, {Illa Laguna}, {Ishikawa}, {Jacobs}, {Jeffrey}, {Jelinsky}, {Jennings}, {Jiang}, {Jimenez}, {Johnson}, {Joyce}, {Jullo}, {Juneau}, {Kama}, {Karcher}, {Karkar}, {Kehoe}, {Kennamer}, {Kent}, {Kilbinger}, {Kim}, {Kirkby}, {Kisner}, {Kitanidis}, {Kneib}, {Koposov}, {Kovacs}, {Koyama}, {Kremin}, {Kron}, {Kronig}, {Kueter-Young}, {Lacey}, {Lafever}, {Lahav}, {Lambert}, {Lampton}, {Landriau}, {Lang}, {Lauer}, {Le Goff}, {Le Guillou}, {Le Van Suu}, {Lee}, {Lee}, {Leitner}, {Lesser}, {Levi}, {L'Huillier}, {Li}, {Liang}, {Lin}, {Linder}, {Loebman}, {Luki{\'c}}, {Ma}, {MacCrann}, {Magneville}, {Makarem}, {Manera}, {Manser}, {Marshall}, {Martini}, {Massey}, {Matheson}, {McCauley},
  {McDonald}, {McGreer}, {Meisner}, {Metcalfe}, {Miller}, {Miquel}, {Moustakas}, {Myers}, {Naik}, {Newman}, {Nichol}, {Nicola}, {Nicolati da Costa}, {Nie}, {Niz}, {Norberg}, {Nord}, {Norman}, {Nugent}, {O'Brien}, {Oh}, {Olsen}, {Padilla}, {Padmanabhan}, {Padmanabhan}, {Palanque-Delabrouille}, {Palmese}, {Pappalardo}, {P{\^a}ris}, {Park}, {Patej}, {Peacock}, {Peiris}, {Peng}, {Percival}, {Perruchot}, {Pieri}, {Pogge}, {Pollack}, {Poppett}, {Prada}, {Prakash}, {Probst}, {Rabinowitz}, {Raichoor}, {Ree}, {Refregier}, {Regal}, {Reid}, {Reil}, {Rezaie}, {Rockosi}, {Roe}, {Ronayette}, {Roodman}, {Ross}, {Ross}, {Rossi}, {Rozo}, {Ruhlmann-Kleider}, {Rykoff}, {Sabiu}, {Samushia}, {Sanchez}, {Sanchez}, {Schlegel}, {Schneider}, {Schubnell}, {Secroun}, {Seljak}, {Seo}, {Serrano}, {Shafieloo}, {Shan}, {Sharples}, {Sholl}, {Shourt}, {Silber}, {Silva}, {Sirk}, {Slosar}, {Smith}, {Smoot}, {Som}, {Song}, {Sprayberry}, {Staten}, {Stefanik}, {Tarle}, {Sien Tie}, {Tinker}, {Tojeiro}, {Valdes}, {Valenzuela}, {Valluri},
  {Vargas-Magana}, {Verde}, {Walker}, {Wang}, {Wang}, {Weaver}, {Weaverdyck}, {Wechsler}, {Weinberg}, {White}, {Yang}, {Yeche}, {Zhang}, {Zhao}, {Zheng}, {Zhou}, {Zhou}, {Zhu}, {Zou}, \& {Zu}}]{DESI2016a}
{DESI Collaboration}, {Aghamousa}, A., {Aguilar}, J., {et~al.} 2016{\natexlab{a}}, arXiv e-prints, arXiv:1611.00036

\bibitem[{{DESI Collaboration} {et~al.}(2016{\natexlab{b}}){DESI Collaboration}, {Aghamousa}, {Aguilar}, {Ahlen}, {Alam}, {Allen}, {Allende Prieto}, {Annis}, {Bailey}, {Balland}, {Ballester}, {Baltay}, {Beaufore}, {Bebek}, {Beers}, {Bell}, {Bernal}, {Besuner}, {Beutler}, {Blake}, {Bleuler}, {Blomqvist}, {Blum}, {Bolton}, {Briceno}, {Brooks}, {Brownstein}, {Buckley-Geer}, {Burden}, {Burtin}, {Busca}, {Cahn}, {Cai}, {Cardiel-Sas}, {Carlberg}, {Carton}, {Casas}, {Castander}, {Cervantes-Cota}, {Claybaugh}, {Close}, {Coker}, {Cole}, {Comparat}, {Cooper}, {Cousinou}, {Crocce}, {Cuby}, {Cunningham}, {Davis}, {Dawson}, {de la Macorra}, {De Vicente}, {Delubac}, {Derwent}, {Dey}, {Dhungana}, {Ding}, {Doel}, {Duan}, {Ealet}, {Edelstein}, {Eftekharzadeh}, {Eisenstein}, {Elliott}, {Escoffier}, {Evatt}, {Fagrelius}, {Fan}, {Fanning}, {Farahi}, {Farihi}, {Favole}, {Feng}, {Fernandez}, {Findlay}, {Finkbeiner}, {Fitzpatrick}, {Flaugher}, {Flender}, {Font-Ribera}, {Forero-Romero}, {Fosalba}, {Frenk}, {Fumagalli}, {Gaensicke},
  {Gallo}, {Garcia-Bellido}, {Gaztanaga}, {Pietro Gentile Fusillo}, {Gerard}, {Gershkovich}, {Giannantonio}, {Gillet}, {Gonzalez-de-Rivera}, {Gonzalez-Perez}, {Gott}, {Graur}, {Gutierrez}, {Guy}, {Habib}, {Heetderks}, {Heetderks}, {Heitmann}, {Hellwing}, {Herrera}, {Ho}, {Holland}, {Honscheid}, {Huff}, {Hutchinson}, {Huterer}, {Hwang}, {Illa Laguna}, {Ishikawa}, {Jacobs}, {Jeffrey}, {Jelinsky}, {Jennings}, {Jiang}, {Jimenez}, {Johnson}, {Joyce}, {Jullo}, {Juneau}, {Kama}, {Karcher}, {Karkar}, {Kehoe}, {Kennamer}, {Kent}, {Kilbinger}, {Kim}, {Kirkby}, {Kisner}, {Kitanidis}, {Kneib}, {Koposov}, {Kovacs}, {Koyama}, {Kremin}, {Kron}, {Kronig}, {Kueter-Young}, {Lacey}, {Lafever}, {Lahav}, {Lambert}, {Lampton}, {Landriau}, {Lang}, {Lauer}, {Le Goff}, {Le Guillou}, {Le Van Suu}, {Lee}, {Lee}, {Leitner}, {Lesser}, {Levi}, {L'Huillier}, {Li}, {Liang}, {Lin}, {Linder}, {Loebman}, {Luki{\'c}}, {Ma}, {MacCrann}, {Magneville}, {Makarem}, {Manera}, {Manser}, {Marshall}, {Martini}, {Massey}, {Matheson}, {McCauley},
  {McDonald}, {McGreer}, {Meisner}, {Metcalfe}, {Miller}, {Miquel}, {Moustakas}, {Myers}, {Naik}, {Newman}, {Nichol}, {Nicola}, {Nicolati da Costa}, {Nie}, {Niz}, {Norberg}, {Nord}, {Norman}, {Nugent}, {O'Brien}, {Oh}, {Olsen}, {Padilla}, {Padmanabhan}, {Padmanabhan}, {Palanque-Delabrouille}, {Palmese}, {Pappalardo}, {P{\^a}ris}, {Park}, {Patej}, {Peacock}, {Peiris}, {Peng}, {Percival}, {Perruchot}, {Pieri}, {Pogge}, {Pollack}, {Poppett}, {Prada}, {Prakash}, {Probst}, {Rabinowitz}, {Raichoor}, {Ree}, {Refregier}, {Regal}, {Reid}, {Reil}, {Rezaie}, {Rockosi}, {Roe}, {Ronayette}, {Roodman}, {Ross}, {Ross}, {Rossi}, {Rozo}, {Ruhlmann-Kleider}, {Rykoff}, {Sabiu}, {Samushia}, {Sanchez}, {Sanchez}, {Schlegel}, {Schneider}, {Schubnell}, {Secroun}, {Seljak}, {Seo}, {Serrano}, {Shafieloo}, {Shan}, {Sharples}, {Sholl}, {Shourt}, {Silber}, {Silva}, {Sirk}, {Slosar}, {Smith}, {Smoot}, {Som}, {Song}, {Sprayberry}, {Staten}, {Stefanik}, {Tarle}, {Sien Tie}, {Tinker}, {Tojeiro}, {Valdes}, {Valenzuela}, {Valluri},
  {Vargas-Magana}, {Verde}, {Walker}, {Wang}, {Wang}, {Weaver}, {Weaverdyck}, {Wechsler}, {Weinberg}, {White}, {Yang}, {Yeche}, {Zhang}, {Zhao}, {Zheng}, {Zhou}, {Zhou}, {Zhu}, {Zou}, \& {Zu}}]{DESI2016b}
{DESI Collaboration}, {Aghamousa}, A., {Aguilar}, J., {et~al.} 2016{\natexlab{b}}, arXiv e-prints, arXiv:1611.00037

\bibitem[{{Dey} {et~al.}(2019){Dey}, {Schlegel}, {Lang}, {Blum}, {Burleigh}, {Fan}, {Findlay}, {Finkbeiner}, {Herrera}, {Juneau}, {Landriau}, {Levi}, {McGreer}, {Meisner}, {Myers}, {Moustakas}, {Nugent}, {Patej}, {Schlafly}, {Walker}, {Valdes}, {Weaver}, {Y{\`e}che}, {Zou}, {Zhou}, {Abareshi}, {Abbott}, {Abolfathi}, {Aguilera}, {Alam}, {Allen}, {Alvarez}, {Annis}, {Ansarinejad}, {Aubert}, {Beechert}, {Bell}, {BenZvi}, {Beutler}, {Bielby}, {Bolton}, {Brice{\~n}o}, {Buckley-Geer}, {Butler}, {Calamida}, {Carlberg}, {Carter}, {Casas}, {Castander}, {Choi}, {Comparat}, {Cukanovaite}, {Delubac}, {DeVries}, {Dey}, {Dhungana}, {Dickinson}, {Ding}, {Donaldson}, {Duan}, {Duckworth}, {Eftekharzadeh}, {Eisenstein}, {Etourneau}, {Fagrelius}, {Farihi}, {Fitzpatrick}, {Font-Ribera}, {Fulmer}, {G{\"a}nsicke}, {Gaztanaga}, {George}, {Gerdes}, {Gontcho}, {Gorgoni}, {Green}, {Guy}, {Harmer}, {Hernand ez}, {Honscheid}, {Huang}, {James}, {Jannuzi}, {Jiang}, {Joyce}, {Karcher}, {Karkar}, {Kehoe}, {Kneib}, {Kueter-Young}, {Lan},
  {Lauer}, {Le Guillou}, {Le Van Suu}, {Lee}, {Lesser}, {Perreault Levasseur}, {Li}, {Mann}, {Marshall}, {Mart{\'\i}nez-V{\'a}zquez}, {Martini}, {du Mas des Bourboux}, {McManus}, {Meier}, {M{\'e}nard}, {Metcalfe}, {Mu{\~n}oz-Guti{\'e}rrez}, {Najita}, {Napier}, {Narayan}, {Newman}, {Nie}, {Nord}, {Norman}, {Olsen}, {Paat}, {Palanque-Delabrouille}, {Peng}, {Poppett}, {Poremba}, {Prakash}, {Rabinowitz}, {Raichoor}, {Rezaie}, {Robertson}, {Roe}, {Ross}, {Ross}, {Rudnick}, {Safonova}, {Saha}, {S{\'a}nchez}, {Savary}, {Schweiker}, {Scott}, {Seo}, {Shan}, {Silva}, {Slepian}, {Soto}, {Sprayberry}, {Staten}, {Stillman}, {Stupak}, {Summers}, {Sien Tie}, {Tirado}, {Vargas-Maga{\~n}a}, {Vivas}, {Wechsler}, {Williams}, {Yang}, {Yang}, {Yapici}, {Zaritsky}, {Zenteno}, {Zhang}, {Zhang}, {Zhou}, \& {Zhou}}]{Dey2018a}
{Dey}, A., {Schlegel}, D.~J., {Lang}, D., {et~al.} 2019, \aj, 157, 168

\bibitem[{{Duncan}(2022)}]{Duncan2022}
{Duncan}, K.~J. 2022, \mnras, 512, 3662

\bibitem[{{Falco} {et~al.}(1985){Falco}, {Gorenstein}, \& {Shapiro}}]{Falco1985}
{Falco}, E.~E., {Gorenstein}, M.~V., \& {Shapiro}, I.~I. 1985, \apjl, 289, L1

\bibitem[{{Foreman-Mackey} {et~al.}(2014){Foreman-Mackey}, {Sick}, \& {Johnson}}]{ForemanMackey2014a}
{Foreman-Mackey}, D., {Sick}, J., \& {Johnson}, B. 2014, {Python-Fsps: Python Bindings To Fsps (V0.1.1)}

\bibitem[{{Fremling} {et~al.}(2020){Fremling}, {Miller}, {Sharma}, {Dugas}, {Perley}, {Taggart}, {Sollerman}, {Goobar}, {Graham}, {Neill}, {Nordin}, {Rigault}, {Walters}, {Andreoni}, {Bagdasaryan}, {Belicki}, {Cannella}, {Bellm}, {Cenko}, {De}, {Dekany}, {Frederick}, {Golkhou}, {Graham}, {Helou}, {Ho}, {Kasliwal}, {Kupfer}, {Laher}, {Mahabal}, {Masci}, {Riddle}, {Rusholme}, {Schulze}, {Shupe}, {Smith}, {van Velzen}, {Yan}, {Yao}, {Zhuang}, \& {Kulkarni}}]{Fremling2020}
{Fremling}, C., {Miller}, A.~A., {Sharma}, Y., {et~al.} 2020, \apj, 895, 32

\bibitem[{{Frye} {et~al.}(2024){Frye}, {Pascale}, {Pierel}, {Chen}, {Foo}, {Leimbach}, {Garuda}, {Cohen}, {Kamieneski}, {Windhorst}, {Koekemoer}, {Kelly}, {Summers}, {Engesser}, {Liu}, {Furtak}, {Polletta}, {Harrington}, {Willner}, {Diego}, {Jansen}, {Coe}, {Conselice}, {Dai}, {Dole}, {D'Silva}, {Driver}, {Grogin}, {Marshall}, {Meena}, {Nonino}, {Ortiz}, {Pirzkal}, {Robotham}, {Ryan}, {Strolger}, {Tompkins}, {Willmer}, {Yan}, {Yun}, \& {Zitrin}}]{Frye2023}
{Frye}, B.~L., {Pascale}, M., {Pierel}, J., {et~al.} 2024, \apj, 961, 171

\bibitem[{{Gal-Yam}(2012)}]{GalYam2012}
{Gal-Yam}, A. 2012, Science, 337, 927

\bibitem[{{Gal-Yam} {et~al.}(2009){Gal-Yam}, {Mazzali}, {Ofek}, {Nugent}, {Kulkarni}, {Kasliwal}, {Quimby}, {Filippenko}, {Cenko}, {Chornock}, {Waldman}, {Kasen}, {Sullivan}, {Beshore}, {Drake}, {Thomas}, {Bloom}, {Poznanski}, {Miller}, {Foley}, {Silverman}, {Arcavi}, {Ellis}, \& {Deng}}]{GalYam2009}
{Gal-Yam}, A., {Mazzali}, P., {Ofek}, E.~O., {et~al.} 2009, \nat, 462, 624

\bibitem[{{Gkini} {et~al.}(2022){Gkini}, {Yan}, {Perley}, {Lunnan}, {Chen}, {Schulze}, {Kangas}, {Sollerman}, \& {Gal-Yam}}]{Gkini2022}
{Gkini}, A., {Yan}, L., {Perley}, D.~A., {et~al.} 2022, Transient Name Server AstroNote, 124, 1

\bibitem[{{Goldstein} \& {Nugent}(2017)}]{Goldstein2017}
{Goldstein}, D.~A. \& {Nugent}, P.~E. 2017, \apjl, 834, L5

\bibitem[{{Goldstein} {et~al.}(2019){Goldstein}, {Nugent}, \& {Goobar}}]{Goldstein2019}
{Goldstein}, D.~A., {Nugent}, P.~E., \& {Goobar}, A. 2019, \apjs, 243, 6

\bibitem[{{Goobar} {et~al.}(2017){Goobar}, {Amanullah}, {Kulkarni}, {Nugent}, {Johansson}, {Steidel}, {Law}, {M{\"o}rtsell}, {Quimby}, {Blagorodnova}, {Brandeker}, {Cao}, {Cooray}, {Ferretti}, {Fremling}, {Hangard}, {Kasliwal}, {Kupfer}, {Lunnan}, {Masci}, {Miller}, {Nayyeri}, {Neill}, {Ofek}, {Papadogiannakis}, {Petrushevska}, {Ravi}, {Sollerman}, {Sullivan}, {Taddia}, {Walters}, {Wilson}, {Yan}, \& {Yaron}}]{Goobar2017}
{Goobar}, A., {Amanullah}, R., {Kulkarni}, S.~R., {et~al.} 2017, Science, 356, 291

\bibitem[{{Goobar} {et~al.}(2024){Goobar}, {Johansson}, \& {Sagu{\'e}s Carracedo}}]{Goobar2024}
{Goobar}, A., {Johansson}, J., \& {Sagu{\'e}s Carracedo}, A. 2024, arXiv e-prints. Accepted for publication in Philosophical Transactions A, arXiv:2406.13519

\bibitem[{{Goobar} {et~al.}(2023){Goobar}, {Johansson}, {Schulze}, {Arendse}, {Carracedo}, {Dhawan}, {M{\"o}rtsell}, {Fremling}, {Yan}, {Perley}, {Sollerman}, {Joseph}, {Hinds}, {Meynardie}, {Andreoni}, {Bellm}, {Bloom}, {Collett}, {Drake}, {Graham}, {Kasliwal}, {Kulkarni}, {Lemon}, {Miller}, {Neill}, {Nordin}, {Pierel}, {Richard}, {Riddle}, {Rigault}, {Rusholme}, {Sharma}, {Stein}, {Stewart}, {Townsend}, {Vinko}, {Wheeler}, \& {Wold}}]{Goobar2023}
{Goobar}, A., {Johansson}, J., {Schulze}, S., {et~al.} 2023, Nature Astronomy, 7, 1098

\bibitem[{{Goobar} {et~al.}(2002){Goobar}, {M{\"o}rtsell}, {Amanullah}, \& {Nugent}}]{Goobar2002}
{Goobar}, A., {M{\"o}rtsell}, E., {Amanullah}, R., \& {Nugent}, P. 2002, \aap, 393, 25

\bibitem[{{Goobar} {et~al.}(2009){Goobar}, {Paech}, {Stanishev}, {Amanullah}, {Dahl{\'e}n}, {J{\"o}nsson}, {Kneib}, {Lidman}, {Limousin}, {M{\"o}rtsell}, {Nobili}, {Richard}, {Riehm}, \& {von Strauss}}]{Goobar2009}
{Goobar}, A., {Paech}, K., {Stanishev}, V., {et~al.} 2009, \aap, 507, 71

\bibitem[{{Graham} {et~al.}(2019){Graham}, {Kulkarni}, {Bellm}, {Adams}, {Barbarino}, {Blagorodnova}, {Bodewits}, {Bolin}, {Brady}, {Cenko}, {Chang}, {Coughlin}, {De}, {Eadie}, {Farnham}, {Feindt}, {Franckowiak}, {Fremling}, {Gezari}, {Ghosh}, {Goldstein}, {Golkhou}, {Goobar}, {Ho}, {Huppenkothen}, {Ivezi{\'c}}, {Jones}, {Juric}, {Kaplan}, {Kasliwal}, {Kelley}, {Kupfer}, {Lee}, {Lin}, {Lunnan}, {Mahabal}, {Miller}, {Ngeow}, {Nugent}, {Ofek}, {Prince}, {Rauch}, {van Roestel}, {Schulze}, {Singer}, {Sollerman}, {Taddia}, {Yan}, {Ye}, {Yu}, {Barlow}, {Bauer}, {Beck}, {Belicki}, {Biswas}, {Brinnel}, {Brooke}, {Bue}, {Bulla}, {Burruss}, {Connolly}, {Cromer}, {Cunningham}, {Dekany}, {Delacroix}, {Desai}, {Duev}, {Feeney}, {Flynn}, {Frederick}, {Gal-Yam}, {Giomi}, {Groom}, {Hacopians}, {Hale}, {Helou}, {Henning}, {Hover}, {Hillenbrand}, {Howell}, {Hung}, {Imel}, {Ip}, {Jackson}, {Kaspi}, {Kaye}, {Kowalski}, {Kramer}, {Kuhn}, {Landry}, {Laher}, {Mao}, {Masci}, {Monkewitz}, {Murphy}, {Nordin}, {Patterson}, {Penprase},
  {Porter}, {Rebbapragada}, {Reiley}, {Riddle}, {Rigault}, {Rodriguez}, {Rusholme}, {van Santen}, {Shupe}, {Smith}, {Soumagnac}, {Stein}, {Surace}, {Szkody}, {Terek}, {Van Sistine}, {van Velzen}, {Vestrand}, {Walters}, {Ward}, {Zhang}, \& {Zolkower}}]{Graham2019}
{Graham}, M.~J., {Kulkarni}, S.~R., {Bellm}, E.~C., {et~al.} 2019, \pasp, 131, 078001

\bibitem[{{Guillochon} {et~al.}(2018){Guillochon}, {Nicholl}, {Villar}, {Mockler}, {Narayan}, {Mandel}, {Berger}, \& {Williams}}]{Guillochon2018}
{Guillochon}, J., {Nicholl}, M., {Villar}, V.~A., {et~al.} 2018, \apjs, 236, 6

\bibitem[{{Guillochon} \& {Ramirez-Ruiz}(2013)}]{Guillochon2013}
{Guillochon}, J. \& {Ramirez-Ruiz}, E. 2013, \apj, 767, 25

\bibitem[{{Guy} {et~al.}(2007){Guy}, {Astier}, {Baumont}, {Hardin}, {Pain}, {Regnault}, {Basa}, {Carlberg}, {Conley}, {Fabbro}, {Fouchez}, {Hook}, {Howell}, {Perrett}, {Pritchet}, {Rich}, {Sullivan}, {Antilogus}, {Aubourg}, {Bazin}, {Bronder}, {Filiol}, {Palanque-Delabrouille}, {Ripoche}, \& {Ruhlmann-Kleider}}]{Guy2007}
{Guy}, J., {Astier}, P., {Baumont}, S., {et~al.} 2007, \aap, 466, 11

\bibitem[{{Guy} {et~al.}(2023){Guy}, {Bailey}, {Kremin}, {Alam}, {Alexander}, {Allende Prieto}, {BenZvi}, {Bolton}, {Brooks}, {Chaussidon}, {Cooper}, {Dawson}, {de la Macorra}, {Dey}, {Dey}, {Dhungana}, {Eisenstein}, {Font-Ribera}, {Forero-Romero}, {Gazta{\~n}aga}, {Gontcho A Gontcho}, {Green}, {Honscheid}, {Ishak}, {Kehoe}, {Kirkby}, {Kisner}, {Koposov}, {Lan}, {Landriau}, {Le Guillou}, {Levi}, {Magneville}, {Manser}, {Martini}, {Meisner}, {Miquel}, {Moustakas}, {Myers}, {Newman}, {Nie}, {Palanque-Delabrouille}, {Percival}, {Poppett}, {Prada}, {Raichoor}, {Ravoux}, {Ross}, {Schlafly}, {Schlegel}, {Schubnell}, {Sharples}, {Tarl{\'e}}, {Weaver}, {Y{\'e}che}, {Zhou}, {Zhou}, \& {Zou}}]{DESIGuy2023}
{Guy}, J., {Bailey}, S., {Kremin}, A., {et~al.} 2023, \aj, 165, 144

\bibitem[{{Hahn} {et~al.}(2023){Hahn}, {Wilson}, {Ruiz-Macias}, {Cole}, {Weinberg}, {Moustakas}, {Kremin}, {Tinker}, {Smith}, {Wechsler}, {Ahlen}, {Alam}, {Bailey}, {Brooks}, {Cooper}, {Davis}, {Dawson}, {Dey}, {Dey}, {Eftekharzadeh}, {Eisenstein}, {Fanning}, {Forero-Romero}, {Frenk}, {Gazta{\~n}aga}, {Gontcho A Gontcho}, {Guy}, {Honscheid}, {Ishak}, {Juneau}, {Kehoe}, {Kisner}, {Lan}, {Landriau}, {Le Guillou}, {Levi}, {Magneville}, {Martini}, {Meisner}, {Myers}, {Nie}, {Norberg}, {Palanque-Delabrouille}, {Percival}, {Poppett}, {Prada}, {Raichoor}, {Ross}, {Safonova}, {Saulder}, {Schlafly}, {Schlegel}, {Sierra-Porta}, {Tarle}, {Weaver}, {Y{\`e}che}, {Zarrouk}, {Zhou}, {Zhou}, \& {Zou}}]{DESIHahn2023}
{Hahn}, C., {Wilson}, M.~J., {Ruiz-Macias}, O., {et~al.} 2023, \aj, 165, 253

\bibitem[{{Howell} {et~al.}(2006){Howell}, {Sullivan}, {Nugent}, {Ellis}, {Conley}, {Le Borgne}, {Carlberg}, {Guy}, {Balam}, {Basa}, {Fouchez}, {Hook}, {Hsiao}, {Neill}, {Pain}, {Perrett}, \& {Pritchet}}]{Howell2006}
{Howell}, D.~A., {Sullivan}, M., {Nugent}, P.~E., {et~al.} 2006, \nat, 443, 308

\bibitem[{{Hsiao} {et~al.}(2020){Hsiao}, {Hoeflich}, {Ashall}, {Lu}, {Contreras}, {Burns}, {Phillips}, {Galbany}, {Anderson}, {Baltay}, {Baron}, {Castell{\'o}n}, {Davis}, {Freedman}, {Gall}, {Gonzalez}, {Graham}, {Hamuy}, {Holoien}, {Karamehmetoglu}, {Krisciunas}, {Kumar}, {Kuncarayakti}, {Morrell}, {Moriya}, {Nugent}, {Perlmutter}, {Persson}, {Piro}, {Rabinowitz}, {Roth}, {Shahbandeh}, {Shappee}, {Stritzinger}, {Suntzeff}, {Taddia}, \& {Uddin}}]{Hsiao2020}
{Hsiao}, E.~Y., {Hoeflich}, P., {Ashall}, C., {et~al.} 2020, \apj, 900, 140

\bibitem[{{Ivezi{\'c}} {et~al.}(2019){Ivezi{\'c}}, {Kahn}, {Tyson}, {Abel}, {Acosta}, {Allsman}, {Alonso}, {AlSayyad}, {Anderson}, {Andrew}, {Angel}, {Angeli}, {Ansari}, {Antilogus}, {Araujo}, {Armstrong}, {Arndt}, {Astier}, {Aubourg}, {Auza}, {Axelrod}, {Bard}, {Barr}, {Barrau}, {Bartlett}, {Bauer}, {Bauman}, {Baumont}, {Bechtol}, {Bechtol}, {Becker}, {Becla}, {Beldica}, {Bellavia}, {Bianco}, {Biswas}, {Blanc}, {Blazek}, {Blandford}, {Bloom}, {Bogart}, {Bond}, {Booth}, {Borgland}, {Borne}, {Bosch}, {Boutigny}, {Brackett}, {Bradshaw}, {Brandt}, {Brown}, {Bullock}, {Burchat}, {Burke}, {Cagnoli}, {Calabrese}, {Callahan}, {Callen}, {Carlin}, {Carlson}, {Chandrasekharan}, {Charles-Emerson}, {Chesley}, {Cheu}, {Chiang}, {Chiang}, {Chirino}, {Chow}, {Ciardi}, {Claver}, {Cohen-Tanugi}, {Cockrum}, {Coles}, {Connolly}, {Cook}, {Cooray}, {Covey}, {Cribbs}, {Cui}, {Cutri}, {Daly}, {Daniel}, {Daruich}, {Daubard}, {Daues}, {Dawson}, {Delgado}, {Dellapenna}, {de Peyster}, {de Val-Borro}, {Digel}, {Doherty}, {Dubois},
  {Dubois-Felsmann}, {Durech}, {Economou}, {Eifler}, {Eracleous}, {Emmons}, {Fausti Neto}, {Ferguson}, {Figueroa}, {Fisher-Levine}, {Focke}, {Foss}, {Frank}, {Freemon}, {Gangler}, {Gawiser}, {Geary}, {Gee}, {Geha}, {Gessner}, {Gibson}, {Gilmore}, {Glanzman}, {Glick}, {Goldina}, {Goldstein}, {Goodenow}, {Graham}, {Gressler}, {Gris}, {Guy}, {Guyonnet}, {Haller}, {Harris}, {Hascall}, {Haupt}, {Hernandez}, {Herrmann}, {Hileman}, {Hoblitt}, {Hodgson}, {Hogan}, {Howard}, {Huang}, {Huffer}, {Ingraham}, {Innes}, {Jacoby}, {Jain}, {Jammes}, {Jee}, {Jenness}, {Jernigan}, {Jevremovi{\'c}}, {Johns}, {Johnson}, {Johnson}, {Jones}, {Juramy-Gilles}, {Juri{\'c}}, {Kalirai}, {Kallivayalil}, {Kalmbach}, {Kantor}, {Karst}, {Kasliwal}, {Kelly}, {Kessler}, {Kinnison}, {Kirkby}, {Knox}, {Kotov}, {Krabbendam}, {Krughoff}, {Kub{\'a}nek}, {Kuczewski}, {Kulkarni}, {Ku}, {Kurita}, {Lage}, {Lambert}, {Lange}, {Langton}, {Le Guillou}, {Levine}, {Liang}, {Lim}, {Lintott}, {Long}, {Lopez}, {Lotz}, {Lupton}, {Lust}, {MacArthur}, {Mahabal},
  {Mandelbaum}, {Markiewicz}, {Marsh}, {Marshall}, {Marshall}, {May}, {McKercher}, {McQueen}, {Meyers}, {Migliore}, {Miller}, {Mills}, {Miraval}, {Moeyens}, {Moolekamp}, {Monet}, {Moniez}, {Monkewitz}, {Montgomery}, {Morrison}, {Mueller}, {Muller}, {Mu{\~n}oz Arancibia}, {Neill}, {Newbry}, {Nief}, {Nomerotski}, {Nordby}, {O'Connor}, {Oliver}, {Olivier}, {Olsen}, {O'Mullane}, {Ortiz}, {Osier}, {Owen}, {Pain}, {Palecek}, {Parejko}, {Parsons}, {Pease}, {Peterson}, {Peterson}, {Petravick}, {Libby Petrick}, {Petry}, {Pierfederici}, {Pietrowicz}, {Pike}, {Pinto}, {Plante}, {Plate}, {Plutchak}, {Price}, {Prouza}, {Radeka}, {Rajagopal}, {Rasmussen}, {Regnault}, {Reil}, {Reiss}, {Reuter}, {Ridgway}, {Riot}, {Ritz}, {Robinson}, {Roby}, {Roodman}, {Rosing}, {Roucelle}, {Rumore}, {Russo}, {Saha}, {Sassolas}, {Schalk}, {Schellart}, {Schindler}, {Schmidt}, {Schneider}, {Schneider}, {Schoening}, {Schumacher}, {Schwamb}, {Sebag}, {Selvy}, {Sembroski}, {Seppala}, {Serio}, {Serrano}, {Shaw}, {Shipsey}, {Sick}, {Silvestri},
  {Slater}, {Smith}, {Smith}, {Sobhani}, {Soldahl}, {Storrie-Lombardi}, {Stover}, {Strauss}, {Street}, {Stubbs}, {Sullivan}, {Sweeney}, {Swinbank}, {Szalay}, {Takacs}, {Tether}, {Thaler}, {Thayer}, {Thomas}, {Thornton}, {Thukral}, {Tice}, {Trilling}, {Turri}, {Van Berg}, {Vanden Berk}, {Vetter}, {Virieux}, {Vucina}, {Wahl}, {Walkowicz}, {Walsh}, {Walter}, {Wang}, {Wang}, {Warner}, {Wiecha}, {Willman}, {Winters}, {Wittman}, {Wolff}, {Wood-Vasey}, {Wu}, {Xin}, {Yoachim}, \& {Zhan}}]{Ivezic2019}
{Ivezi{\'c}}, {\v{Z}}., {Kahn}, S.~M., {Tyson}, J.~A., {et~al.} 2019, \apj, 873, 111

\bibitem[{{Johansson} {et~al.}(2021){Johansson}, {Goobar}, {Price}, {Sagu{\'e}s Carracedo}, {Della Bruna}, {Nugent}, {Dhawan}, {M{\"o}rtsell}, {Papadogiannakis}, {Amanullah}, {Goldstein}, {Cenko}, {De}, {Dugas}, {Kasliwal}, {Kulkarni}, \& {Lunnan}}]{Johansson2021}
{Johansson}, J., {Goobar}, A., {Price}, S.~H., {et~al.} 2021, \mnras, 502, 510

\bibitem[{{Johnson} {et~al.}(2021){Johnson}, {Leja}, {Conroy}, \& {Speagle}}]{Johnson2021a}
{Johnson}, B.~D., {Leja}, J., {Conroy}, C., \& {Speagle}, J.~S. 2021, \apjs, 254, 22

\bibitem[{{Kasen} \& {Bildsten}(2010)}]{Kasen2010}
{Kasen}, D. \& {Bildsten}, L. 2010, \apj, 717, 245

\bibitem[{{Kelly} {et~al.}(2015){Kelly}, {Rodney}, {Treu}, {Foley}, {Brammer}, {Schmidt}, {Zitrin}, {Sonnenfeld}, {Strolger}, {Graur}, {Filippenko}, {Jha}, {Riess}, {Bradac}, {Weiner}, {Scolnic}, {Malkan}, {von der Linden}, {Trenti}, {Hjorth}, {Gavazzi}, {Fontana}, {Merten}, {McCully}, {Jones}, {Postman}, {Dressler}, {Patel}, {Cenko}, {Graham}, \& {Tucker}}]{Kelly2015}
{Kelly}, P.~L., {Rodney}, S.~A., {Treu}, T., {et~al.} 2015, Science, 347, 1123

\bibitem[{{Lang}(2014)}]{Lang2014a}
{Lang}, D. 2014, \aj, 147, 108

\bibitem[{{Leloudas} {et~al.}(2015){Leloudas}, {Schulze}, {Kr{\"u}hler}, {Gorosabel}, {Christensen}, {Mehner}, {de Ugarte Postigo}, {Amor{\'\i}n}, {Th{\"o}ne}, {Anderson}, {Bauer}, {Gallazzi}, {He{\l}miniak}, {Hjorth}, {Ibar}, {Malesani}, {Morell}, {Vinko}, \& {Wheeler}}]{Leloudas2015}
{Leloudas}, G., {Schulze}, S., {Kr{\"u}hler}, T., {et~al.} 2015, \mnras, 449, 917

\bibitem[{{Levi} {et~al.}(2013){Levi}, {Bebek}, {Beers}, {Blum}, {Cahn}, {Eisenstein}, {Flaugher}, {Honscheid}, {Kron}, {Lahav}, {McDonald}, {Roe}, {Schlegel}, \& {representing the DESI collaboration}}]{DESILevi2013}
{Levi}, M., {Bebek}, C., {Beers}, T., {et~al.} 2013, arXiv e-prints, arXiv:1308.0847

\bibitem[{{Lu} {et~al.}(2021){Lu}, {Ashall}, {Hsiao}, {Hoeflich}, {Galbany}, {Baron}, {Phillips}, {Contreras}, {Burns}, {Suntzeff}, {Stritzinger}, {Anais}, {Anderson}, {Brown}, {Busta}, {Castell{\'o}n}, {Davis}, {Diamond}, {Falco}, {Gonzalez}, {Hamuy}, {Holmbo}, {Holoien}, {Krisciunas}, {Kirshner}, {Kumar}, {Kuncarayakti}, {Marion}, {Morrell}, {Persson}, {Piro}, {Prieto}, {Sand}, {Shahbandeh}, {Shappee}, \& {Taddia}}]{Lu2021}
{Lu}, J., {Ashall}, C., {Hsiao}, E.~Y., {et~al.} 2021, \apj, 920, 107

\bibitem[{{Lunnan} {et~al.}(2014){Lunnan}, {Chornock}, {Berger}, {Laskar}, {Fong}, {Rest}, {Sanders}, {Challis}, {Drout}, {Foley}, {Huber}, {Kirshner}, {Leibler}, {Marion}, {McCrum}, {Milisavljevic}, {Narayan}, {Scolnic}, {Smartt}, {Smith}, {Soderberg}, {Tonry}, {Burgett}, {Chambers}, {Flewelling}, {Hodapp}, {Kaiser}, {Magnier}, {Price}, \& {Wainscoat}}]{Lunnan2014}
{Lunnan}, R., {Chornock}, R., {Berger}, E., {et~al.} 2014, \apj, 787, 138

\bibitem[{{Magee} {et~al.}(2023){Magee}, {Sainz de Murieta}, {Collett}, \& {Enzi}}]{Magee2023}
{Magee}, M.~R., {Sainz de Murieta}, A., {Collett}, T.~E., \& {Enzi}, W. 2023, \mnras, 525, 542

\bibitem[{{Martin} {et~al.}(2005){Martin}, {Fanson}, {Schiminovich}, {Morrissey}, {Friedman}, {Barlow}, {Conrow}, {Grange}, {Jelinsky}, {Milliard}, {Siegmund}, {Bianchi}, {Byun}, {Donas}, {Forster}, {Heckman}, {Lee}, {Madore}, {Malina}, {Neff}, {Rich}, {Small}, {Surber}, {Szalay}, {Welsh}, \& {Wyder}}]{Martin2005a}
{Martin}, D.~C., {Fanson}, J., {Schiminovich}, D., {et~al.} 2005, \apj, 619, L1

\bibitem[{Masci {et~al.}(2023)Masci, Laher, Rusholme, Shupe, Paladini, Groom, Wold, Miller, \& Drake}]{Masci2023}
Masci, F.~J., Laher, R.~R., Rusholme, B., {et~al.} 2023, A New Forced Photometry Service for the Zwicky Transient Facility

\bibitem[{{Masci} {et~al.}(2019){Masci}, {Laher}, {Rusholme}, {Shupe}, {Groom}, {Surace}, {Jackson}, {Monkewitz}, {Beck}, {Flynn}, {Terek}, {Landry}, {Hacopians}, {Desai}, {Howell}, {Brooke}, {Imel}, {Wachter}, {Ye}, {Lin}, {Cenko}, {Cunningham}, {Rebbapragada}, {Bue}, {Miller}, {Mahabal}, {Bellm}, {Patterson}, {Juri{\'c}}, {Golkhou}, {Ofek}, {Walters}, {Graham}, {Kasliwal}, {Dekany}, {Kupfer}, {Burdge}, {Cannella}, {Barlow}, {Van Sistine}, {Giomi}, {Fremling}, {Blagorodnova}, {Levitan}, {Riddle}, {Smith}, {Helou}, {Prince}, \& {Kulkarni}}]{Masci2019}
{Masci}, F.~J., {Laher}, R.~R., {Rusholme}, B., {et~al.} 2019, \pasp, 131, 018003

\bibitem[{{Miller} {et~al.}(2024){Miller}, {Doel}, {Gutierrez}, {Besuner}, {Brooks}, {Gallo}, {Heetderks}, {Jelinsky}, {Kent}, {Lampton}, {Levi}, {Liang}, {Meisner}, {Sholl}, {Silber}, {Sprayberry}, {Aguilar}, {de la Macorra}, {Eisenstein}, {Fanning}, {Font-Ribera}, {Gazta{\~n}aga}, {Gontcho A Gontcho}, {Honscheid}, {Jimenez}, {Joyce}, {Kehoe}, {Kisner}, {Kremin}, {Landriau}, {Le Guillou}, {Magneville}, {Martini}, {Miquel}, {Moustakas}, {Nie}, {Percival}, {Poppett}, {Prada}, {Rossi}, {Schlegel}, {Schubnell}, {Seo}, {Sharples}, {Tarl{\'e}}, {Vargas-Maga{\~n}a}, {Zhou}, \& {the DESI Collaboration}}]{DESIMiller2023}
{Miller}, T.~N., {Doel}, P., {Gutierrez}, G., {et~al.} 2024, \aj, 168, 95

\bibitem[{{Millon} {et~al.}(2020){Millon}, {Galan}, {Courbin}, {Treu}, {Suyu}, {Ding}, {Birrer}, {Chen}, {Shajib}, {Sluse}, {Wong}, {Agnello}, {Auger}, {Buckley-Geer}, {Chan}, {Collett}, {Fassnacht}, {Hilbert}, {Koopmans}, {Motta}, {Mukherjee}, {Rusu}, {Sonnenfeld}, {Spiniello}, \& {Van de Vyvere}}]{Millon2020}
{Millon}, M., {Galan}, A., {Courbin}, F., {et~al.} 2020, \aap, 639, A101

\bibitem[{{Mockler} {et~al.}(2019){Mockler}, {Guillochon}, \& {Ramirez-Ruiz}}]{Mockler2019}
{Mockler}, B., {Guillochon}, J., \& {Ramirez-Ruiz}, E. 2019, \apj, 872, 151

\bibitem[{{M{\"o}rtsell} {et~al.}(2020){M{\"o}rtsell}, {Johansson}, {Dhawan}, {Goobar}, {Amanullah}, \& {Goldstein}}]{Mortsell2020}
{M{\"o}rtsell}, E., {Johansson}, J., {Dhawan}, S., {et~al.} 2020, \mnras, 496, 3270

\bibitem[{{Nordin} {et~al.}(2019){Nordin}, {Brinnel}, {van Santen}, {Bulla}, {Feindt}, {Franckowiak}, {Fremling}, {Gal-Yam}, {Giomi}, {Kowalski}, {Mahabal}, {Miranda}, {Rauch}, {Reusch}, {Rigault}, {Schulze}, {Sollerman}, {Stein}, {Yaron}, {van Velzen}, \& {Ward}}]{Nordin2019}
{Nordin}, J., {Brinnel}, V., {van Santen}, J., {et~al.} 2019, \aap, 631, A147

\bibitem[{{Nordin} {et~al.}(2014){Nordin}, {Rubin}, {Richard}, {Rykoff}, {Aldering}, {Amanullah}, {Atek}, {Barbary}, {Deustua}, {Fakhouri}, {Fruchter}, {Goobar}, {Hook}, {Hsiao}, {Huang}, {Kneib}, {Lidman}, {Meyers}, {Perlmutter}, {Saunders}, {Spadafora}, {Suzuki}, \& {Supernova Cosmology Project}}]{Nordin2014}
{Nordin}, J., {Rubin}, D., {Richard}, J., {et~al.} 2014, \mnras, 440, 2742

\bibitem[{{Patel} {et~al.}(2014){Patel}, {McCully}, {Jha}, {Rodney}, {Jones}, {Graur}, {Merten}, {Zitrin}, {Riess}, {Matheson}, {Sako}, {Holoien}, {Postman}, {Coe}, {Bartelmann}, {Balestra}, {Ben{\'\i}tez}, {Bouwens}, {Bradley}, {Broadhurst}, {Cenko}, {Donahue}, {Filippenko}, {Ford}, {Garnavich}, {Grillo}, {Infante}, {Jouvel}, {Kelson}, {Koekemoer}, {Lahav}, {Lemze}, {Maoz}, {Medezinski}, {Melchior}, {Meneghetti}, {Molino}, {Moustakas}, {Moustakas}, {Nonino}, {Rosati}, {Seitz}, {Strolger}, {Umetsu}, \& {Zheng}}]{Patel2014}
{Patel}, B., {McCully}, C., {Jha}, S.~W., {et~al.} 2014, \apj, 786, 9

\bibitem[{{Patterson} {et~al.}(2019){Patterson}, {Bellm}, {Rusholme}, {Masci}, {Juric}, {Krughoff}, {Golkhou}, {Graham}, {Kulkarni}, {Helou}, \& {Zwicky Transient Facility Collaboration}}]{Patterson2019}
{Patterson}, M.~T., {Bellm}, E.~C., {Rusholme}, B., {et~al.} 2019, \pasp, 131, 018001

\bibitem[{{Perley} {et~al.}(2020){Perley}, {Fremling}, {Sollerman}, {Miller}, {Dahiwale}, {Sharma}, {Bellm}, {Biswas}, {Brink}, {Bruch}, {De}, {Dekany}, {Drake}, {Duev}, {Filippenko}, {Gal-Yam}, {Goobar}, {Graham}, {Graham}, {Ho}, {Irani}, {Kasliwal}, {Kim}, {Kulkarni}, {Mahabal}, {Masci}, {Modak}, {Neill}, {Nordin}, {Riddle}, {Soumagnac}, {Strotjohann}, {Schulze}, {Taggart}, {Tzanidakis}, {Walters}, \& {Yan}}]{Perley2020}
{Perley}, D.~A., {Fremling}, C., {Sollerman}, J., {et~al.} 2020, \apj, 904, 35

\bibitem[{{Perley} {et~al.}(2016){Perley}, {Quimby}, {Yan}, {Vreeswijk}, {De Cia}, {Lunnan}, {Gal-Yam}, {Yaron}, {Filippenko}, {Graham}, {Laher}, \& {Nugent}}]{Perley2016}
{Perley}, D.~A., {Quimby}, R.~M., {Yan}, L., {et~al.} 2016, \apj, 830, 13

\bibitem[{{Petrushevska} {et~al.}(2017){Petrushevska}, {Amanullah}, {Bulla}, {Kromer}, {Ferretti}, {Goobar}, \& {Papadogiannakis}}]{Petrushevska2017}
{Petrushevska}, T., {Amanullah}, R., {Bulla}, M., {et~al.} 2017, \aap, 603, A136

\bibitem[{{Petrushevska} {et~al.}(2016){Petrushevska}, {Amanullah}, {Goobar}, {Fabbro}, {Johansson}, {Kjellsson}, {Lidman}, {Paech}, {Richard}, {Dahle}, {Ferretti}, {Kneib}, {Limousin}, {Nordin}, \& {Stanishev}}]{Petrushevska2016}
{Petrushevska}, T., {Amanullah}, R., {Goobar}, A., {et~al.} 2016, \aap, 594, A54

\bibitem[{{Pierel} {et~al.}(2023){Pierel}, {Arendse}, {Ertl}, {Huang}, {Moustakas}, {Schuldt}, {Shajib}, {Shu}, {Birrer}, {Bronikowski}, {Hjorth}, {Suyu}, {Agarwal}, {Agnello}, {Bolton}, {Chakrabarti}, {Cold}, {Courbin}, {Della Costa}, {Dhawan}, {Engesser}, {Fox}, {Gall}, {Gomez}, {Goobar}, {Jha}, {Jimenez}, {Johansson}, {Larison}, {Li}, {Marques-Chaves}, {Mao}, {Mazzali}, {Perez-Fournon}, {Petrushevska}, {Poidevin}, {Rest}, {Sheu}, {Shirley}, {Silver}, {Storfer}, {Strolger}, {Treu}, {Wojtak}, \& {Zenati}}]{Pierel2023}
{Pierel}, J.~D.~R., {Arendse}, N., {Ertl}, S., {et~al.} 2023, \apj, 948, 115

\bibitem[{{Pierel} {et~al.}(2024){Pierel}, {Newman}, {Dhawan}, {Gu}, {Joshi}, {Li}, {Schuldt}, {Strolger}, {Suyu}, {Caminha}, {Cohen}, {Diego}, {D{\'S}ilva}, {Ertl}, {Frye}, {Granata}, {Grillo}, {Koekemoer}, {Li}, {Robotham}, {Summers}, {Treu}, {Windhorst}, {Zitrin}, {Agarwal}, {Agrawal}, {Arendse}, {Belli}, {Burns}, {Ca{\~n}ameras}, {Chakrabarti}, {Chen}, {Collett}, {Coulter}, {Ellis}, {Engesser}, {Foo}, {Fox}, {Gall}, {Garuda}, {Gezari}, {Gomez}, {Glazebrook}, {Hjorth}, {Huang}, {Jha}, {Kamieneski}, {Kelly}, {Larison}, {Moustakas}, {Pascale}, {P{\'e}rez-Fournon}, {Petrushevska}, {Poidevin}, {Rest}, {Shahbandeh}, {Shajib}, {Siebert}, {Storfer}, {Talbot}, {Wang}, {Wevers}, \& {Zenati}}]{Pierel2024}
{Pierel}, J.~D.~R., {Newman}, A.~B., {Dhawan}, S., {et~al.} 2024, \apjl, 967, L37

\bibitem[{{Planck Collaboration} {et~al.}(2020){Planck Collaboration}, {Aghanim}, {Akrami}, {Ashdown}, {Aumont}, {Baccigalupi}, {Ballardini}, {Banday}, {Barreiro}, {Bartolo}, {Basak}, {Battye}, {Benabed}, {Bernard}, {Bersanelli}, {Bielewicz}, {Bock}, {Bond}, {Borrill}, {Bouchet}, {Boulanger}, {Bucher}, {Burigana}, {Butler}, {Calabrese}, {Cardoso}, {Carron}, {Challinor}, {Chiang}, {Chluba}, {Colombo}, {Combet}, {Contreras}, {Crill}, {Cuttaia}, {de Bernardis}, {de Zotti}, {Delabrouille}, {Delouis}, {Di Valentino}, {Diego}, {Dor{\'e}}, {Douspis}, {Ducout}, {Dupac}, {Dusini}, {Efstathiou}, {Elsner}, {En{\ss}lin}, {Eriksen}, {Fantaye}, {Farhang}, {Fergusson}, {Fernandez-Cobos}, {Finelli}, {Forastieri}, {Frailis}, {Fraisse}, {Franceschi}, {Frolov}, {Galeotta}, {Galli}, {Ganga}, {G{\'e}nova-Santos}, {Gerbino}, {Ghosh}, {Gonz{\'a}lez-Nuevo}, {G{\'o}rski}, {Gratton}, {Gruppuso}, {Gudmundsson}, {Hamann}, {Handley}, {Hansen}, {Herranz}, {Hildebrandt}, {Hivon}, {Huang}, {Jaffe}, {Jones}, {Karakci}, {Keih{\"a}nen},
  {Keskitalo}, {Kiiveri}, {Kim}, {Kisner}, {Knox}, {Krachmalnicoff}, {Kunz}, {Kurki-Suonio}, {Lagache}, {Lamarre}, {Lasenby}, {Lattanzi}, {Lawrence}, {Le Jeune}, {Lemos}, {Lesgourgues}, {Levrier}, {Lewis}, {Liguori}, {Lilje}, {Lilley}, {Lindholm}, {L{\'o}pez-Caniego}, {Lubin}, {Ma}, {Mac{\'\i}as-P{\'e}rez}, {Maggio}, {Maino}, {Mandolesi}, {Mangilli}, {Marcos-Caballero}, {Maris}, {Martin}, {Martinelli}, {Mart{\'\i}nez-Gonz{\'a}lez}, {Matarrese}, {Mauri}, {McEwen}, {Meinhold}, {Melchiorri}, {Mennella}, {Migliaccio}, {Millea}, {Mitra}, {Miville-Desch{\^e}nes}, {Molinari}, {Montier}, {Morgante}, {Moss}, {Natoli}, {N{\o}rgaard-Nielsen}, {Pagano}, {Paoletti}, {Partridge}, {Patanchon}, {Peiris}, {Perrotta}, {Pettorino}, {Piacentini}, {Polastri}, {Polenta}, {Puget}, {Rachen}, {Reinecke}, {Remazeilles}, {Renzi}, {Rocha}, {Rosset}, {Roudier}, {Rubi{\~n}o-Mart{\'\i}n}, {Ruiz-Granados}, {Salvati}, {Sandri}, {Savelainen}, {Scott}, {Shellard}, {Sirignano}, {Sirri}, {Spencer}, {Sunyaev}, {Suur-Uski}, {Tauber}, {Tavagnacco},
  {Tenti}, {Toffolatti}, {Tomasi}, {Trombetti}, {Valenziano}, {Valiviita}, {Van Tent}, {Vibert}, {Vielva}, {Villa}, {Vittorio}, {Wandelt}, {Wehus}, {White}, {White}, {Zacchei}, \& {Zonca}}]{Planck2020}
{Planck Collaboration}, {Aghanim}, N., {Akrami}, Y., {et~al.} 2020, \aap, 641, A6

\bibitem[{{Quimby} {et~al.}(2014){Quimby}, {Oguri}, {More}, {More}, {Moriya}, {Werner}, {Tanaka}, {Folatelli}, {Bersten}, {Maeda}, \& {Nomoto}}]{Quimby2014}
{Quimby}, R.~M., {Oguri}, M., {More}, A., {et~al.} 2014, Science, 344, 396

\bibitem[{{Quimby} {et~al.}(2013){Quimby}, {Werner}, {Oguri}, {More}, {More}, {Tanaka}, {Nomoto}, {Moriya}, {Folatelli}, {Maeda}, \& {Bersten}}]{Quimby2013}
{Quimby}, R.~M., {Werner}, M.~C., {Oguri}, M., {et~al.} 2013, \apjl, 768, L20

\bibitem[{{Refsdal}(1964)}]{Refsdal1964}
{Refsdal}, S. 1964, \mnras, 128, 307

\bibitem[{Reusch(2023)}]{Reusch2023}
Reusch, S. 2023, simeonreusch/fpbot: Release 1.1.2

\bibitem[{{Reusch} {et~al.}(2022){Reusch}, {Stein}, {Kowalski}, {van Velzen}, {Franckowiak}, {Lunardini}, {Murase}, {Winter}, {Miller-Jones}, {Kasliwal}, {Gilfanov}, {Garrappa}, {Paliya}, {Ahumada}, {Anand}, {Barbarino}, {Bellm}, {Brinnel}, {Buson}, {Cenko}, {Coughlin}, {De}, {Dekany}, {Frederick}, {Gal-Yam}, {Gezari}, {Giroletti}, {Graham}, {Karambelkar}, {Kimura}, {Kong}, {Kool}, {Laher}, {Medvedev}, {Necker}, {Nordin}, {Perley}, {Rigault}, {Rusholme}, {Schulze}, {Schweyer}, {Singer}, {Sollerman}, {Strotjohann}, {Sunyaev}, {van Santen}, {Walters}, {Zhang}, \& {Zimmerman}}]{Reusch2022}
{Reusch}, S., {Stein}, R., {Kowalski}, M., {et~al.} 2022, \prl, 128, 221101

\bibitem[{{Rodney} {et~al.}(2021){Rodney}, {Brammer}, {Pierel}, {Richard}, {Toft}, {O'Connor}, {Akhshik}, \& {Whitaker}}]{Rodney2021}
{Rodney}, S.~A., {Brammer}, G.~B., {Pierel}, J. D.~R., {et~al.} 2021, Nature Astronomy, 5, 1118

\bibitem[{{Rodney} {et~al.}(2015){Rodney}, {Patel}, {Scolnic}, {Foley}, {Molino}, {Brammer}, {Jauzac}, {Brada{\v{c}}}, {Broadhurst}, {Coe}, {Diego}, {Graur}, {Hjorth}, {Hoag}, {Jha}, {Johnson}, {Kelly}, {Lam}, {McCully}, {Medezinski}, {Meneghetti}, {Merten}, {Richard}, {Riess}, {Sharon}, {Strolger}, {Treu}, {Wang}, {Williams}, \& {Zitrin}}]{Rodney2015}
{Rodney}, S.~A., {Patel}, B., {Scolnic}, D., {et~al.} 2015, \apj, 811, 70

\bibitem[{{Ryu} {et~al.}(2020){Ryu}, {Krolik}, \& {Piran}}]{Ryu2020}
{Ryu}, T., {Krolik}, J., \& {Piran}, T. 2020, \apj, 904, 73

\bibitem[{{Sagu{\'e}s Carracedo} {et~al.}(2024){Sagu{\'e}s Carracedo}, {Goobar}, {M{\"o}rtsell}, {Arendse}, {Johansson}, {Townsend}, {Dhawan}, {Nordin}, {Sollerman}, \& {Schulze}}]{SaguesCarracedo2024}
{Sagu{\'e}s Carracedo}, A., {Goobar}, A., {M{\"o}rtsell}, E., {et~al.} 2024, arXiv e-prints. Submitted to A\&A., arXiv:2406.00052

\bibitem[{{Saini} {et~al.}(2000){Saini}, {Raychaudhury}, \& {Shchekinov}}]{Saini2000}
{Saini}, T.~D., {Raychaudhury}, S., \& {Shchekinov}, Y.~A. 2000, \aap, 363, 349

\bibitem[{{Sainz de Murieta} {et~al.}(2023){Sainz de Murieta}, {Collett}, {Magee}, {Weisenbach}, {Krawczyk}, \& {Enzi}}]{SainzdeMurieta2023}
{Sainz de Murieta}, A., {Collett}, T.~E., {Magee}, M.~R., {et~al.} 2023, \mnras, 526, 4296

\bibitem[{{Sarin} {et~al.}(2024){Sarin}, {H{\"u}bner}, {Omand}, {Setzer}, {Schulze}, {Adhikari}, {Sagu{\'e}s-Carracedo}, {Galaudage}, {Wallace}, {Lamb}, \& {Lin}}]{Sarin2024}
{Sarin}, N., {H{\"u}bner}, M., {Omand}, C. M.~B., {et~al.} 2024, \mnras, 531, 1203

\bibitem[{{Schulze} {et~al.}(2024){Schulze}, {Fransson}, {Kozyreva}, {Chen}, {Yaron}, {Jerkstrand}, {Gal-Yam}, {Sollerman}, {Yan}, {Kangas}, {Leloudas}, {Omand}, {Smartt}, {Yang}, {Nicholl}, {Sarin}, {Yao}, {Brink}, {Sharon}, {Rossi}, {Chen}, {Chen}, {Cikota}, {De}, {Drake}, {Filippenko}, {Fremling}, {Fr{\'e}our}, {Fynbo}, {Ho}, {Inserra}, {Irani}, {Kuncarayakti}, {Lunnan}, {Mazzali}, {Ofek}, {Palazzi}, {Perley}, {Pursiainen}, {Rothberg}, {Shingles}, {Smith}, {Taggart}, {Tartaglia}, {Zheng}, {Anderson}, {Cassara}, {Christensen}, {George Djorgovski}, {Galbany}, {Gkini}, {Graham}, {Gromadzki}, {Groom}, {Hiramatsu}, {Andrew Howell}, {Kasliwal}, {McCully}, {M{\"u}ller-Bravo}, {Paiano}, {Paraskeva}, {Pessi}, {Polishook}, {Rau}, {Rigault}, \& {Rusholme}}]{Schulze2024}
{Schulze}, S., {Fransson}, C., {Kozyreva}, A., {et~al.} 2024, \aap, 683, A223

\bibitem[{{Schulze} {et~al.}(2018){Schulze}, {Kr{\"u}hler}, {Leloudas}, {Gorosabel}, {Mehner}, {Buchner}, {Kim}, {Ibar}, {Amor{\'\i}n}, {Herrero-Illana}, {Anderson}, {Bauer}, {Christensen}, {de Pasquale}, {de Ugarte Postigo}, {Gallazzi}, {Hjorth}, {Morrell}, {Malesani}, {Sparre}, {Stalder}, {Stark}, {Th{\"o}ne}, \& {Wheeler}}]{Schulze2018}
{Schulze}, S., {Kr{\"u}hler}, T., {Leloudas}, G., {et~al.} 2018, \mnras, 473, 1258

\bibitem[{{Schulze} {et~al.}(2021){Schulze}, {Yaron}, {Sollerman}, {Leloudas}, {Gal}, {Wright}, {Lunnan}, {Gal-Yam}, {Ofek}, {Perley}, {Filippenko}, {Kasliwal}, {Kulkarni}, {Neill}, {Nugent}, {Quimby}, {Sullivan}, {Strotjohann}, {Arcavi}, {Ben-Ami}, {Bianco}, {Bloom}, {De}, {Fraser}, {Fremling}, {Horesh}, {Johansson}, {Kelly}, {Kne{\v{z}}evi{\'c}}, {Kne{\v{z}}evi{\'c}}, {Maguire}, {Nyholm}, {Papadogiannakis}, {Petrushevska}, {Rubin}, {Yan}, {Yang}, {Adams}, {Bufano}, {Clubb}, {Foley}, {Green}, {Harmanen}, {Ho}, {Hook}, {Hosseinzadeh}, {Howell}, {Kong}, {Kotak}, {Matheson}, {McCully}, {Milisavljevic}, {Pan}, {Poznanski}, {Shivvers}, {van Velzen}, \& {Verbeek}}]{Schulze2021a}
{Schulze}, S., {Yaron}, O., {Sollerman}, J., {et~al.} 2021, \apjs, 255, 29

\bibitem[{{Schwarz}(1978)}]{Schwarz1978}
{Schwarz}, G. 1978, Annals of Statistics, 6, 461

\bibitem[{{Shajib} {et~al.}(2020){Shajib}, {Birrer}, {Treu}, {Agnello}, {Buckley-Geer}, {Chan}, {Christensen}, {Lemon}, {Lin}, {Millon}, {Poh}, {Rusu}, {Sluse}, {Spiniello}, {Chen}, {Collett}, {Courbin}, {Fassnacht}, {Frieman}, {Galan}, {Gilman}, {More}, {Anguita}, {Auger}, {Bonvin}, {McMahon}, {Meylan}, {Wong}, {Abbott}, {Annis}, {Avila}, {Bechtol}, {Brooks}, {Brout}, {Burke}, {Carnero Rosell}, {Carrasco Kind}, {Carretero}, {Castander}, {Costanzi}, {da Costa}, {De Vicente}, {Desai}, {Dietrich}, {Doel}, {Drlica-Wagner}, {Evrard}, {Finley}, {Flaugher}, {Fosalba}, {Garc{\'\i}a-Bellido}, {Gerdes}, {Gruen}, {Gruendl}, {Gschwend}, {Gutierrez}, {Hollowood}, {Honscheid}, {Huterer}, {James}, {Jeltema}, {Krause}, {Kuropatkin}, {Li}, {Lima}, {MacCrann}, {Maia}, {Marshall}, {Melchior}, {Miquel}, {Ogando}, {Palmese}, {Paz-Chinch{\'o}n}, {Plazas}, {Romer}, {Roodman}, {Sako}, {Sanchez}, {Santiago}, {Scarpine}, {Schubnell}, {Scolnic}, {Serrano}, {Sevilla-Noarbe}, {Smith}, {Soares-Santos}, {Suchyta}, {Tarle}, {Thomas},
  {Walker}, \& {Zhang}}]{Shajib2020}
{Shajib}, A.~J., {Birrer}, S., {Treu}, T., {et~al.} 2020, \mnras, 494, 6072

\bibitem[{{Sharma} {et~al.}(2023){Sharma}, {Sollerman}, {Fremling}, {Kulkarni}, {De}, {Irani}, {Schulze}, {Strotjohann}, {Gal-Yam}, {Maguire}, {Perley}, {Bellm}, {Kool}, {Brink}, {Bruch}, {Deckers}, {Dekany}, {Dugas}, {Filippenko}, {Goldwasser}, {Graham}, {Graham}, {Groom}, {Hankins}, {Jencson}, {Johansson}, {Karambelkar}, {Kasliwal}, {Masci}, {Medford}, {Neill}, {Nir}, {Riddle}, {Rigault}, {Schweyer}, {Terwel}, {Yan}, {Yang}, \& {Yao}}]{Sharma2023}
{Sharma}, Y., {Sollerman}, J., {Fremling}, C., {et~al.} 2023, \apj, 948, 52

\bibitem[{{Shu} {et~al.}(2018){Shu}, {Bolton}, {Mao}, {Kang}, {Li}, \& {Soraisam}}]{Shu2018}
{Shu}, Y., {Bolton}, A.~S., {Mao}, S., {et~al.} 2018, \apj, 864, 91

\bibitem[{{Silber} {et~al.}(2023){Silber}, {Fagrelius}, {Fanning}, {Schubnell}, {Aguilar}, {Ahlen}, {Ameel}, {Ballester}, {Baltay}, {Bebek}, {Benton Beard}, {Besuner}, {Cardiel-Sas}, {Casas}, {Castander}, {Claybaugh}, {Dobson}, {Duan}, {Dunlop}, {Edelstein}, {Emmet}, {Elliott}, {Evatt}, {Gershkovich}, {Guy}, {Harris}, {Heetderks}, {Heetderks}, {Honscheid}, {Illa}, {Jelinsky}, {Jelinsky}, {Jimenez}, {Karcher}, {Kent}, {Kirkby}, {Kneib}, {Lambert}, {Lampton}, {Leitner}, {Levi}, {McCauley}, {Meisner}, {Miller}, {Miquel}, {Mundet}, {Poppett}, {Rabinowitz}, {Reil}, {Roman}, {Schlegel}, {Serrano}, {Van Shourt}, {Sprayberry}, {Tarl{\'e}}, {Tie}, {Weaverdyck}, {Zhang}, {Azzaro}, {Bailey}, {Becerril}, {Blackwell}, {Bouri}, {Brooks}, {Buckley-Geer}, {Castro}, {Derwent}, {Dey}, {Dhungana}, {Doel}, {Eisenstein}, {Fahim}, {Garcia-Bellido}, {Gazta{\~n}aga}, {A Gontcho}, {Gutierrez}, {H{\"o}rler}, {Kehoe}, {Kisner}, {Kremin}, {Kronig}, {Landriau}, {Le Guillou}, {Martini}, {Moustakas}, {Palanque-Delabrouille}, {Peng},
  {Percival}, {Prada}, {Allende Prieto}, {de Rivera}, {Sanchez}, {Sanchez}, {Sharples}, {Soares-Santos}, {Schlafly}, {Weaver}, {Zhou}, {Zhu}, {Zou}, \& {DESI Collaboration}}]{DESISilber2023}
{Silber}, J.~H., {Fagrelius}, P., {Fanning}, K., {et~al.} 2023, \aj, 165, 9

\bibitem[{{Skrutskie} {et~al.}(2006){Skrutskie}, {Cutri}, {Stiening}, {Weinberg}, {Schneider}, {Carpenter}, {Beichman}, {Capps}, {Chester}, {Elias}, {Huchra}, {Liebert}, {Lonsdale}, {Monet}, {Price}, {Seitzer}, {Jarrett}, {Kirkpatrick}, {Gizis}, {Howard}, {Evans}, {Fowler}, {Fullmer}, {Hurt}, {Light}, {Kopan}, {Marsh}, {McCallon}, {Tam}, {Van Dyk}, \& {Wheelock}}]{Skrutskie2006a}
{Skrutskie}, M.~F., {Cutri}, R.~M., {Stiening}, R., {et~al.} 2006, \aj, 131, 1163

\bibitem[{{Smith} {et~al.}(2007){Smith}, {Li}, {Foley}, {Wheeler}, {Pooley}, {Chornock}, {Filippenko}, {Silverman}, {Quimby}, {Bloom}, \& {Hansen}}]{Smith2007}
{Smith}, N., {Li}, W., {Foley}, R.~J., {et~al.} 2007, \apj, 666, 1116

\bibitem[{{Sonnenfeld} {et~al.}(2020){Sonnenfeld}, {Verma}, {More}, {Beaten}, {MacMillan}, {Wong}, {Chan}, {Jaelani}, {Lee}, {Oguri}, {Rusu}, {Veldthuis}, {Trouille}, {Marshall}, {Hutchings}, {Allen}, {O'Donnell}, {Cornen}, {Davis}, {McMaster}, {Lintott}, \& {Miller}}]{Sonnenfeld2020}
{Sonnenfeld}, A., {Verma}, A., {More}, A., {et~al.} 2020, VizieR Online Data Catalog, J/A+A/642/A148

\bibitem[{{Soumagnac} \& {Ofek}(2018)}]{Soumagnac2018}
{Soumagnac}, M.~T. \& {Ofek}, E.~O. 2018, \pasp, 130, 075002

\bibitem[{{Speagle}(2020)}]{Speagle2020a}
{Speagle}, J.~S. 2020, \mnras, 493, 3132

\bibitem[{{Sullivan} {et~al.}(2000){Sullivan}, {Ellis}, {Nugent}, {Smail}, \& {Madau}}]{Sullivan2000}
{Sullivan}, M., {Ellis}, R., {Nugent}, P., {Smail}, I., \& {Madau}, P. 2000, \mnras, 319, 549

\bibitem[{{Taubenberger} {et~al.}(2011){Taubenberger}, {Benetti}, {Childress}, {Pakmor}, {Hachinger}, {Mazzali}, {Stanishev}, {Elias-Rosa}, {Agnoletto}, {Bufano}, {Ergon}, {Harutyunyan}, {Inserra}, {Kankare}, {Kromer}, {Navasardyan}, {Nicolas}, {Pastorello}, {Prosperi}, {Salgado}, {Sollerman}, {Stritzinger}, {Turatto}, {Valenti}, \& {Hillebrandt}}]{Taubenberger2011}
{Taubenberger}, S., {Benetti}, S., {Childress}, M., {et~al.} 2011, \mnras, 412, 2735

\bibitem[{{Wevers} {et~al.}(2017){Wevers}, {van Velzen}, {Jonker}, {Stone}, {Hung}, {Onori}, {Gezari}, \& {Blagorodnova}}]{Wevers2017}
{Wevers}, T., {van Velzen}, S., {Jonker}, P.~G., {et~al.} 2017, \mnras, 471, 1694

\bibitem[{Williams(2021)}]{Williamsnessai}
Williams, M.~J. 2021, nessai: Nested Sampling with Artificial Intelligence

\bibitem[{Williams {et~al.}(2021)Williams, Veitch, \& Messenger}]{Williams2021}
Williams, M.~J., Veitch, J., \& Messenger, C. 2021, Phys. Rev. D, 103, 103006

\bibitem[{Williams {et~al.}(2023)Williams, Veitch, \& Messenger}]{Williams2023}
Williams, M.~J., Veitch, J., \& Messenger, C. 2023 [\eprint[arXiv]{2302.08526}]

\bibitem[{{Wojtak} {et~al.}(2019){Wojtak}, {Hjorth}, \& {Gall}}]{Wojtak2019}
{Wojtak}, R., {Hjorth}, J., \& {Gall}, C. 2019, \mnras, 487, 3342

\bibitem[{{Wong} {et~al.}(2020){Wong}, {Suyu}, {Chen}, {Rusu}, {Millon}, {Sluse}, {Bonvin}, {Fassnacht}, {Taubenberger}, {Auger}, {Birrer}, {Chan}, {Courbin}, {Hilbert}, {Tihhonova}, {Treu}, {Agnello}, {Ding}, {Jee}, {Komatsu}, {Shajib}, {Sonnenfeld}, {Blandford}, {Koopmans}, {Marshall}, \& {Meylan}}]{Wong2020}
{Wong}, K.~C., {Suyu}, S.~H., {Chen}, G. C.~F., {et~al.} 2020, \mnras, 498, 1420

\bibitem[{{Woosley}(2010)}]{Woosley2010}
{Woosley}, S.~E. 2010, \apjl, 719, L204

\bibitem[{{Wright} {et~al.}(2016){Wright}, {Robotham}, {Bourne}, {Driver}, {Dunne}, {Maddox}, {Alpaslan}, {Andrews}, {Bauer}, {Bland-Hawthorn}, {Brough}, {Brown}, {Clarke}, {Cluver}, {Davies}, {Grootes}, {Holwerda}, {Hopkins}, {Jarrett}, {Kafle}, {Lange}, {Liske}, {Loveday}, {Moffett}, {Norberg}, {Popescu}, {Smith}, {Taylor}, {Tuffs}, {Wang}, \& {Wilkins}}]{Wright2016a}
{Wright}, A.~H., {Robotham}, A.~S.~G., {Bourne}, N., {et~al.} 2016, \mnras, 460, 765

\bibitem[{{Wright} {et~al.}(2010){Wright}, {Eisenhardt}, {Mainzer}, {Ressler}, {Cutri}, {Jarrett}, {Kirkpatrick}, {Padgett}, {McMillan}, {Skrutskie}, {Stanford}, {Cohen}, {Walker}, {Mather}, {Leisawitz}, {Gautier}, {McLean}, {Benford}, {Lonsdale}, {Blain}, {Mendez}, {Irace}, {Duval}, {Liu}, {Royer}, {Heinrichsen}, {Howard}, {Shannon}, {Kendall}, {Walsh}, {Larsen}, {Cardon}, {Schick}, {Schwalm}, {Abid}, {Fabinsky}, {Naes}, \& {Tsai}}]{Wright2010a}
{Wright}, E.~L., {Eisenhardt}, P.~R.~M., {Mainzer}, A.~K., {et~al.} 2010, \aj, 140, 1868

\bibitem[{{Yang} {et~al.}(2022){Yang}, {Wang}, {Suntzeff}, {Hu}, {Aldoroty}, {Brown}, {Krisciunas}, {Arcavi}, {Burke}, {Galbany}, {Hiramatsu}, {Hosseinzadeh}, {Howell}, {McCully}, {Pellegrino}, \& {Valenti}}]{Yang2022}
{Yang}, J., {Wang}, L., {Suntzeff}, N., {et~al.} 2022, \apj, 938, 83

\end{thebibliography}

\begin{appendix} 

\section{Data for candidates that passed selection criteria}
Table~\ref{table:photz_err} shows the photometric redshift errors from the LS and PS1-STRM catalogues for the 27 candidates from selection method S2 (the photometric sample). As is discussed in Sect.~\ref{subsection:pipeline}, the redshifts in our analysis are averaged from a selection of photometric galaxy catalogues. All our candidates had a photometeric redshift value from either the LS or the PS1-STRM catalogues, and these were the catalogues with the lowest uncertainties typically.  Therefore, we selected the lowest photometric redshift value from either catalogue to propagate to peak absolute magnitude error.

Table A.2 details the measured parameters of the 125 candidates that passed the selection criteria in Sects. \ref{subsection:DESI} and \ref{subsection:cuts}, before the final visual inspection phase. This includes the ZTF ID, right ascension (RA), declination (dec.), photometric redshift (phot. $z$), DESI redshift, angular separation from the catalogue counterpart (ang. sep.) for both the photometric and DESI redshifts, peak absolute $B$-band magnitude from SALT2 (for both the photometric and DESI redshifts), SALT2 $x_1$ and associated error, SALT2 $c$ and associated error, whether it passed selection method 1 with DESI cross-matching (detailed in Sect. \ref{subsection:DESI}) and/or selection method 2 with additional cuts (detailed in Sect. \ref{subsection:cuts}), and the reason it was rejected from the final candidate sample (if any).

\begin{table}[h!]
\caption{Errors on photometric redshifts for the S2 sample}
\label{table:photz_err}
\centering
\begin{tabular}{c | c c c}
\hline\hline
ZTF ID & PS1 phot. $z$ error & LS phot. $z$ error  \\
\hline
ZTF19abamkfs & 0.0359    & 0.2248         \\
ZTF19abctwkl & -         & 0.0538         \\
ZTF19abdkdze & 0.0932    & 0.0303         \\
ZTF19abheyzp & 0.1219    & -              \\
ZTF19abpfedt & 0.0191    & 0.0186         \\
ZTF19abpjicm & 0.0517    & 0.0375         \\
ZTF19adbmdsa & 0.0324    & 0.0137         \\
ZTF20aahgirj & 0.0889    & 0.0232         \\
ZTF20aaiqpgv & 0.0593    & 0.1632         \\
ZTF20abjyrxf & 0.0360    & 0.0327         \\
ZTF20ablmtxz & -         & 0.0825         \\
ZTF20aceekkd & 0.0322    & 0.0325         \\
ZTF20achutix & 0.0895    & 0.0371         \\
ZTF20acotfan & -         & 0.0922         \\
ZTF21aablrfe & 0.0743    & 0.0902         \\
ZTF21aaxvxbu & 0.0232    & 0.0200         \\
ZTF21abawmyd & 0.1088    & 0.0921         \\
ZTF21abcwuhh & 0.0194    & 0.0076         \\
ZTF21acdypeu & 0.0725    & 0.0332         \\
ZTF21acmqvwv & 0.1135    & 0.1735         \\
ZTF22aabfojs & -         & 0.0564         \\
ZTF22aabifrp & -         & 0.0525         \\
ZTF22aadeqlh & 0.0485    & 0.0310         \\
ZTF22aahmovu & 0.0378    & 0.0266         \\
ZTF22aamltry & 0.0573    & -              \\
ZTF22aaoncel & 0.0184    & 0.0259         \\
ZTF22abakgnp & 0.0477    & 0.0293         \\
\hline
\end{tabular}
\tablefoot{The full table that includes the photometric redshift from AMPEL is given in Table A.2 (available at the CDS).}
\end{table}

\onecolumn
\section{Two-image SALT2 fits to gold sample candidates}

\begin{figure*}[h!]
        \centering
        \begin{subfigure}[t]{0.48\textwidth}
            \centering
            \includegraphics[width=\textwidth]{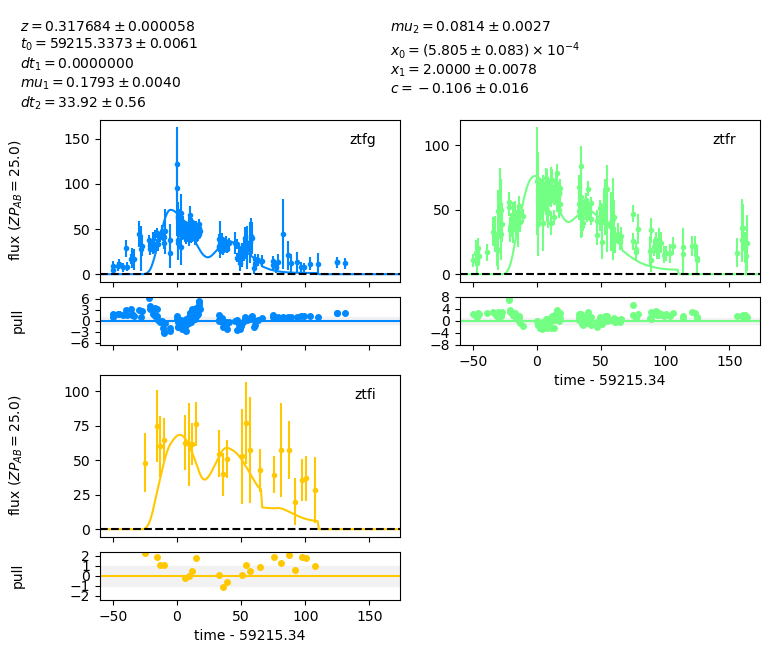}   
        \end{subfigure}
        \begin{subfigure}[t]{0.48\textwidth}  
            \centering 
            \includegraphics[width=\textwidth]{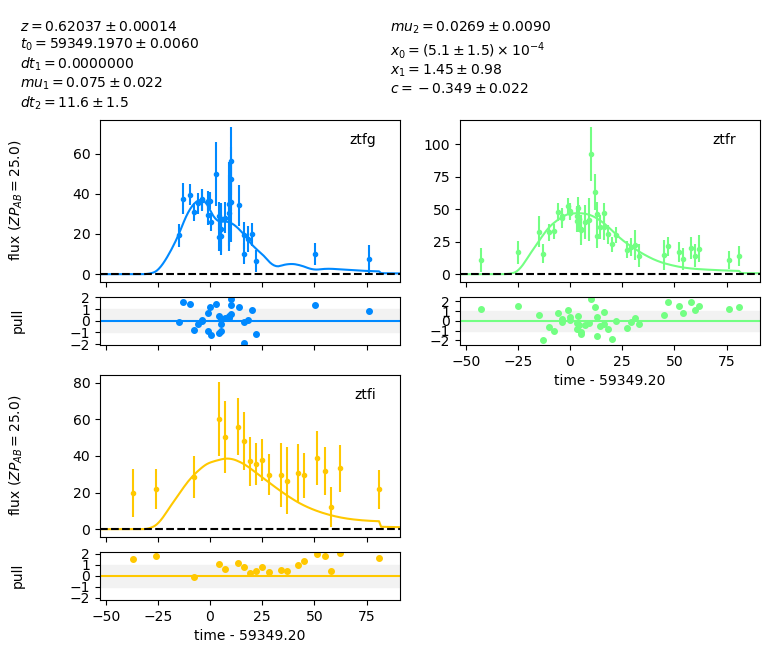}  
        \end{subfigure}
        \begin{subfigure}[t]{0.48\textwidth}  
            \centering 
            \includegraphics[width=\textwidth]{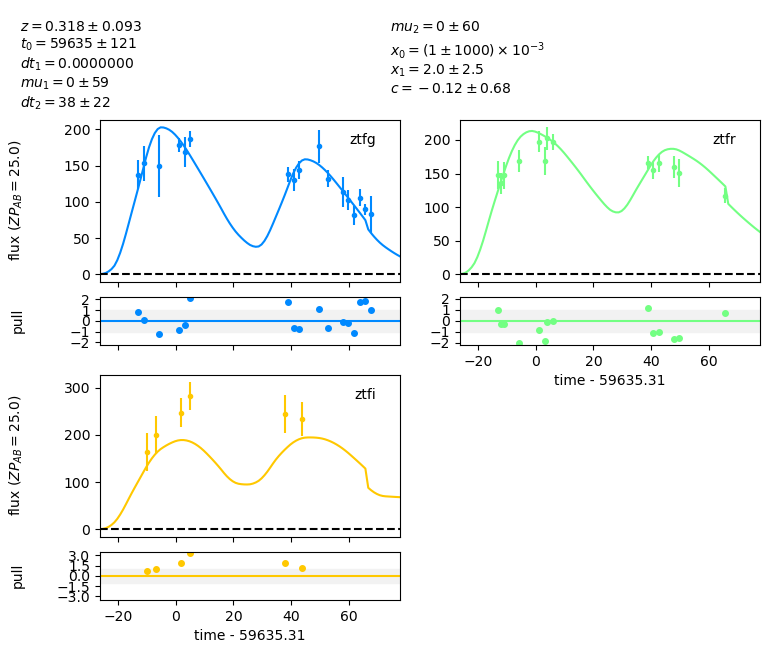}  
        \end{subfigure}
        \begin{subfigure}[t]{0.48\textwidth}  
            \centering 
            \includegraphics[width=\textwidth]{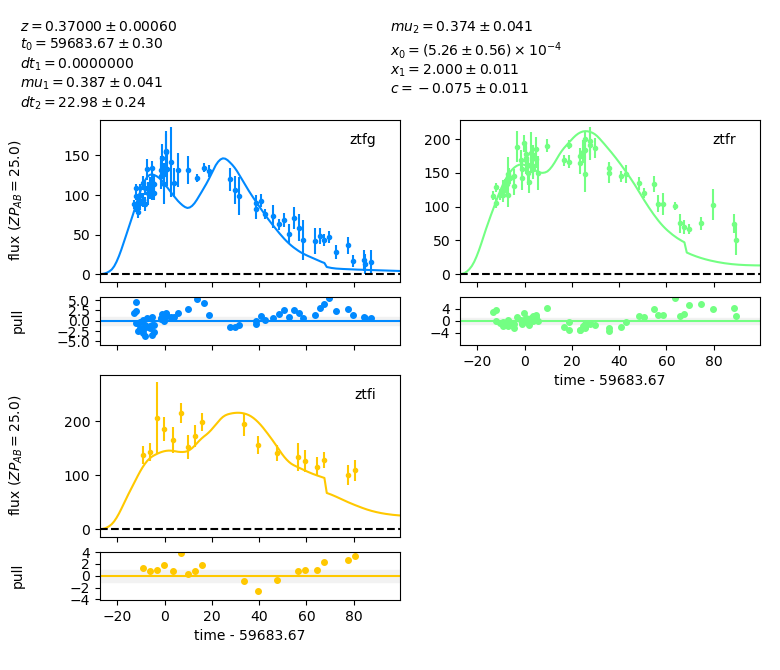}  
        \end{subfigure}
        \caption
        {Two-image combined SALT2 model to forced photometry light curves of ZTF21aablrfe (upper left), ZTF21abcwuhh (upper right), ZTF22aabifrp (lower left), and ZTF22aadeqlh (lower right).
        } 
        \label{figure:badgold_salt2}
    \end{figure*}

As is discussed in Sect.~\ref{subsection:salt2}, we performed fits of the two-image combined SALT2 model to forced photometry light curves of all seven of the gold sample candidates. Three of them converged to reasonable fits: ZTF19abpjicm, ZTF20abjyrxf, and ZTF22aahmovu. The remaining four objects in the gold sample have poor combined two-image SALT2 fits because they are too wide to
only display a single peak (ZTF21aablrfe, ZTF22aabifrp, and ZTF22aadeqlh) or they are too noisy to provide a convincing
fit (ZTF21abcwuhh). Figure~\ref{figure:badgold_salt2} shows the fits for these four objects.

\FloatBarrier 

\onecolumn
\section{Two-image SALT2 fits to SLSNe and TDEs in BTS}
\begin{figure*}[h!]
        \centering
        \begin{subfigure}[t]{0.48\textwidth}
            \centering
            \includegraphics[width=\textwidth]{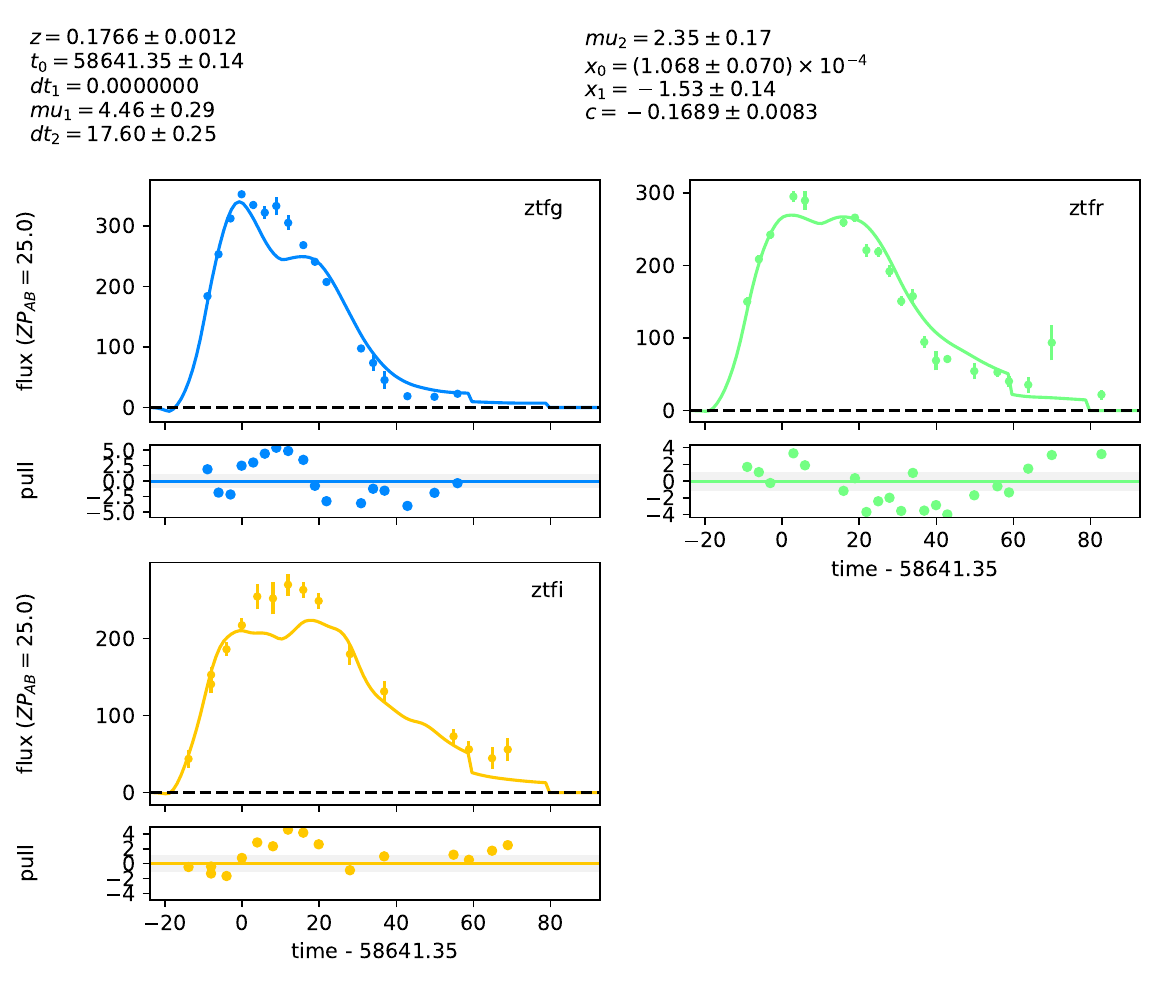}   
        \end{subfigure}
        \begin{subfigure}[t]{0.48\textwidth}  
            \centering 
            \raisebox{18.5ex}{ 
                \includegraphics[width=\textwidth]{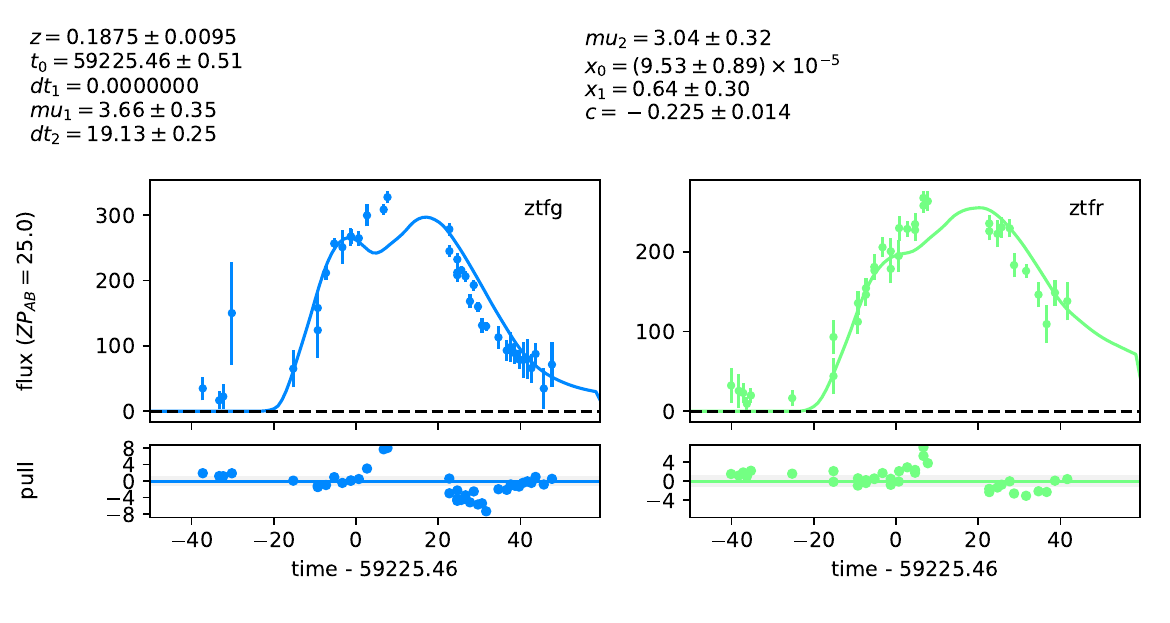}
            }  
        \end{subfigure}
        \caption
        {Two-image combined SALT2 model to forced photometry light curves of the SLSNe-I, ZTF19aavouyw (left) and ZTF21aaarmti (right).
        } 
        \label{figure:btsslsn_salt2}
    \end{figure*}

\begin{figure}[h!]
   \centering
   \includegraphics[width=0.48\hsize]{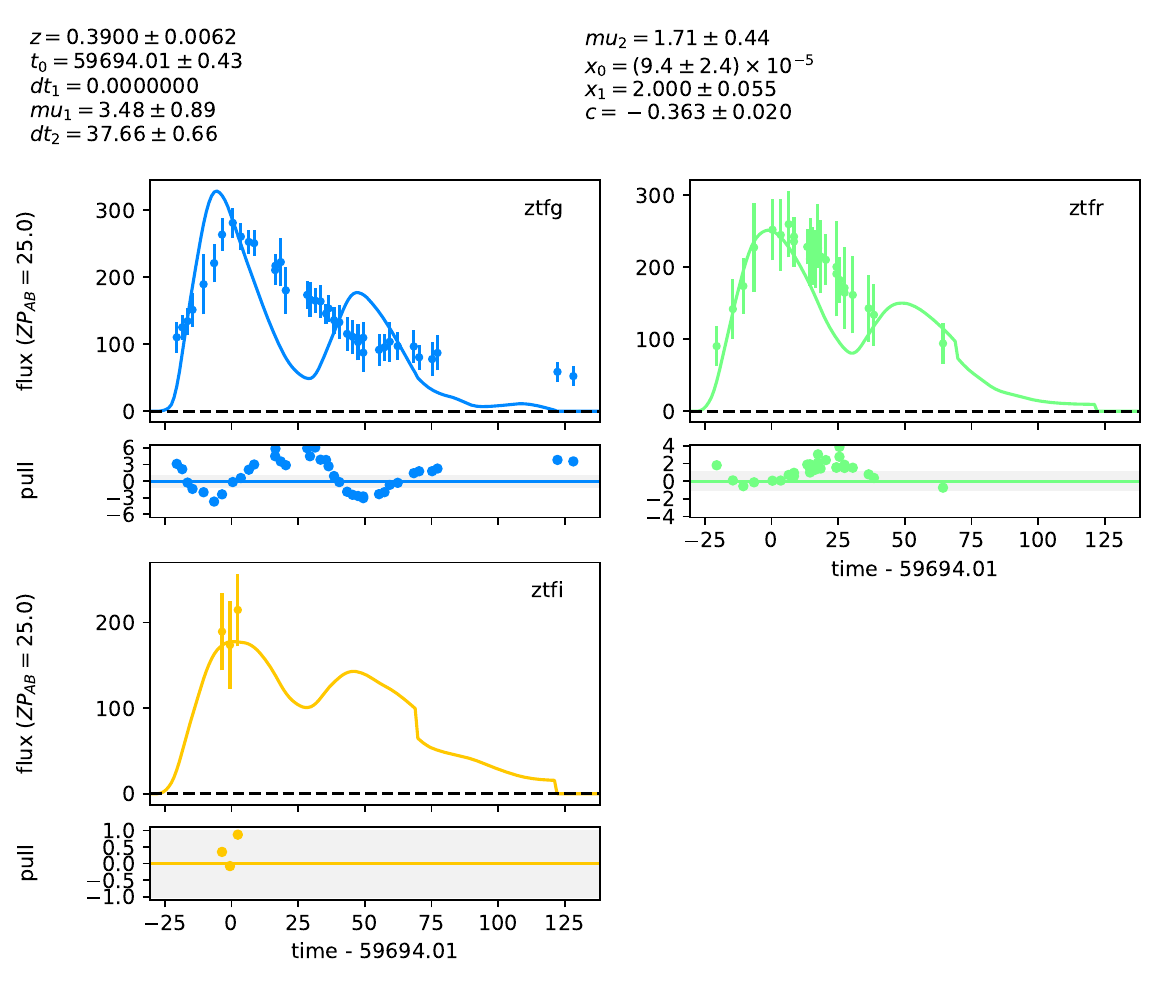}
      \caption{Two-image combined SALT2 model to the forced photometry light curve of the TDE, ZTF22aagvrlq.
              }
         \label{figure:btstde_salt2}
   \end{figure}

As is discussed in Sect.~\ref{subsubsection:twoimage_SLSN}, we performed fits of the two-image combined SALT2 model to forced photometry light curves of SLSNe and TDEs from the BTS. Figure~\ref{figure:btsslsn_salt2} shows the fits for ZTF19aavouyw and ZTF21aaarmti, the two SLSNe-I with a reduced $\chi^2$ of less than five. Figure~\ref{figure:btstde_salt2} shows the fit for ZTF22aagvrlq, the only TDE with a reduced $\chi^2$ of less than five.

\FloatBarrier

\section{SLSN and TDE model fit with \texttt{redback}}
\begin{figure*}[h!]
        \centering
        \begin{subfigure}[b]{0.9\textwidth}
            \centering
            \includegraphics[width=\textwidth]{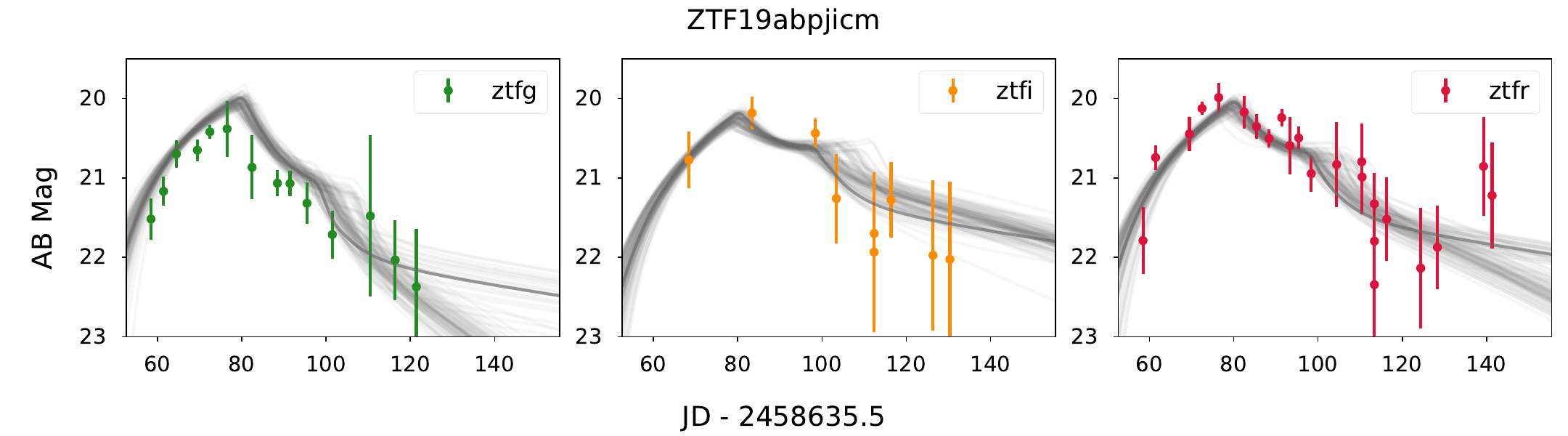}   
        \end{subfigure}
        \begin{subfigure}[b]{0.9\textwidth}
            \centering
            \includegraphics[width=\textwidth]{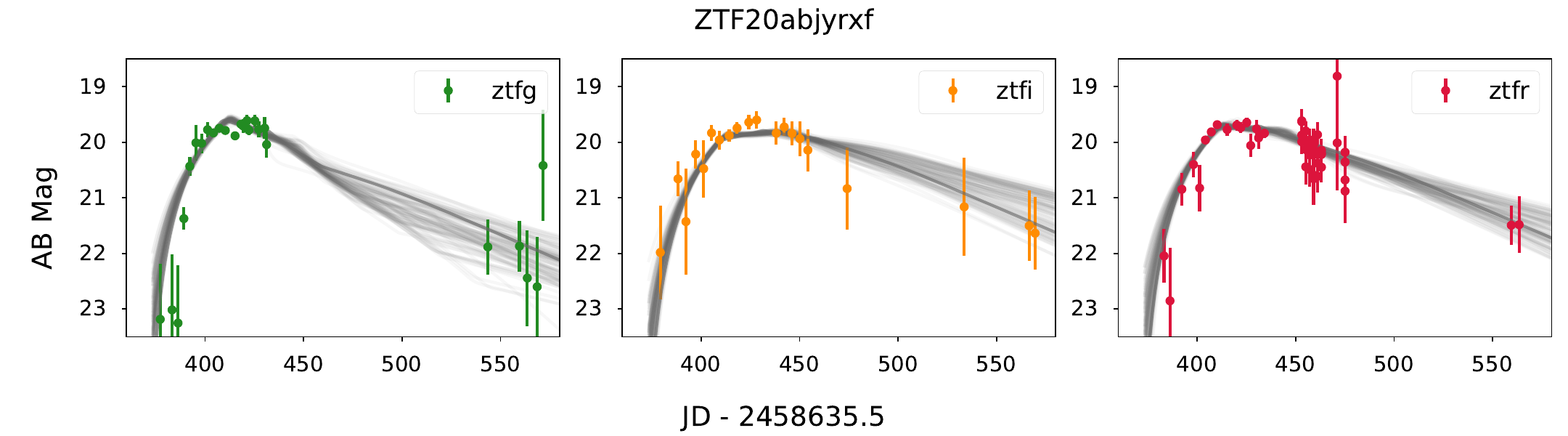}   
        \end{subfigure}
        \begin{subfigure}[b]{0.9\textwidth}
            \centering
            \includegraphics[width=\textwidth]{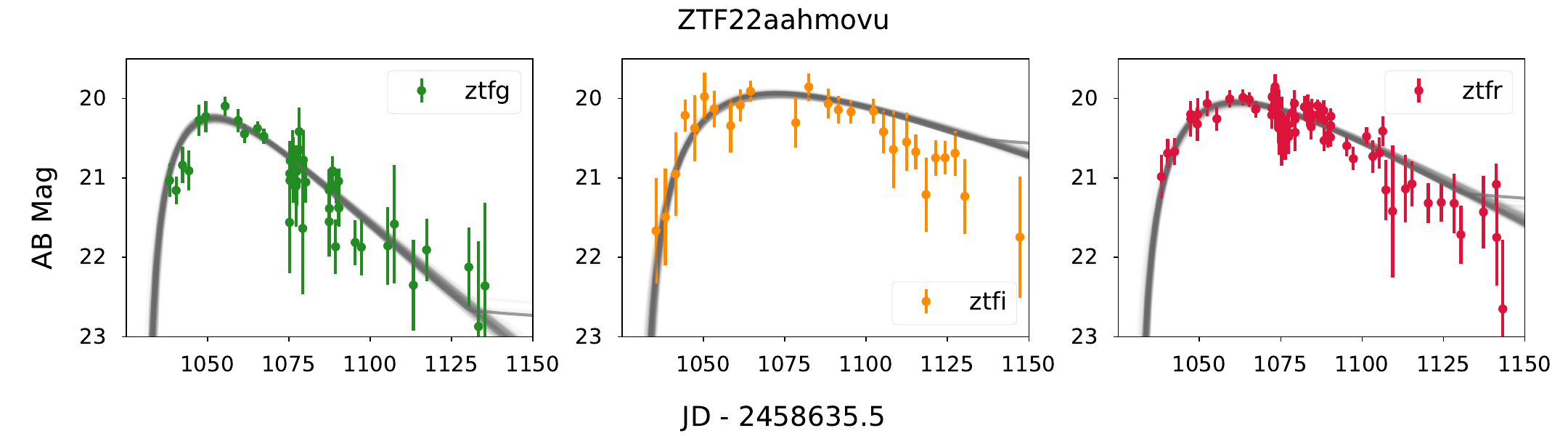}   
        \end{subfigure}
        \caption
        {CSM and nickel model fits to ZTF19abpjicm (top), ZTF20abjyrxf (middle), and ZTF22aahmovu (bottom) with \texttt{redback}. The grey lines correspond to one hundred random models randomly drawn from the posterior and the maximum likelihood model.
        } 
        \label{figure:gold_slsn}
    \end{figure*}

\begin{figure*}[h!]
        \centering
        \begin{subfigure}[b]{0.9\textwidth}
            \centering
            \includegraphics[width=\textwidth]{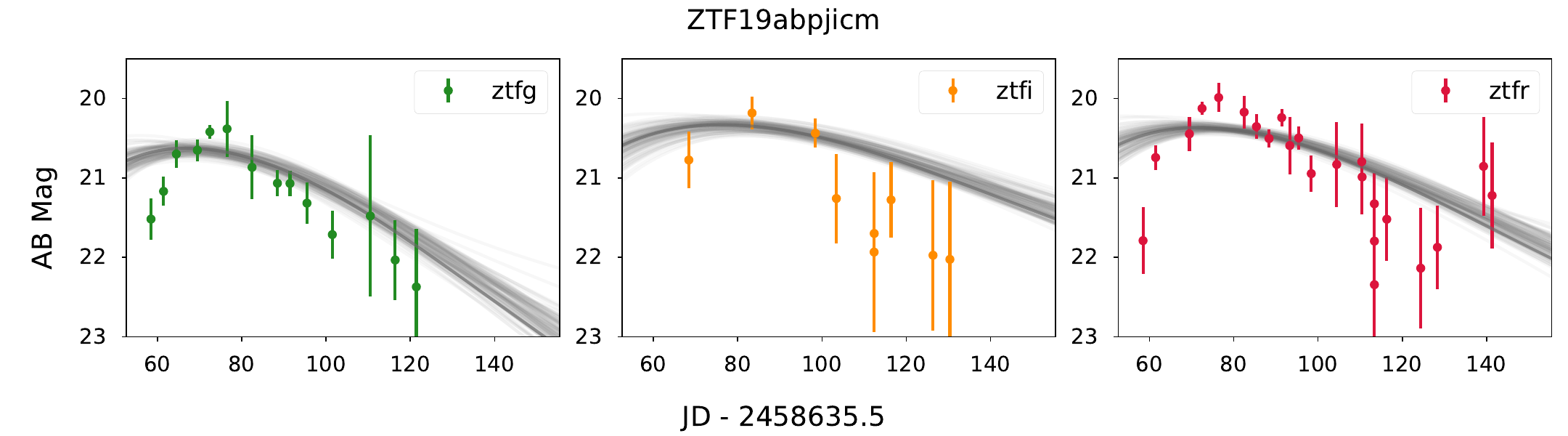}   
        \end{subfigure}
        \begin{subfigure}[b]{0.9\textwidth}
            \centering
            \includegraphics[width=\textwidth]{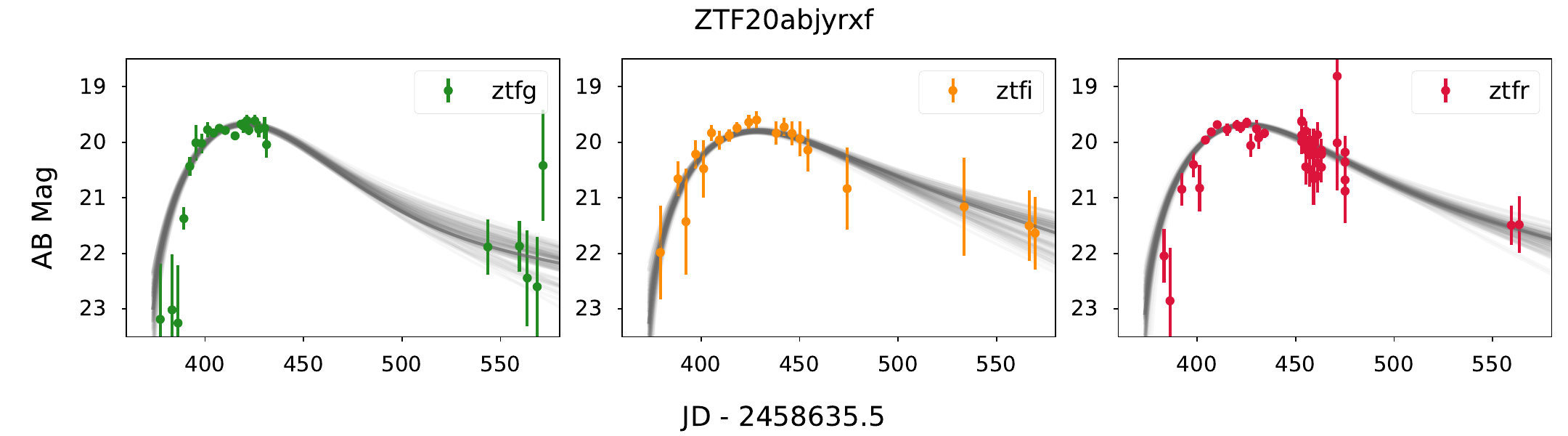}   
        \end{subfigure}
        \begin{subfigure}[b]{0.9\textwidth}
            \centering
            \includegraphics[width=\textwidth]{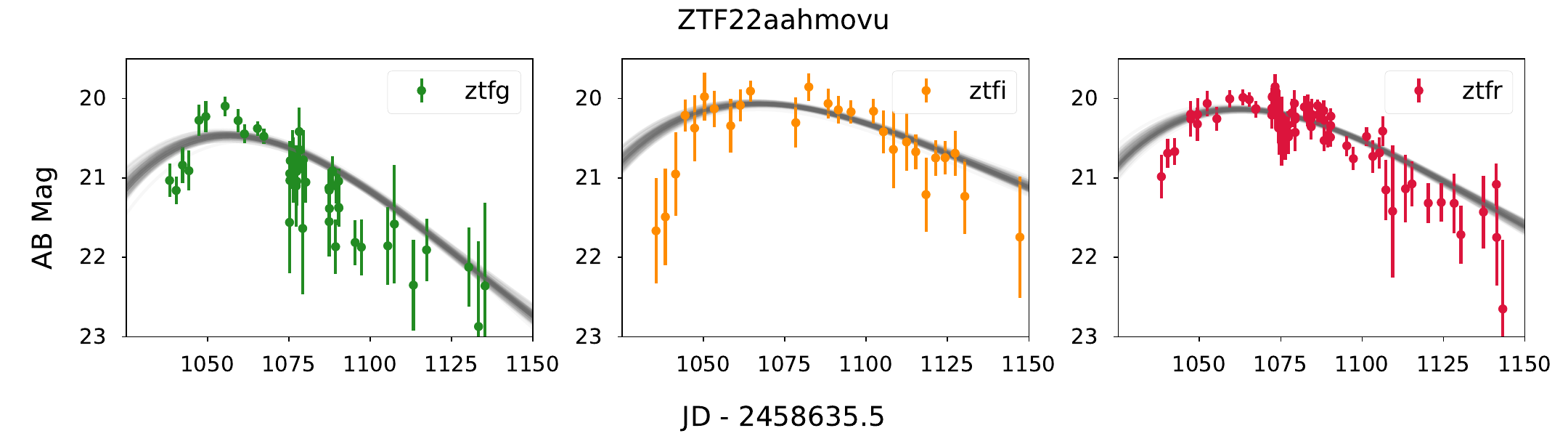}   
        \end{subfigure}
        \caption
        {TDE model fits to ZTF19abpjicm (top), ZTF20abjyrxf (middle), and ZTF22aahmovu (bottom) with \texttt{redback}. The grey lines correspond to one hundred random models randomly drawn from the posterior and the maximum likelihood model.
        } 
        \label{figure:gold_tde}
    \end{figure*}

As is described in Sect.~\ref{subsubsection:csm_tde_model_fit}, we performed fits to a SLSN-II model (with a powering mechanism that combines CSI and nickel) and a TDE model for the three candidates that had good fits to the two-image SALT2 model. The model fit light curves are shown in Fig.~\ref{figure:gold_slsn} and Fig.~\ref{figure:gold_tde} for the SLSN-II and TDE models, respectively.

\FloatBarrier
\twocolumn

\section{Galaxy photometry of gold sample candidates}

Tables \ref{table:host_phot1} to \ref{table:host_phot7} show the photometry we acquired for the potential lens or host galaxies of the seven candidates in the gold sample (as is described in Sect. \ref{subsection:host_obs}). From this photometry, we calculated the stellar masses and star formation rate for each candidate.

\begin{table}[h!]
  \caption{Host photometry of ZTF19abpjicm\label{table:host_phot1}}
  \centering
  \begin{tabular}{lcc}
\hline\hline
ZTF19abpjicm\\
\hline
GALEX/$FUV$	&$22.92\pm0.42$\\
GALEX/$NUV$	&$21.35\pm0.14$\\
SDSS/$u$	  &$21.15\pm0.60$\\
SDSS/$g$	  &$20.14\pm0.06$\\
SDSS/$i$	  &$18.56\pm0.05$\\
SDSS/$r$	  &$19.10\pm0.03$\\
SDSS/$z$	  &$18.07\pm0.10$\\
PS1/$g$	    &$19.97\pm0.15$\\
PS1/$r$	    &$19.09\pm0.06$\\
PS1/$i$	    &$18.49\pm0.03$\\
PS1/$z$	    &$18.21\pm0.04$\\
PS1/$y$	    &$17.93\pm0.15$\\
LS/$g$	    &$19.89\pm0.02$\\
LS/$r$	    &$18.86\pm0.01$\\
LS/$z$	    &$18.01\pm0.02$\\
2MASS/$J$	  &$17.82\pm0.18$\\
2MASS/$H$	  &$16.77\pm0.11$\\
2MASS/$K$	  &$17.00\pm0.16$\\
WISE/$W1$	&$17.36\pm0.04$\\
WISE/$W2$	&$17.47\pm0.03$\\
\hline
\end{tabular}
\tablefoot{All measurements are in the AB system and are not corrected for reddening.}
\end{table}

\begin{table}[h!]
  \caption{Host photometry of ZTF20abjyrxf\label{table:host_phot2}}
  \centering
\begin{tabular}{lcc}
\hline\hline
ZTF20abjyrxf\\
\hline
SDSS/$u$	  &$21.29\pm0.91$\\
SDSS/$g$	  &$19.99\pm0.08$\\
SDSS/$r$	  &$19.09\pm0.06$\\
SDSS/$i$	  &$18.67\pm0.06$\\
SDSS/$z$	  &$18.24\pm0.16$\\
PS1/$g$	    &$19.92\pm0.12$\\
PS1/$r$	    &$19.22\pm0.08$\\
PS1/$i$	    &$18.72\pm0.04$\\
PS1/$z$	    &$18.46\pm0.06$\\
PS1/$y$	    &$18.24\pm0.17$\\
LS/$g$	    &$19.84\pm0.03$\\
LS/$r$	    &$18.94\pm0.03$\\
LS/$z$	    &$18.32\pm0.02$\\
WISE/$W1$	&$18.36\pm0.06$\\
WISE/$W2$	&$18.87\pm0.09$\\
\hline
\end{tabular}
\tablefoot{All measurements are in the AB system and are not corrected for reddening.}
\end{table}

\begin{table}[h!]
  \caption{Host photometry of ZTF21aablrfe\label{table:host_phot3}}
  \centering
\begin{tabular}{lcc}
\hline\hline
ZTF21aablrfe		\\
\hline
GALEX/$FUV$	&$23.20\pm0.27$\\
GALEX/$NUV$	&$22.83\pm0.20$\\
SDSS/$u$	  &$22.33\pm0.54$\\
SDSS/$g$	  &$20.89\pm0.09$\\
SDSS/$r$	  &$19.72\pm0.05$\\
SDSS/$i$	  &$19.16\pm0.04$\\
SDSS/$z$	  &$18.55\pm0.07$\\
PS1/$g$	    &$20.63\pm0.14$\\
PS1/$r$	    &$19.63\pm0.06$\\
PS1/$i$	    &$19.15\pm0.04$\\
PS1/$z$	    &$18.81\pm0.04$\\
PS1/$y$	    &$18.77\pm0.16$\\
LS/$g$	    &$20.61\pm0.03$\\
LS/$r$	    &$19.51\pm0.03$\\
LS/$z$	    &$18.61\pm0.02$\\
2MASS/$J$	  &$18.28\pm0.19$\\
2MASS/$H$	  &$18.03\pm0.25$\\
2MASS/$K$	  &$17.49\pm0.13$\\
WISE/$W1$	&$17.61\pm0.06$\\
WISE/$W2$	&$17.73\pm0.05$\\
\hline
\end{tabular}
\tablefoot{All measurements are in the AB system and are not corrected for reddening.}
\end{table}

\begin{table}[h!]
  \caption{Host photometry of ZTF21abcwuhh\label{table:host_phot4}}
  \centering
\begin{tabular}{lcc}
\hline\hline
ZTF21abcwuhh		\\
\hline
SDSS/$u$	  &$20.12\pm0.30$\\
SDSS/$g$	  &$19.40\pm0.08$\\
SDSS/$r$	  &$18.17\pm0.04$\\
SDSS/$i$	  &$17.62\pm0.04$\\
SDSS/$z$	  &$17.07\pm0.07$\\
PS1/$g$	    &$19.65\pm0.19$\\
PS1/$r$	    &$18.22\pm0.06$\\
PS1/$i$	    &$17.62\pm0.05$\\
PS1/$z$	    &$17.39\pm0.04$\\
PS1/$y$	    &$17.11\pm0.05$\\
LS/$g$	    &$19.66\pm0.08$\\
LS/$r$	    &$17.98\pm0.05$\\
LS/$z$	    &$17.28\pm0.03$\\
2MASS/$H$	  &$16.15\pm0.06$\\
2MASS/$K$	  &$16.01\pm0.06$\\
WISE/$W1$	&$16.81\pm0.03$\\
WISE/$W2$	&$17.29\pm0.03$\\
\hline
\end{tabular}
\tablefoot{All measurements are in the AB system and are not corrected for reddening.}
\end{table}

\begin{table}[h!]
  \caption{Host photometry of ZTF22aabifrp\label{table:host_phot5}}
  \centering
\begin{tabular}{lcc}
\hline\hline
ZTF22aabifrp		\\
\hline
PS1/$g$	    &$21.24\pm0.18$\\
PS1/$r$	    &$20.22\pm0.13$\\
PS1/$i$	    &$19.89\pm0.15$\\
PS1/$z$	    &$19.56\pm0.11$\\
PS1/$y$	    &$19.37\pm0.15$\\
LS/$g$	    &$21.04\pm0.18$\\
LS/$r$	    &$20.01\pm0.14$\\
LS/$z$	    &$19.34\pm0.10$\\
2MASS/$J$	  &$18.87\pm0.40$\\
2MASS/$H$	  &$18.08\pm0.20$\\
2MASS/$K$	  &$17.96\pm0.28$\\
\hline
\end{tabular}
\tablefoot{All measurements are in the AB system and are not corrected for reddening.}
\end{table}

\begin{table}[h!]
  \caption{Host photometry of ZTF22aadeqlh\label{table:host_phot6}}
  \centering
\begin{tabular}{lcc}
\hline\hline
ZTF22aadeqlh		\\
\hline
GALEX/$FUV$	&$21.96\pm0.59$\\
GALEX/$NUV$	&$20.53\pm0.16$\\
SDSS/$u$	  &$20.57\pm0.21$\\
SDSS/$g$	  &$19.62\pm0.03$\\
SDSS/$r$	  &$18.84\pm0.02$\\
SDSS/$i$	  &$18.49\pm0.04$\\
SDSS/$z$	  &$18.03\pm0.09$\\
PS1/$g$	    &$19.52\pm0.05$\\
PS1/$r$	    &$18.81\pm0.03$\\
PS1/$i$	    &$18.57\pm0.03$\\
PS1/$z$	    &$18.30\pm0.06$\\
PS1/$y$	    &$18.24\pm0.06$\\
LS/$g$	    &$19.51\pm0.01$\\
LS/$r$	    &$18.67\pm0.01$\\
LS/$z$	    &$18.14\pm0.02$\\
\hline
\end{tabular}
\tablefoot{All measurements are in the AB system and are not corrected for reddening.}
\end{table}

\begin{table}[h!]
  \caption{Host photometry of ZTF22aahmovu\label{table:host_phot7}}
  \centering
\begin{tabular}{lcc}
\hline\hline
ZTF22aahmovu\\
\hline
SDSS/$u$	  &$21.23\pm0.35$\\
SDSS/$g$	  &$19.56\pm0.05$\\
SDSS/$r$	  &$18.68\pm0.02$\\
SDSS/$i$	  &$18.26\pm0.03$\\
SDSS/$z$	  &$18.00\pm0.10$\\
PS1/$g$	    &$19.64\pm0.08$\\
PS1/$r$	    &$18.74\pm0.05$\\
PS1/$i$	    &$18.27\pm0.04$\\
PS1/$z$	    &$17.98\pm0.05$\\
PS1/$y$	    &$17.79\pm0.14$\\
LS/$g$	    &$19.54\pm0.02$\\
LS/$r$	    &$18.53\pm0.02$\\
LS/$z$	    &$17.87\pm0.02$\\
2MASS/$J$	  &$17.81\pm0.21$\\
2MASS/$H$	  &$17.71\pm0.32$\\
WISE/$W1$	&$17.65\pm0.04$\\
WISE/$W2$	&$17.93\pm0.04$\\
\hline
\end{tabular}
\tablefoot{All measurements are in the AB system and are not corrected for reddening.}
\end{table}

\end{appendix}

\end{document}